\documentclass{aa}
\usepackage{float}
\usepackage{natbib}
\usepackage{graphicx}
\usepackage[varg]{txfonts}
\usepackage{color}
\usepackage{times}
\usepackage{hyperref}
\hypersetup{colorlinks=true, urlcolor=blue, linkcolor=blue,  citecolor=blue}
\usepackage{amsmath}
\usepackage{bm}
\usepackage{multirow}
\usepackage[warn]{textcomp}

\usepackage{newtxtext}

\begin{document}

  \title{Active galactic nucleus activity in nearby disk galaxies: The roles of bar strength and spiral arm morphology}

  \author{Zh.~R.~Martirosyan\inst{\ref{inst1}}\email{jaklin@ysu.am}
          \and
          A.~A.~Hakobyan\inst{\ref{inst2},}\corrauth{artur.hakobyan@yerphi.am}
          \and
          V.~Adibekyan\inst{\ref{inst3},\ref{inst4}}\email{vardan.adibekyan@astro.up.pt}
          \and
          L.~V.~Barkhudaryan\inst{\ref{inst2}}\email{l.barkhudaryan@yerphi.am}
          \and
          M.~H.~Gevorgyan\inst{\ref{inst2}}\email{m.gevorgyan@yerphi.am}
          \and
          A.~G.~Karapetyan\inst{\ref{inst2}}\email{a.karapetyan@yerphi.am}
          }

  \institute{
          Institute of Physics, Yerevan State University, 1 Alex Manoogian Str., 0025 Yerevan, Armenia\label{inst1}
          \and
          Center for Cosmology and Astrophysics, Alikhanian National Science Laboratory, 2 Alikhanian Brothers Str., 0036 Yerevan, Armenia\label{inst2}
          \and
          Instituto de Astrof\'isica e Ci\^encias do Espa\c{c}o, Universidade do Porto, CAUP, Rua das Estrelas, 4150-762 Porto, Portugal\label{inst3}
          \and
          Departamento de F\'{\i}sica e Astronomia, Faculdade de Ci\^encias, Universidade do Porto, Rua do Campo  Alegre, 4169-007 Porto, Portugal\label{inst4}
          }

  \date{Received 12 June 2026 /
        Accepted 02 September 2026}

  \abstract
   {The transport of gas toward galactic centers and its connection to active galactic nucleus (AGN) activity remain key open questions in galaxy evolution. In particular, the relative roles of stellar bars and spiral arms and their possible combined influence on angular momentum redistribution and supermassive black hole fueling are not yet fully understood.}
   {We investigate the dependence of AGN activity on bar strength and spiral arm morphology in nearby spiral galaxies, with a particular emphasis on disentangling their individual and joint effects while controlling for galaxy stellar mass and color.}
   {We constructed a sample of 843 morphologically undisturbed Sa--Sd galaxies from the Mapping Nearby Galaxies at Apache Point Observatory survey in the redshift range $0.026 \leq z \leq 0.1$. Bar strengths, stellar masses, colors, and AGN classes were adopted from the literature, while spiral arm classes (ACs), i.e., flocculent (FL), multi-armed (MA), and grand design (GD) systems, were visually determined for the entire sample using optical imaging together with bulge--disk decomposition residual maps.}
   {The AGN fraction increases systematically with bar strength, from $0.10^{+0.02}_{-0.02}$ in unbarred galaxies to $0.34^{+0.03}_{-0.03}$ in strongly barred systems. After controlling for stellar mass and color, a statistically significant enhancement remains detectable primarily at intermediate stellar masses ($10.5 \lesssim \log(M_\star/M_\odot) \lesssim 11.0$), whereas no significant dependence is found at higher masses. In contrast, spiral arm morphology alone does not show a robust independent connection with AGN activity once stellar mass and color are controlled, although a weak enhancement is present in intermediate-mass GD and MA systems. When bar strength and spiral AC are considered jointly, the highest AGN fraction is observed in strongly barred GD$+$MA galaxies $(0.39^{+0.04}_{-0.04})$, while the lowest AGN fractions are found in unbarred FL systems $(0.09^{+0.03}_{-0.02})$.}
   {Our results suggest that stellar mass and color drive the primary trend in AGN activity in nearby disk galaxies, while stellar bars and spiral arms act as secondary structural drivers. The effect of spiral arm morphology is weaker and becomes most apparent when considered jointly with bars. These findings highlight the importance of simultaneously accounting for stellar mass, color, and the combined large-scale morphological structure of galaxies when investigating the triggering of nuclear activity.}

  \keywords{galaxies: spiral -- galaxies: structure -- galaxies: photometry --
          galaxies: active -- galaxies: statistics}

\titlerunning{Exploring AGN activity}
\authorrunning{Martirosyan et al.}

\maketitle
\nolinenumbers

\section{Introduction}
\label{intro}

Active galactic nuclei (AGNs) are powered by the accretion of gas onto supermassive black holes (SMBHs) residing at the centers of galaxies.
Although SMBHs are thought to be ubiquitous in massive galaxies, only a fraction exhibit significant nuclear activity.
This suggests that AGN phases are regulated by the efficiency with which gas loses angular momentum and is transported from kiloparsec scales to the immediate vicinity of the SMBH.

In disk galaxies, large-scale non-axisymmetric structures are among the most efficient mechanisms for redistributing angular momentum.
In particular, stellar bars can drive substantial gas inflows through gravitational torques, funneling material from the outer disk toward the central kiloparsec and thereby contributing to secular evolution \citep[e.g.,][]{1989Natur.338...45S,2004ARA&A..42..603K}.
From a theoretical perspective, gas inflow may proceed through a cascade of instabilities operating across multiple spatial scales, ultimately enabling material to reach the nuclear region and fuel the SMBH \citep[e.g.,][]{2010MNRAS.407.1529H}.

Observationally, however, the connection between bars and AGN activity
remains debated. Early studies reported enhanced AGN fractions in
barred galaxies \citep[e.g.,][]{2000ApJ...529...93K,2002ApJ...567...97L}.
Similarly, \citet{2013A&A...549A.141A} found that barred AGN hosts tend
to exhibit enhanced nuclear activity and higher accretion indicators
than unbarred systems. In contrast, analyses based on larger and more
carefully controlled samples have shown that the apparent bar--AGN
correlation weakens or disappears once host galaxy properties such as
stellar mass, color, morphology, and environment are taken into account
\citep[e.g.,][]{2012ApJ...750..141L,2015MNRAS.448.3442G}. These results suggest
that the presence of a bar alone is insufficient to determine whether
a galaxy hosts an AGN and that additional galaxy properties may play
an important role in regulating SMBH fueling.

Recent studies have further demonstrated that the impact of bars
depends not only on their presence but also on their strength and
dynamical properties. Strong bars are associated with
more efficient gas transport toward galaxy centers through
angular momentum redistribution and with more rapid
quenching of star formation in disk galaxies
\citep[e.g.,][]{2021MNRAS.507.4389G}. Spatially resolved analyses have revealed
systematic differences in the stellar and gas kinematics of weakly and
strongly barred galaxies, indicating that stronger bars induce larger
departures from axisymmetric motions and more efficient radial transport
processes \citep[e.g.,][]{2023MNRAS.521.1775G,2024ApJ...973..129G}. Consistent
with this picture, recent studies suggest that bars primarily increase
the probability of AGN occurrence rather than directly regulating the
accretion rate itself, with the strongest effects observed in specific
stellar-mass regimes \citep[e.g.,][]{2024MNRAS.532.2320G,2025A&A...699A.204M,2026A&A...707A.152L}.
Spatially resolved kinematic studies have likewise
demonstrated that barred galaxies exhibit prominent bisymmetric velocity
perturbations and organized gas flows, highlighting the dynamical role
of bars in driving secular evolution and central gas transport
\citep[e.g.,][]{2024ApJ...973..116D}.

Spiral arms provide an additional and potentially complementary
mechanism for angular momentum transport in disk galaxies.
According to classical density wave theory, spiral perturbations can
induce shocks, orbit crowding, and angular momentum loss in the
interstellar medium, thereby facilitating radial gas transport toward
smaller galactocentric radii
\citep[e.g.,][]{1969ApJ...158..123R,1972ApL....11...41K,2016ARA&A..54..667S}.
Observational studies have shown that spiral structure plays an
important role in regulating molecular gas distributions, central gas
concentrations, and star formation efficiency across galactic disks
\citep[e.g.,][]{2022A&A...666A.175Y,2024A&A...687A.293Q,2024AJ....168...12S,2025A&A...698A.296R}.
Spiral galaxies span a broad morphological sequence, from
grand design (GD) to flocculent (FL) systems, with multi-armed (MA)
spirals representing an intermediate class
\citep[e.g.,][]{1987ApJ...314....3E,1990NYASA.596...40E}.
These spiral arm classes (ACs) are thought to reflect different dynamical
regimes and levels of global organization within galactic disks,
which may influence the efficiency of gas redistribution and the
delivery of material toward the central regions. Despite their
potential importance for SMBH fueling, the connection between spiral
structure and AGN activity remains significantly less explored than
that of stellar bars.
In addition, disk instabilities may also contribute to gas inflow
and the fueling of nuclear activity.

Importantly, bars and spiral arms are not independent structures
but dynamically coupled components of disk galaxies.
Theoretical models and observations both suggest that bars can excite
or reinforce spiral density waves, while spiral arms may regulate the
efficiency of bar-driven inflows by redistributing angular momentum
throughout the disk \citep[e.g.,][]{2009AJ....137.4487B,2024AJ....168...12S}.
Results from the Extragalactic Database for Galaxy Evolution--Calar Alto
Legacy Integral Field Area (EDGE--CALIFA) survey further indicate that both bars
and spiral arms contribute to the buildup of central molecular gas
concentrations, although their relative importance depends on the
properties and evolutionary state of the host galaxy
\citep{2022A&A...666A.175Y}.
More recently, studies of nearby galaxies have shown that bars and
spiral structure jointly influence the redistribution of gas and metals
across galactic disks, supporting a picture in which secular evolution
is governed by the interplay of multiple non-axisymmetric structures
rather than by bars alone \citep[e.g.,][]{2025ApJ...983...57W}.
This coupling raises the possibility that the effectiveness of
bar-driven fueling depends on the surrounding spiral structure,
potentially contributing to the diversity of results reported in
previous bar--AGN studies.

The advent of large integral-field unit (IFU) surveys has
enabled direct links to be established between galaxy morphology,
stellar populations, gas kinematics, and nuclear activity.
In particular, the Mapping Nearby Galaxies at Apache Point Observatory (MaNGA) survey \citep{2015ApJ...798....7B} has provided spatially resolved spectroscopic data for
thousands of nearby galaxies, opening new opportunities to investigate
the physical processes governing galaxy evolution
\citep{2019MNRAS.489.1338P}.
Combined with modern imaging data and detailed morphological
classifications, these observations provide an ideal framework for
exploring how large-scale structural components, such as bars and
spiral arms, influence gas transport and AGN activity.

Despite these advances, a systematic investigation of the combined
influence of bar strength and spiral arm morphology on AGN activity
remains lacking. Most previous studies have considered bars and
spiral structure separately, making it difficult to disentangle their
individual and joint contributions to SMBH fueling. In this work,
we investigate the connection between AGN activity, bar strength,
and spiral structure in nearby disk galaxies using spatially resolved
spectroscopy from the MaNGA survey. We combine literature-based
measurements of bar strength, stellar mass, color, and nuclear
activity with new visual classifications of spiral ACs
performed for the entire galaxy sample. This enabled us to examine,
for the first time, the joint dependence of AGN activity on bar
strength and spiral arm morphology while explicitly controlling
for stellar mass and color.

The paper is organized as follows.
In Sect.~\ref{sec:sample} we describe the sample selection and data reduction and the
morphological, bar, AGN, and arm classifications.
Sect.~\ref{sec:results} presents the analysis of the AGN fraction
as a function of bar strength and spiral AC, and we discuss
the implications for SMBH fueling.
Finally, Sect.~\ref{sec:concl} presents a summary of our main conclusions.

\section{Sample selection and data reduction}
\label{sec:sample}
\defcitealias{2025MNRAS.544.1056V}{V25}
\defcitealias{2022ApJS..262...36S}{S22}
\defcitealias{2023A&A...674A..85A}{A23}

Our primary sample was drawn from the MaNGA survey \citep[][hereafter S22]{2022ApJS..262...36S}, which provides IFU spectroscopy for a large number of nearby galaxies. This data set was supplemented with high-resolution imaging, allowing robust morphological decomposition \citep[][hereafter V25]{2025MNRAS.544.1056V} and spiral arm classification  (see Sect.~\ref{SSandR2}).
To assess nuclear activity, we adopted AGN classifications (Sect.~\ref{SSandR3}) from \citet[][hereafter A23]{2023A&A...674A..85A}. After establishing the base sample, a visual classification of spiral arm morphologies was carried out for all galaxies (Sect.~\ref{SSandR4}).
In Sect.~\ref{SSandR5}, we summarize the main physical and morphological properties of the resulting sample, which provide a foundation for the analysis presented in the following sections.

\subsection{Morphological classification of galaxies}
\label{SSandR2}

The MaNGA survey is one of the three core programs of Sloan Digital Sky Survey (SDSS) - IV \citep{2017AJ....154...28B}.
MaNGA is an IFU spectroscopic survey that provides spatially resolved spectroscopy over the wavelength range $3622$--$10354$\,\AA\,with a spectral resolution of $R \sim 2000$. Observations are conducted with the 2.5~m Sloan Telescope \citep{2006AJ....131.2332G}. The IFU field of view varies from 12 to 32~arcsec in diameter, allowing coverage to reach at least $1.5$ effective radii ($R_{\rm e}$) for $\sim 80\%$ of the targets and up to $2.5\,R_{\rm e}$ for a subset of galaxies.

The SDSS Data Release 17 \citep{2022ApJS..259...35A} contains $\sim 10,000$ high-quality MaNGA data cubes with unique IDs, within the redshift range $0.01 \lesssim z \lesssim 0.15$ and stellar masses $M_\star \gtrsim 10^{9}\,M_\odot$. This constitutes a statistically significant sample for detailed investigations of galaxy structure and evolution in the local Universe. MaNGA's IFU spectroscopy enables spatially resolved analyses of key physical processes, providing a powerful tool to study the interplay between morphology, stellar populations, star formation, gas content, and nuclear activity \citepalias{2022ApJS..262...36S}.

Reliable morphological classifications are crucial for this study. We adopted the visual classifications of \citetalias{2025MNRAS.544.1056V}, who performed a uniform Hubble-type analysis based on high-quality SDSS and Dark Energy Spectroscopic Instrument
(DESI) Legacy imaging data \citep{2019AJ....157..168D}. Their Value-Added Catalog includes $\sim 10,059$ galaxies. We adopted this catalog because it combines comprehensive morphological coverage, systematic comparisons with previous classifications, and high-quality image mosaics that allow verification and detailed spiral arm classification (Sect.~\ref{SSandR4}).

The \citetalias{2025MNRAS.544.1056V} study classified galaxies along the Hubble sequence using
$r$-band images from SDSS and DESI. The images were post-processed to enhance internal features (bars, spiral arms, rings) and external low-surface-brightness structures (faint spiral arms, tidal features). The resulting mosaics\footnote{\href{https://data.sdss.org/sas/dr17/env/MANGA_MORPHOLOGY/manga_visual_morpho/2.0.1/images/}{https://data.sdss.org/sas/dr17/env/MANGA\_MORPHOLOGY/} \href{https://data.sdss.org/sas/dr17/env/MANGA_MORPHOLOGY/manga_visual_morpho/2.0.1/images/}{manga\_visual\_morpho/2.0.1/images/}} include RGB images and residuals obtained by subtracting the best-fit bulge and disk surface brightness models \citepalias{2025MNRAS.544.1056V}.

Focusing on spiral structure, we selected 6,136 galaxies classified as Sa--Sd.
We adopted inclinations for 5,826 galaxies from \citetalias{2022ApJS..262...36S},
and derived inclinations for the remaining 310 galaxies, for which such measurements are unavailable, using \texttt{Inclinet}.\footnote{\href{https://edd.ifa.hawaii.edu/inclinet/}{https://edd.ifa.hawaii.edu/inclinet/}}
Applying the $i > 70^\circ$ cut
to minimize projection and internal extinction effects
removed 1,879 galaxies, resulting in a preliminary sample of 4,257.
Visual inspection led to the exclusion of 13 poor-quality images and 101 galaxies affected by tidal features or bright foreground stars, yielding a sample of 4,143 galaxies.

Bars, which play a critical role in disk dynamics, were classified following \citetalias{2025MNRAS.544.1056V} on a graded scale: unbarred (0), weakly barred (0.25--0.5), and strongly barred (0.75--1), corresponding to morphological types S, SAB, and SB. We visually inspected all 4,143 mosaics to verify bar classifications, and reclassified $\sim 4\%$ of galaxies ($\sim 1\%$ for morphology, $\sim 3\%$ for bar class), primarily due to limitations in image quality or saturated/bright central regions.

\subsection{AGN classification}
\label{SSandR3}

An important advantage of using the MaNGA survey to study AGN activity is the availability of well-defined nuclear activity classifications. In particular, \citetalias{2023A&A...674A..85A} provides a comprehensive spectroscopic classification for all MaNGA galaxies, based on standard emission-line diagnostics using Baldwin--Phillips--Terlevich \citep[BPT;][]{1981PASP...93....5B} diagrams. This framework enables a homogeneous distinction between different ionization sources, including star formation, AGN (Seyfert or low-ionization nuclear emission-line region [LINER]), and composite mechanisms.

The aperture-dependent analysis of \citetalias{2023A&A...674A..85A} highlights that optical AGN selection is highly sensitive to methodology, with IFU spectroscopy providing a more reliable approach than traditional single-fiber observations \citep[e.g.,][]{2017MNRAS.472.4382R,2018RMxAA..54..217S}. To mitigate contamination from ionization by evolved stellar populations, single-fiber studies typically adopt an equivalent width (EW) threshold of EW(H$\alpha$) $>$ 1.5\,\AA\ \citep{2010MNRAS.403.1036C}. For comparison with previous work, \citetalias{2023A&A...674A..85A} also considered a stricter criterion of EW(H$\alpha$) $>$ 3\,\AA.

In the \citetalias{2023A&A...674A..85A} catalog, galaxies are classified into five nuclear activity types: Seyferts, LINERs, star-forming, composite, and ambiguous. Seyfert and LINER galaxies with EW(H$\alpha$) $>$ 3\,\AA\,are conventionally considered AGN hosts. For statistical robustness in our subsamples, we adopted a more inclusive threshold of EW(H$\alpha$) $>$ 1.5\,\AA.
We also tested the robustness of our results using the stricter
criterion of EW(H$\alpha$) $>$ 3\,\AA\,and presented the corresponding analysis
in Appendix~\ref{app:stricterEW}.

We cross-matched our sample of 4,143 galaxies with the \citetalias{2023A&A...674A..85A} catalog. Galaxies classified as Seyfert or LINER with EW(H$\alpha$) $>$ 1.5\,\AA\,were considered AGN hosts, while non-AGN galaxies include star-forming and quiescent (absorption-line-dominated) systems. To ensure a clear separation, composite and ambiguous cases as well as
Seyfert and LINER galaxies with EW(H$\alpha$) $\leq$ 1.5\,\AA\,were excluded. This selection yielded a refined sample of 2,942 galaxies for further analysis.

\subsection{Spiral arm classification}
\label{SSandR4}

\begin{figure*}
\begin{center}$
\begin{array}{@{\hspace{0mm}}c@{\hspace{0mm}}}
\includegraphics[width=\hsize]{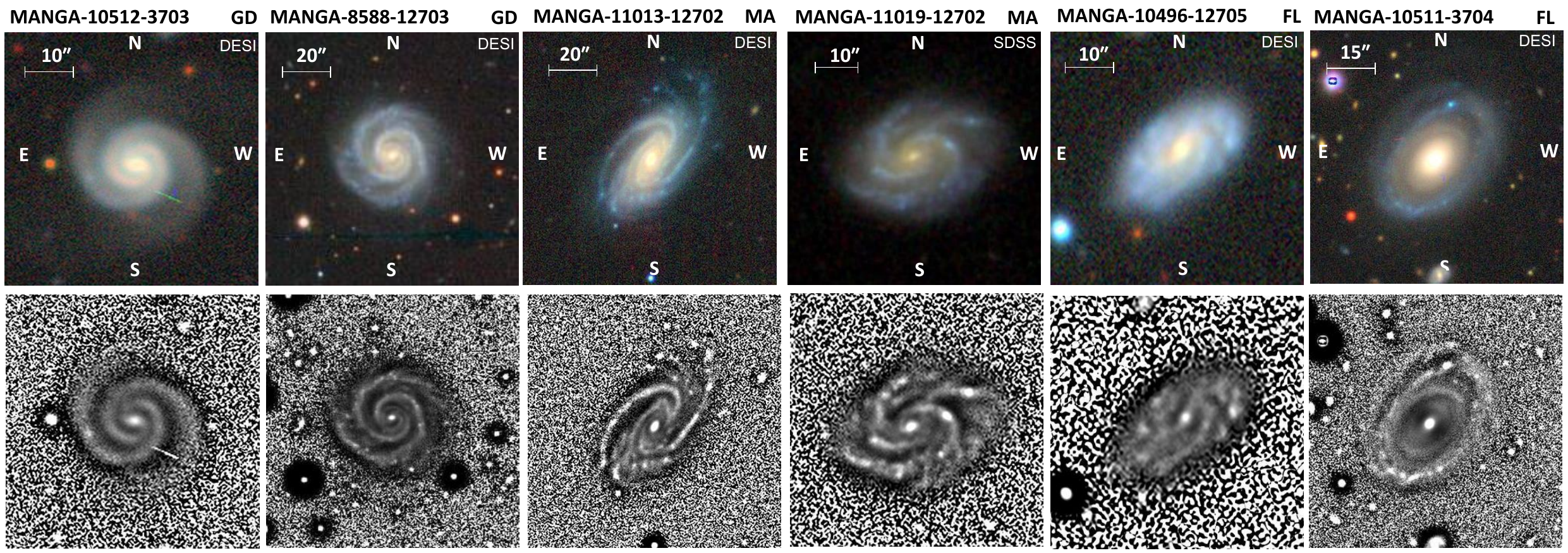}
\end{array}$
\end{center}
\caption{Examples of galaxies from our sample. For each galaxy, the upper panel shows the DESI or SDSS RGB image, while the bottom panel displays the corresponding residual image from \citetalias{2025MNRAS.544.1056V}. Two representative examples are shown for each AC. The MaNGA ID, AC, and spatial scale are indicated. In all images, north is up and east is to the left.}
\label{galaxExampls}
\end{figure*}

Spiral galaxies are the most common morphological type in the local universe \citep{2006MNRAS.373.1389C}. Their observed spiral structure depends on both the wavelength of observation \citep[e.g.,][]{2002ApJS..143...73E, 2015ApJS..217...32B} and the photometric depth of the imaging data \citep[e.g.,][]{2023A&A...671A.141M}. Automated spiral arm classification methods have been developed to process large datasets efficiently \citep[e.g.,][]{2023JCAP...08..044S}, but they remain limited in capturing detailed morphological parameters. Common approaches include Fourier decomposition \citep[e.g.,][]{2011ApJ...737...32E, 2018ApJ...869...29Y, 2020ApJ...900..150Y} and citizen-science classifications such as Galaxy Zoo 2 \citep{2013MNRAS.435.2835W}, with more recent methods proposed by \citet{2021MNRAS.507.3923M} and \citet{2024AJ....168...12S}, as well as modern machine-learning models such as ZooBot \citep{2022MNRAS.509.3966W}.

Spiral galaxies are often categorized based on arm structure into three main types: GD, MA, and FL \citep{1990NYASA.596...40E}. Specifically, GD spirals exhibit two prominent symmetric arms; FL spirals have numerous short, fragmented arms; and MA galaxies have three or more moderately extended arms with less symmetry \citep[e.g.,][]{2011ApJ...737...32E, 2015ApJS..217...32B}.

We performed a detailed visual classification of spiral arms following \citet{1987ApJ...314....3E}, using the SDSS and DESI Legacy image mosaics employed for morphological and bar strength assessments \citepalias{2025MNRAS.544.1056V}. The classification was applied to the 2,942 galaxies defined in Sect.~\ref{SSandR3}.
In 1,742 cases, spiral
structure could not be reliably determined due to low surface
brightness or poor contrast, leaving 1,200 galaxies with secure
AC determinations.
A residual limitation is that very low-mass, faint galaxies may be
more readily assigned to the FL class because their lower surface
brightness and weaker arm contrast make coherent spiral features more
difficult to identify.

The initial classification was performed by the first coauthor,
and its reliability was assessed through an independent reclassification of all 2,942 galaxies by two other coauthors
(AAH and AGK), drawing on their previous experience in
the visual classification of galaxy morphology and
structural components \citep{2018MNRAS.481..566K,2022MNRAS.517L.132K}.
The agreement between the classifications reached $\sim 98\%$.
For the remaining 2\% of cases, the most frequently assigned AC was adopted.

As an external validation of our classification,
we used the Galaxy Zoo arm-multiplicity fractions from \citet{2021MNRAS.507.3923M}
to define approximate pseudo-ACs, with
$P_{\rm GD}=P({\rm 2~arms})$,
$P_{\rm MA}=P({\rm 3~arms})+P({\rm 4~arms})+P({\rm >4~arms})$, and
$P_{\rm FL}=P({\rm 1~arm})+P({\rm cannot~tell})$.
Of the 1,200 galaxies with visual spiral arm classifications,
1,144 have usable Galaxy Zoo arm-multiplicity fractions.
For these galaxies, the highest-probability pseudo-AC agrees with our visual classification in
70.4\% of cases: 94.0\% for GD, 64.3\% for MA, and 66.4\% for FL galaxies.
The corresponding probability distributions differ significantly among the three visual classes
(Kruskal--Wallis tests, all $P<2\times10^{-90}$; Appendix~\ref{app:validation}).
This demonstrates the statistical consistency of the two classification schemes,
while the lower object-by-object agreement for MA and FL reflects the fact that
arm multiplicity is not an exact proxy for Elmegreen arm morphology,
which additionally incorporates arm continuity, symmetry, and global organization.
In addition, our classifications were based on visual inspection of deep SDSS and DESI image mosaics, supplemented by bulge--disk decomposition residual maps, which reveal low-surface-brightness spiral features and finer structural details that are often difficult to identify in shallower imaging data.
The complete comparison is presented in Appendix~\ref{app:validation}.
Examples of galaxies with different ACs, including DESI/SDSS RGB images and residual maps from \citetalias{2025MNRAS.544.1056V}, are shown in Fig.~\ref{galaxExampls}.

\subsection{Biases and sample statistics}
\label{SSandR5}

Before analyzing the statistical properties of the final sample,
we examined potential selection effects associated with the availability of reliable
spiral arm classifications. We assessed the representativeness of the 1,200 galaxies
with secure classifications by comparing their stellar masses, rest-frame colors,
bar properties, and AGN fractions with those of the 1,742 unclassified galaxies and
the full parent sample of 2,942 galaxies.
Importantly, the classified sample was found to be broadly representative of
the parent galaxy population (see Appendix~\ref{app:representative}).

\begin{figure}
\begin{center}$
\begin{array}{@{\hspace{0mm}}c@{\hspace{0mm}}}
\includegraphics[width=0.9\hsize]{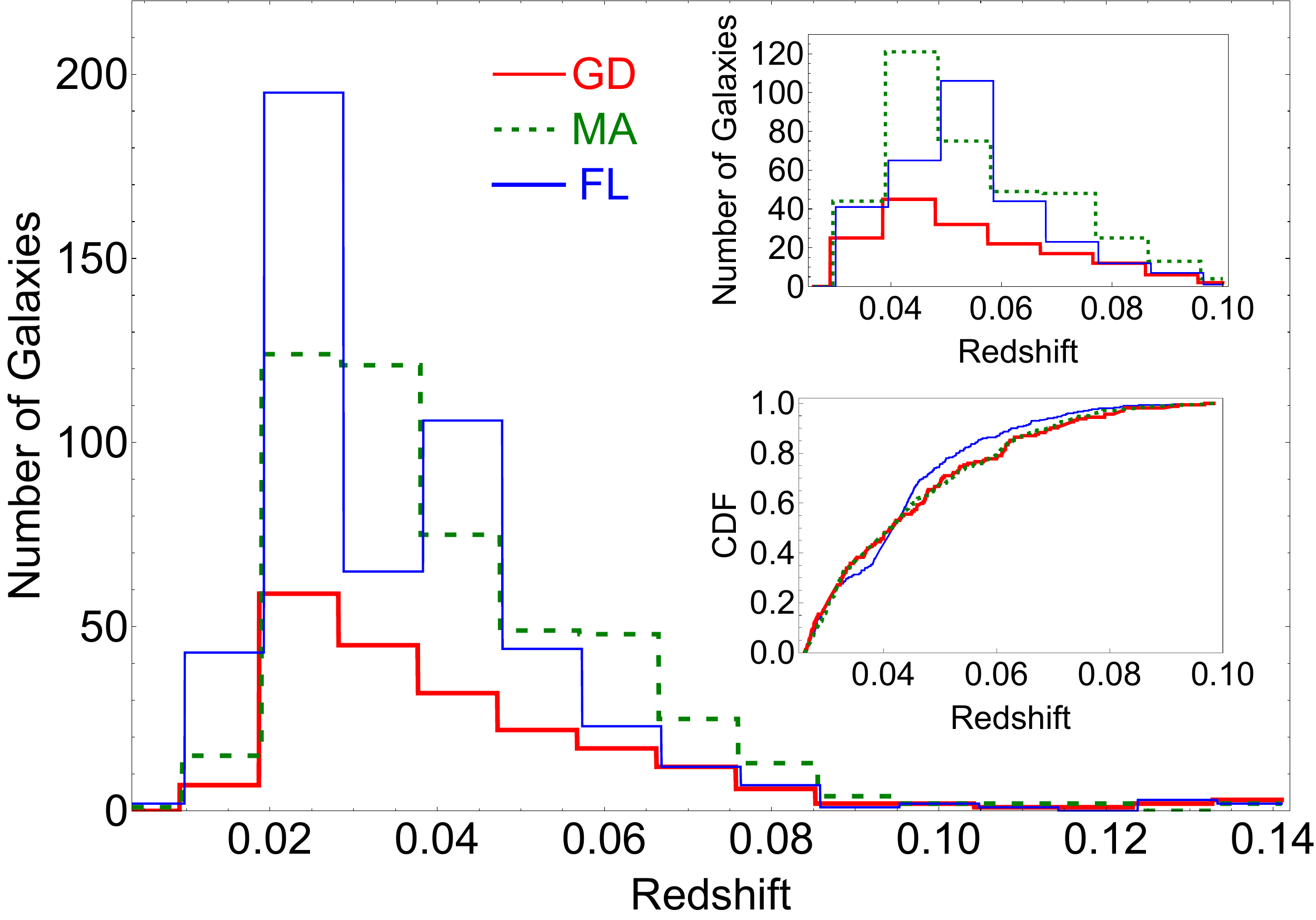}
\end{array}$
\end{center}
\caption{Redshift histogram of the initial sample of 1,200 galaxies, separated by AC. The top and bottom insets show histograms and the corresponding cumulative distribution functions for the three classes, limited to a restricted redshift range.}
\label{redshift_distribution}
\end{figure}

It is essential to verify that galaxies with different ACs (GD, MA, and FL) are represented with comparable probabilities across the redshift range $0.0037 < z < 0.139$ of our sample (see Fig.~\ref{redshift_distribution}).
To assess potential selection biases, we applied two-sample Kolmogorov--Smirnov (KS) and Anderson--Darling (AD) tests to the redshift distributions of subsamples grouped by AC. Comparisons between GD and FL galaxies, as well as between MA and FL galaxies, yield $P$ values below 0.05, indicating that the null hypothesis (i.e., the two samples are drawn from the same parent distribution) is rejected and that their redshift distributions differ significantly. In contrast, the comparison between GD and MA galaxies results in a $P$ value above 0.1, consistent with the null hypothesis and suggesting no significant differences between their redshift distributions.

This behavior can be attributed to the classification challenges of FL galaxies, which become increasingly difficult to identify at higher redshifts due to their lower surface brightness and fragmented structure. At lower redshifts, however, their classification rate increases sharply (see Fig.~\ref{redshift_distribution}). In contrast, GD and MA galaxies exhibit more gradual changes in their classification rate across the same redshift range.

To mitigate this bias, we progressively narrowed the redshift interval by adjusting its lower and upper limits in steps of 0.001, guided by the redshift distribution and the KS and AD test results, until the $P$ values exceeded 0.05 (see Table~\ref{compz}). This procedure led to the exclusion of 357 galaxies with $z < 0.026$ and $z > 0.1$ from the initial sample of 1,200. The top and bottom insets of Fig.~\ref{redshift_distribution} show the histograms and corresponding cumulative distribution functions for the GD, MA, and FL subsamples, based on the redshift-limited subset of 843 galaxies.

\begin{table*}
  \caption{Comparison of the redshift distributions $(0.026 \leq z \leq 0.1)$ between 843 galaxies with different parameters.}
  \label{compz}
  \centering
    \begin{tabular}{lcccccccc}
    \hline\hline
  \multicolumn{3}{c}{Subsample~1} & \multicolumn{1}{c}{vs.}&\multicolumn{3}{c}{Subsample~2}&\multicolumn{1}{c}{$P_{\rm KS}^{\rm MC}$}&\multicolumn{1}{c}{$P_{\rm AD}^{\rm MC}$}\\
   \multicolumn{1}{c}{} & \multicolumn{1}{c}{$N$}&\multicolumn{1}{c}{$\langle z \rangle$}&\multicolumn{1}{c}{} &\multicolumn{1}{c}{}&\multicolumn{1}{c}{$N$}&\multicolumn{1}{c}{$\langle z \rangle$}&&\\
   \hline
     GD & 162 & 0.045 & vs.& FL & 300 & 0.043 & 0.231 &  0.161\\
      MA & 381 & 0.045 & vs.& FL & 300 & 0.043 & 0.069 &  0.058\\
       GD & 162 & 0.045 & vs.& MA & 381 & 0.045 & 0.954 &  0.890 \\
   Sa--Sb & 330 & 0.044 & vs.& Sbc--Sd & 513 & 0.044 & 0.270 &  0.637 \\
   Barred & 527 & 0.044 & vs.& Unbarred & 316 & 0.045 & 0.159 &  0.161 \\
   AGN & 165 & 0.045 & vs.& Non-AGN & 678 & 0.044 & 0.763 &  0.580 \\
  \hline
   \end{tabular}
  \tablefoot{For the two-sample KS and AD tests, Monte Carlo (MC) simulations with $10^5$ iterations were used to provide
  $P_{\rm KS}^{\rm MC}$ and $P_{\rm AD}^{\rm MC}$ probabilities.}
\end{table*}

We also performed analogous analyses by separating the redshift-limited sample of 843 galaxies according to morphology (early-type: Sa--Sb; late-type: Sbc--Sd), the presence of bars, and AGN versus non-AGN classification. As reported in Table~\ref{compz}, the KS and AD tests yield $P > 0.1$ for all these comparisons, indicating no significant differences in the redshift distributions between the corresponding subsets. This suggests that the final sample of 843 galaxies is largely free from biases associated with redshift-dependent selection effects.

For the final sample, we carried out comparative statistical analyses by further dividing the galaxies into early-type and late-type subsamples, each simultaneously subdivided according to spiral arm and bar classifications. The resulting counts and distributions are summarized in Table~\ref{numberofgal}.

\begin{table}
  \caption{Number of galaxies by AC and bar presence separated into early- and late-type morphologies.}
  \label{numberofgal}
  \centering
    \begin{tabular}{lccc}
    \hline
    \hline
    \multicolumn{1}{l}{ } & \multicolumn{1}{c}{Early-type} & \multicolumn{1}{l}{Late-type} & \multicolumn{1}{c}{All} \\
    \hline
    \multicolumn{1}{c}{ } & \multicolumn{2}{c}{Unbar} &\\
    GD & 14 & 9 & 23 \\
    MA & 48 & 101 & 149 \\
    FL & 41 & 103 & 144 \\
    All & 103 & 213 & 316 \\
    \hline
    & \multicolumn{2}{c}{Weak bar} &\\
     GD & 57 & 18 & 75 \\
    MA & 38 & 91 & 129 \\
    FL & 12 & 86 & 98 \\
    All & 107 & 195 & 302 \\
    \hline
     & \multicolumn{2}{c}{Strong bar} &\\
     GD & 52 & 12 & 64 \\
    MA & 40 & 63 & 103 \\
    FL & 28 & 30 & 58 \\
    All & 120 & 105 & 225 \\
     \hline
    ALL & 330 & 513 & 843 \\
     \hline
  \end{tabular}
 \end{table}

Overall, $39^{+2}_{-2}\%$ of the galaxies in our sample are classified as early-type, while $61^{+2}_{-2}\%$ are late-type. Across the full sample, $63^{+2}_{-2}\%$ of galaxies exhibit bars, with $36^{+2}_{-2}\%$ hosting weak bars and $27^{+2}_{-2}\%$ strongly barred; the remaining $37^{+2}_{-2}\%$ are unbarred. Taking into account the ACs, $19^{+2}_{-1}\%$ of galaxies are GD, $36^{+2}_{-2}\%$ are FL, and $45^{+2}_{-2}\%$ are MA.

A comparison with previous studies revealed broad consistency. \citet{2013JKAS...46..141A} reported $20^{+1}_{-1}\%$ GD, $42^{+1}_{-1}\%$ FL, and $38^{+1}_{-1}\%$ MA spirals in a local sample ($z \leq 0.02$, 1,635 galaxies), which agrees well with our AC distribution. Using the Spitzer Survey of Stellar Structure in Galaxies, \citet{2015ApJS..217...32B} provided benchmark morphological and bar classifications. Restricting their sample to Sa--Sd spirals with full morphological, bar, and AC information yields a subsample of 736 galaxies: $36^{+2}_{-2}\%$ early-type, $64^{+2}_{-2}\%$ late-type; $75^{+2}_{-2}\%$ barred ($\sim 35\%$ weak, $\sim 40\%$ strong), and $25^{+2}_{-2}\%$ unbarred. Their AC distribution ($f_{\rm GD}=24^{+2}_{-2}\%$, $f_{\rm FL}=30^{+2}_{-2}\%$, and $f_{\rm MA}=47^{+2}_{-2}\%$) is broadly consistent with ours within the $2\sigma$ limits. The uncertainties quoted correspond to $1\sigma$ errors.

Pairwise analyses of morphology, AC, and bar presence reveal clear trends, with the role of the most probable underlying drivers, stellar mass and color, examined in the following analysis:
\begin{itemize}
    \item[(a)] Early-type spirals have an overall bar fraction of $69^{+3}_{-3}\%$ ($32^{+3}_{-3}\%$ weak, $37^{+3}_{-3}\%$ strong), with $31^{+3}_{-3}\%$ unbarred. Late-type spirals show a lower bar fraction of $58^{+2}_{-2}\%$ ($38^{+2}_{-2}\%$ weak, $20^{+2}_{-2}\%$ strong), with $42^{+2}_{-2}\%$ unbarred.
    \item[(b)] Among the ACs, GD galaxies are most frequently barred ($86^{+3}_{-3}\%$: $\sim 46\%$ weak, $\sim 40\%$ strong), FL galaxies are the least ($52^{+3}_{-3}\%$: $\sim 33\%$ weak, $\sim 19\%$ strong), and MA galaxies are intermediate ($61^{+3}_{-3}\%$: $\sim 34\%$ weak, $\sim 27\%$ strong).
    \item[(c)] Morphological distribution across ACs shows that GD galaxies are predominantly early-type ($76^{+4}_{-4}\%$ early, $24^{+4}_{-4}\%$ late), FL galaxies are mainly late-type ($27^{+3}_{-3}\%$ early, $73^{+3}_{-3}\%$ late), and MA galaxies exhibit a mixed distribution ($33^{+3}_{-3}\%$ early, $67^{+3}_{-3}\%$ late).
\end{itemize}

These statistics are consistent with previous findings: early-type spirals are more likely to host bars than late-types (\citealt{2011MNRAS.418.1055M}: $\sim 66\%$ versus $\sim 44\%$; \citealt{2010ApJS..186..427N}: $\sim 74\%$ versus $\sim 50\%$), and GD galaxies exhibit the highest bar fraction ($\sim 85\%$), FL the lowest ($\sim 50\%$), and MA an intermediate value ($\sim 60\%$; \citealt{2020IAUS..353..140B}). Bar strength also correlates with morphology: strong bars dominate early-types, while weak bars are more common in late-types (\citealt{2019ApJ...872...97L}).

The distribution of ACs is closely linked to morphological type. As reported by \citealt{1987ApJ...314....3E}, \citealt{1989ApJ...342..677E}, \citealt{2016A&A...590A..44G}, and \citealt{2017MNRAS.471.1070B}, GD galaxies are predominantly early-type ($\sim 55$--$65\%$), MA galaxies show a roughly even mix ($\sim 25$--$35\%$ early, $\sim 25$--$35\%$ late), and FL galaxies are mostly late-type ($\sim 45$--$65\%$).

\begin{figure*}
\centering
\includegraphics[width=0.9\hsize]{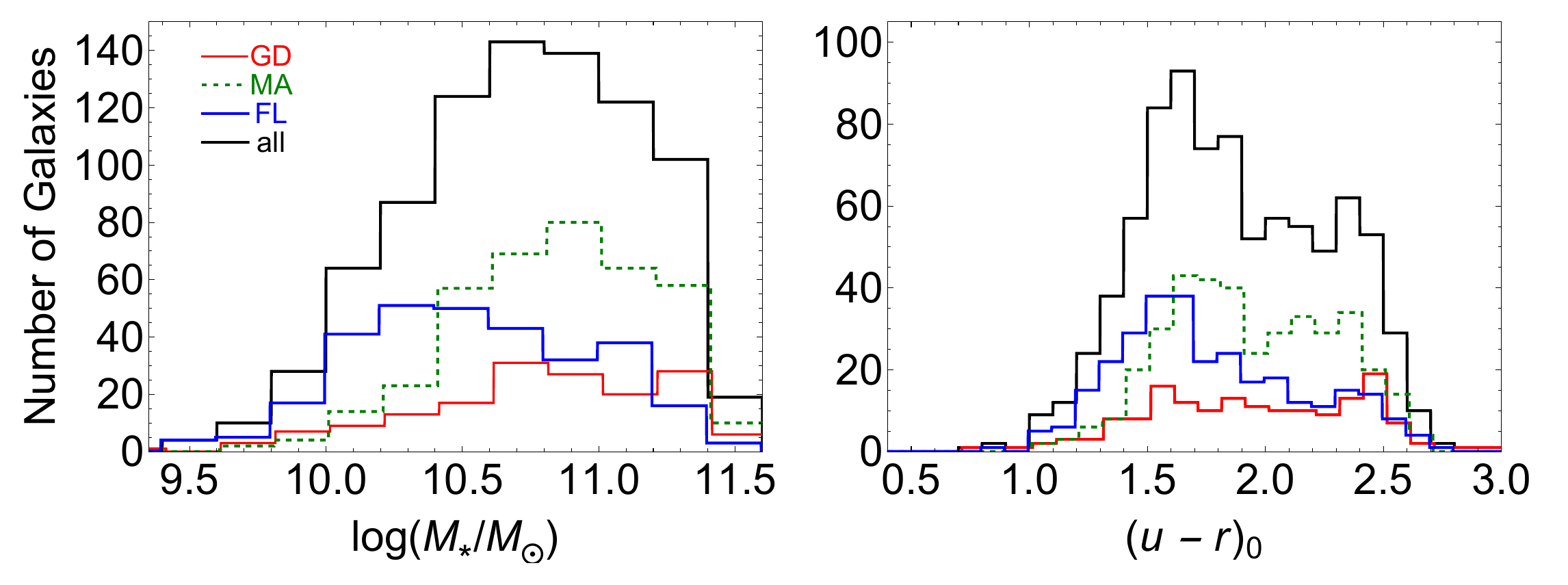}
\caption{Left: Stellar-mass distributions for the full sample and for
         galaxies in different spiral ACs. Right: Same as the left panel
         but for the rest-frame $(u-r)_0$ color distributions.
        }
\label{mass_color_distribution}
\end{figure*}

In addition,
Fig.~\ref{mass_color_distribution} presents the distributions of stellar mass and
rest-frame $(u-r)_0$ color for the full galaxy sample, as well as separately for
galaxies in different spiral ACs.
The stellar masses and $u-r$ colors were adopted from \citetalias{2022ApJS..262...36S} and \citet{2022ApJS..259...35A},
respectively. Note that the $(u-r)_0$ colors are corrected
for Galactic foreground extinction and have also been K-corrected following \cite{2010MNRAS.405.1409C}.
Pairwise comparisons of the stellar-mass and color distributions using the KS and AD tests show that GD and MA galaxies have statistically indistinguishable distributions in both stellar mass and color ($P>0.1$), while FL galaxies systematically exhibit lower stellar masses and significantly different colors ($P<0.001$ when compared to both GD and MA systems). This implies that GD and MA subsamples are consistent with being drawn from the same parent distributions, while FL galaxies constitute a distinct population in both mass and color. These trends are consistent with previous studies \citep[e.g.,][]{2017MNRAS.471.1070B}, showing that FL spirals tend to be less massive and bluer, while GD spirals are typically more massive and redder.

Overall, our study provides a detailed and quantitative characterization of the relationships among galaxy morphology, spiral AC, and bar properties.
The observed trends are consistent with previous findings, lending additional confidence to our classifications and reinforcing the close connection between bar strength, morphological type, and spiral arm structure.

\section{Results and discussion}
\label{sec:results}

As mentioned in the Introduction,
the dynamical nature of stellar bars, together with their ability to funnel
gas toward the central regions of galaxies, motivates the hypothesis that
they may act as triggers of AGN activity \citep[e.g.,][]{2003ASPC..290..411C}.
However, observational studies provide mixed evidence of a direct connection
between bars and AGNs, and their detailed interplay remains a subject of active investigation and debate
\citep[e.g.,][]{2000ApJ...529...93K,2002ApJ...567...97L,2012ApJ...750..141L,2013A&A...549A.141A,2015MNRAS.448.3442G,2019MNRAS.489.1338P,2023MNRAS.521.1775G,2024ApJ...973..129G,2024MNRAS.532.2320G,2024ApJ...973..116D,2024A&A...692A.159S,2024MNRAS.527.3366K,2025A&A...699A.204M,2026A&A...707A.152L}.
In a similar vein, the configuration of spiral arms may also regulate
the inflow of gas toward the central SMBH
\citep[e.g.,][]{2003ApJ...589..774M,2010MNRAS.407.1529H,2022A&A...666A.175Y}.
Investigating these structural components is therefore essential for
understanding the physical mechanisms that drive nuclear activity
\citep[e.g.,][]{2004ARA&A..42..603K,2014RvMP...86....1S,2024MNRAS.532.2320G}.

To quantitatively explore the connection between AGN activity and
dynamical properties, such as spiral ACs and the presence of bars,
we first compare the AGN fraction in barred and unbarred galaxies
in Sect.~\ref{sec:bar}.
We then examine, in Sect.~\ref{sec:DW}, the AGN fraction among galaxies with
different spiral ACs. Finally, in Sect.~\ref{sec:combined},
we analyze the combined as well as the separate influence of bars and
spiral arm structure on the AGN fraction.
Table~\ref{tab:agn_distribution} presents the distribution of AGN and non-AGN galaxies in the final sample of 843 galaxies, subdivided according to bar strength and spiral AC.

\begin{table}
\centering
\caption{Distribution of AGN and non-AGN galaxies as a function of bar strength
         and spiral AC.}
\label{tab:agn_distribution}
\begin{tabular}{llcccc}
\hline\hline
 & Bar class & FL & MA & GD & All \\
\hline
\multirow{4}{*}{AGN}
 & Unbarred   & 13 & 18 & 2  & 33 \\
 & Weak bar   & 9  & 31 & 16 & 56 \\
 & Strong bar & 11 & 39 & 26 & 76 \\
 & All        & 33 & 88 & 44 & 165 \\
\hline
\multirow{4}{*}{non-AGN}
 & Unbarred   & 131 & 131 & 21 & 283 \\
 & Weak bar   & 89  & 98  & 59 & 246 \\
 & Strong bar & 47  & 64  & 38 & 149 \\
 & All       & 267 & 293 & 118 & 678 \\
\hline
 & ALL & 300 & 381 & 162 & 843 \\
\hline
\end{tabular}
\end{table}

\subsection{Influence of bars}
\label{sec:bar}

As shown in Table~\ref{tab:fisher_baronly}, we compared the AGN fractions among
galaxies with different bar classes, namely unbarred, weakly barred, and strongly barred systems.
Unbarred galaxies exhibit a relatively low AGN fraction of
$f_{\rm AGN} = 0.10_{-0.02}^{+0.02}$.
This fraction increases to
$f_{\rm AGN} = 0.18_{-0.02}^{+0.02}$ for galaxies hosting weak bars
and reaches $f_{\rm AGN} = 0.34_{-0.03}^{+0.03}$ for strongly barred galaxies.
This increase in AGN fraction with bar strength indicates a correlation between bar presence and AGN activity in our sample.

\begin{table}
\centering
\caption{Pairwise comparisons between different bar classes.}
\label{tab:fisher_baronly}

\tabcolsep=4.5pt

\begin{tabular}{l c c l c c c}
\hline\hline
\multicolumn{2}{c}{Subsample I} &
vs.&
\multicolumn{2}{c}{Subsample II} &
$P_{\rm B}$ &
$P_{\rm F}$ \\
\cline{1-2} \cline{4-5}
Bar & $f_{\rm AGN}$ &
&
Bar & $f_{\rm AGN}$ &
&
\\
\hline
Unbar &
$0.10^{+0.02}_{-0.02}$ &
vs.&
Weak &
$0.18^{+0.02}_{-0.02}$ &
\textbf{0.004} &
\textbf{0.006}
\\

Weak &
$0.18^{+0.02}_{-0.02}$ &
vs.&
Strong &
$0.34^{+0.03}_{-0.03}$ &
$< \textbf{0.001}$ &
$< \textbf{0.001}$
\\

Unbar &
$0.10^{+0.02}_{-0.02}$ &
vs.&
Strong &
$0.34^{+0.03}_{-0.03}$ &
$< \textbf{0.001}$ &
$< \textbf{0.001}$
\\
\hline
\end{tabular}

\tablefoot{
The uncertainties on the AGN fractions, $f_{\rm AGN} = N_{\rm AGN}/(N_{\rm AGN}+N_{\rm non\mbox{-}AGN})$, correspond to
binomial confidence intervals calculated following the approach of
\citet{2011PASA...28..128C}.
The terms $P_{\rm B}$ and $P_{\rm F}$ denote the $P$ values obtained from Barnard's
test and Fisher's exact test, respectively.
Statistically significant differences are highlighted in bold.
}
\end{table}

The statistical significance of these differences was assessed using
Fisher's exact test \citep{Fisher1922} and Barnard's test \citep{1945Natur.156..177B},
both of which are widely applied to $2 \times 2$ contingency tables.
In this section, we examine the relationship between bar presence
(barred versus unbarred galaxies) and nuclear activity (AGN versus non-AGN).
Accordingly, each contingency table consists of four categories:
barred AGN galaxies, barred non-AGN galaxies, unbarred AGN galaxies,
and unbarred non-AGN galaxies.
To evaluate whether bar presence and AGN activity are statistically associated,
both tests compute the probability of obtaining the observed distribution under
the null hypothesis that the two variables are independent.
A probability ($P$ value) below 0.05 indicates rejection of the null hypothesis
and implies a statistically significant association.\footnote{Unlike
Fisher's exact test, Barnard's test does not assume fixed
marginal distributions and is generally more powerful, particularly
in cases of unbalanced samples or small-number statistics.}
In an analogous manner, we also investigate the relationship between bar strength
(weakly barred versus strongly barred galaxies) and nuclear activity.

In Table~\ref{tab:fisher_baronly},
the $P$ values obtained from Fisher's exact test,
$P_{\rm F} = 0.006$ for the comparison between weakly barred and
unbarred galaxies, and $P_{\rm F} < 0.001$ for comparisons involving
strongly barred galaxies (relative to both unbarred and weakly barred systems),
indicate that the observed differences are statistically significant and
unlikely to arise from random fluctuations.
Barnard's test yields similarly low $P$ values, with $P_{\rm B} = 0.004$
for weakly barred versus unbarred galaxies and $P_{\rm B} < 0.001$
for strongly barred galaxies compared to either weakly barred or unbarred systems.
These results further support the statistical significance and robustness of
the observed differences.
Both tests yield consistent results, supporting a robust association between bar presence, bar strength, and AGN activity.

The increase in the AGN fraction with increasing bar strength is consistent with
a scenario in which bars play an important role in regulating gas inflow
toward the central regions of galaxies and, consequently,
in enhancing nuclear activity.
In this context, previous studies have reported mixed results.
While some works \citep[e.g.,][]{2003ApJ...589..774M, 2012ApJ...750..141L} found no significant dependence of AGN activity on strong bars when controlling for stellar mass and color, others \citep[e.g.,][]{2009ASPC..419..402H, 2012ApJS..198....4O, 2013A&A...549A.141A, 2015MNRAS.448.3442G, 2024MNRAS.532.2320G,2026ApJ..1003...25L} reported enhanced AGN fractions in barred systems.

For comparison, \citet{2012ApJ...750..141L} analyzed a homogeneous sample of 8,655
SDSS spiral galaxies within the redshift range $0.01 < z < 0.05$.
As described in Sect.~\ref{SSandR3}, galaxies in our sample classified as
Seyferts or LINERs with EW(H$\alpha$) $>$ 1.5\,\AA\,are considered AGNs,
while the non-AGN category includes star-forming and quiescent galaxies.
Composite and ambiguous cases were excluded from our analysis.
To enable a direct comparison with the results of \citet{2012ApJ...750..141L},
we recalibrated the sample definitions for both datasets.
Specifically, we excluded quiescent galaxies from our non-AGN sample,
as such systems are not included in \citet{2012ApJ...750..141L}.
In addition, based on their table~3, we retained only star-forming galaxies from
the \citet{2012ApJ...750..141L} sample and exclude composite galaxies,
which are not present in our dataset.

The recalculated AGN fractions for our subsample of 562 galaxies are
$f_{\rm AGN} = 0.14_{-0.03}^{+0.03}$ for unbarred galaxies,
$f_{\rm AGN} = 0.21_{-0.03}^{+0.03}$ for weakly barred galaxies, and
$f_{\rm AGN} = 0.49_{-0.04}^{+0.04}$ for strongly barred galaxies.
Similarly, for the recalculated sample of 6,682 galaxies from
\citet{2012ApJ...750..141L}, the corresponding AGN fractions are
$f_{\rm AGN} = 0.19_{-0.01}^{+0.01}$ for unbarred galaxies,
$f_{\rm AGN} = 0.24_{-0.02}^{+0.02}$ for weakly barred galaxies, and
$f_{\rm AGN} = 0.49_{-0.01}^{+0.01}$ for strongly barred galaxies.

In a recent study, \citet{2024MNRAS.532.2320G} analyzed a sample of 48,871 disk
galaxies in a volume-limited sample from Galaxy Zoo DESI, investigating
the AGN fraction in strongly barred, weakly barred, and unbarred galaxies
up to $z = 0.1$ across a range of stellar masses and colors.
To allow a direct comparison with \citet{2024MNRAS.532.2320G},
we again recalibrated the sample definitions for both datasets,
following the same procedure applied in our comparison with \citet{2012ApJ...750..141L}.
For the recalculation, we used the data presented in table~A1 of \citet{2024MNRAS.532.2320G}.
For the recalculated sample of 36,814 galaxies from \citet{2024MNRAS.532.2320G},
the corresponding AGN fractions are
$f_{\rm AGN} = 0.15_{-0.01}^{+0.01}$ for unbarred systems,
$f_{\rm AGN} = 0.24_{-0.01}^{+0.01}$ for weakly barred galaxies,
and $f_{\rm AGN} = 0.52_{-0.01}^{+0.01}$ for strongly barred galaxies.

Importantly, within approximately 1$\sigma$ uncertainties,
our recalculated AGN fractions agree well with those derived from both of the mentioned studies.
However, as discussed above, the stellar-mass and color distributions of
sample galaxies can significantly affect comparisons of AGN fractions across
different bins of bar strength \citep[see e.g.,][]{2012ApJ...750..141L}.

\begin{figure}
\centering
\includegraphics[width=0.9\hsize]{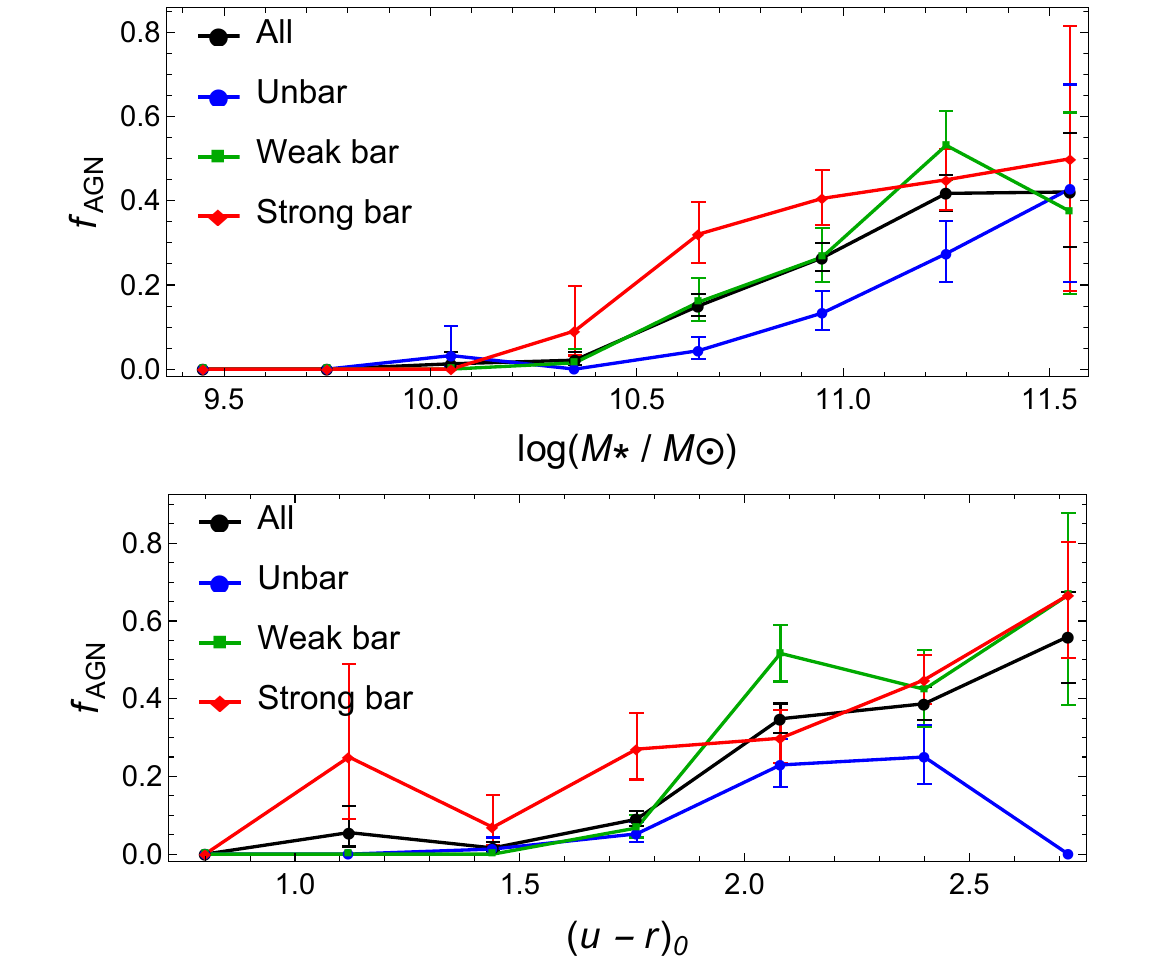}
\caption{Fraction of AGNs as a function of stellar mass (upper panel) and rest-frame $(u-r)_0$ color (bottom panel) for different bar classes.
The black, blue, green, and red curves correspond to all, unbarred, weakly barred, and strongly barred galaxies, respectively.
Error bars represent binomial confidence intervals on the AGN fraction in each bin.}

\label{AGN_mass_color1}
\end{figure}

In particular, strong bars are well known to be more common in early-type,
massive and redder disks, while weak bars are found more frequently in late-type,
lower-mass and bluer spirals \citep[e.g.,][]{2010ApJ...714L.260N,2011A&A...532A..75D}.
Consequently, differences in AGN fractions across bins of bar strength may
partly reflect underlying dependencies on stellar mass and galaxy color.
To investigate these trends, we first examine how the AGN fraction varies separately with stellar mass and color for different bar classes (Fig.~\ref{AGN_mass_color1}).

The AGN fraction increases with stellar mass for all bar classes,
exhibiting a pronounced rise at $\log(M_\ast/M_\odot) \gtrsim 10.5$,
which suggests the presence of a characteristic stellar-mass scale above which nuclear activity becomes significantly more common, although part of this trend may reflect the lower detectability of weak AGNs in lower mass galaxies.
At fixed stellar mass, a systematic offset between bar classes is mildly evident,
with strongly barred galaxies generally hosting higher AGN fractions than weakly barred or unbarred systems.
Overall, stellar mass appears to drive the primary trend in AGN activity,
while bar structure introduces a secondary modulation without altering the underlying mass dependence.
A qualitatively similar, although more scattered, behavior is observed as a function of the rest-frame $(u - r)_0$ color.

However, it is important to note that stellar mass and color are themselves strongly correlated,
and therefore a simultaneous control of both parameters is required for a consistent comparison between different bar classes.
For example, \citet{2012ApJ...750..141L} adopted bin widths of 0.5 dex in stellar mass
and 0.3 mag in the Galactic-extinction-corrected $(u - r)_0$ color.
In \citet{2024MNRAS.532.2320G}, stellar mass and color are divided into 15 equal-width
intervals, spanning $7.0 \leq \log(M_*/M_\odot) \leq 12.0$ in stellar mass and
$0.4 \leq (g - r)_0 \leq 2.0$ in color.
When computing AGN fractions, \citet{2024MNRAS.532.2320G} adopted a more conservative
lower stellar-mass limit of $\log(M_*/M_\odot) \geq 10.0$, following
\citet{2012ApJ...746...90A}, to reduce selection biases associated with
AGN detectability.

\begin{figure}
\centering
\includegraphics[width=0.9\hsize]{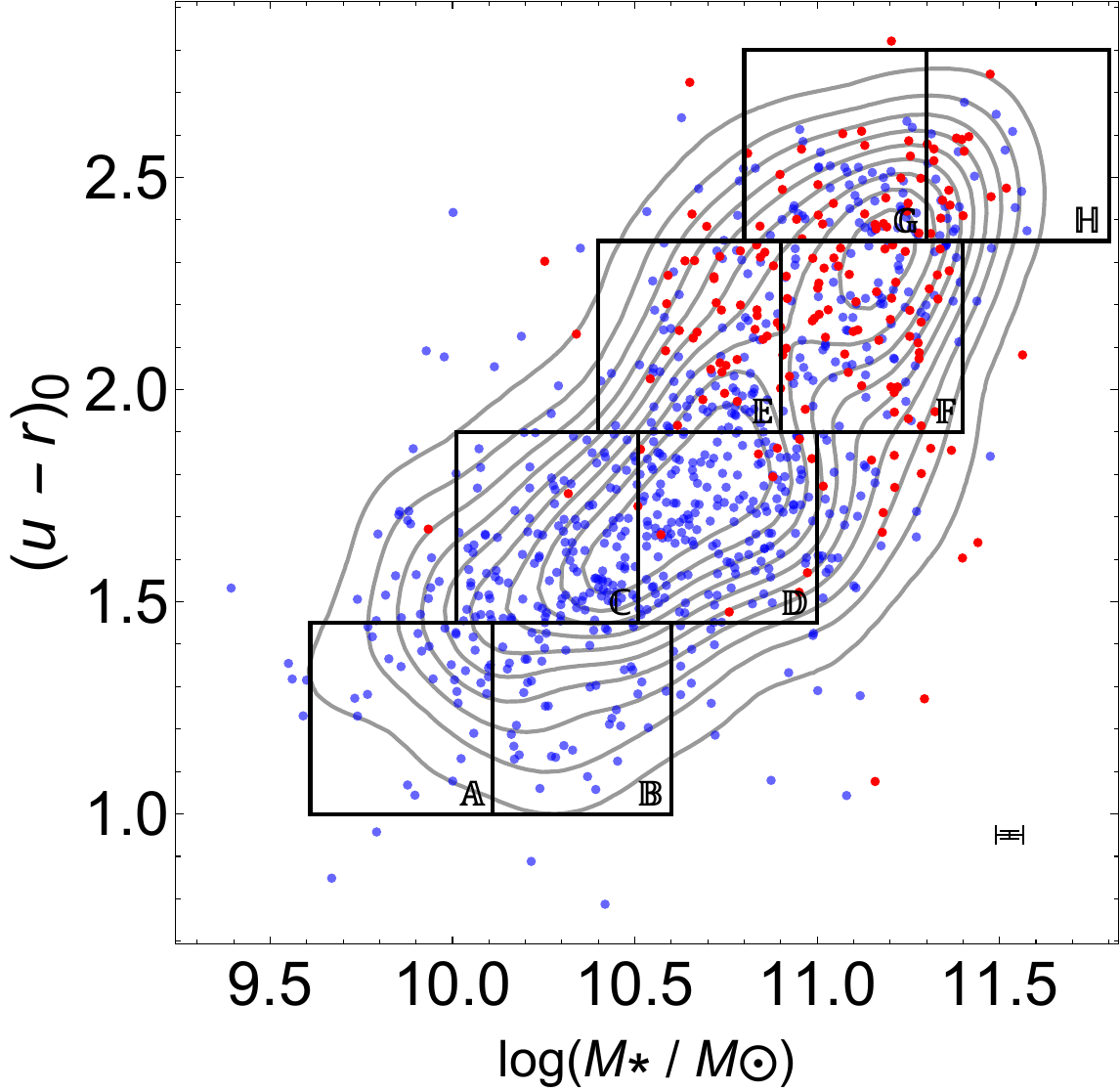}
\caption{
Distribution of Sa--Sd galaxies in the $(u-r)_0$ versus $\log(M_*/M_\odot)$ plane for our sample.
The AGN and non-AGN hosts are shown in red and blue, respectively. Black contours represent the density distribution of the full sample.
Black boxes with 0.45 mag $\times$ 0.5 dex sizes indicate the bins used to compute AGN fractions for different bar classes. To avoid overcrowding the figure with individual error bars, a representative uncertainty is shown in the bottom-right corner, corresponding to median errors of $\sim0.02$ mag in $(u-r)_0$ and $\sim0.07$ dex in stellar mass.
}
\label{mass_color_diagram}
\end{figure}

For this purpose, we used the color--mass diagram to define bins with comparable stellar-mass (and color) distributions.
In Fig.~\ref{mass_color_diagram}, we show the distribution of sample galaxies on the $(u-r)_0$ versus $\log(M_*/M_\odot)$ plane, overlaid with density contours.
Recall that the sample consists exclusively of spiral galaxies spanning the morphological types Sa to Sd (AGN host galaxies highlighted by red points in Fig.~\ref{mass_color_diagram}). The overall distribution exhibits the well-known bimodality \citep[e.g.,][]{2014MNRAS.440..889S}, with a blue star-forming galaxies at low stellar masses and a red, more quiescent spiral galaxies at higher masses. The density contours trace a continuous, tilted structure, reflecting the correlation between galaxy color and stellar mass.
AGN host galaxies are predominantly found at stellar masses above $\log(M_*/M_\odot) > 10.5$, while they are largely absent at lower masses (as  in Fig.~\ref{AGN_mass_color1}). In the high-mass regime, galaxies preferentially occupy intermediate-to-red colors, consistent with more evolved spiral systems \citep[e.g.,][]{2014MNRAS.440..889S}.

In Fig.~\ref{mass_color_diagram}, we divided the $(u-r)_0$ -- $\log(M_*/M_\odot)$ plane into equal-sized bins that cover the full galaxy distribution while keeping bin sizes as large as possible. The binning was further constrained so that two-sample KS and AD tests do not reject the null hypothesis of identical underlying stellar mass and color distributions across bar classes. However, in bins $\mathbb{A}$, $\mathbb{B}$, and $\mathbb{E}$, these tests indicate significant differences between at least one pair of bar classes. On the other hand, bins $\mathbb{A, B,}$ and $\mathbb{C}$ contain almost no AGN hosts.
Therefore, $\mathbb{A, B, C,}$ and $\mathbb{E}$ bins were excluded from the AGN analysis.

Thus, we computed AGN fractions only in bins that (\textit{i}) contain a sufficient number of galaxies (at least ten)  and (\textit{ii}) exhibit no statistically significant differences in stellar-mass (and color) distributions between bar classes. This ensured that any observed trends are not driven by selection effects in parameter space but reflect intrinsic differences between populations.

Accordingly, we focused on bins $\mathbb{D}$, $\mathbb{F}$, $\mathbb{G}$, and $\mathbb{H}$, where both AGN and non-AGN spirals are present and statistically consistent comparisons between bar classes are possible.
To compare AGN fractions between bar classes, we applied only Barnard's exact test for $2 \times 2$ contingency tables as our primary statistical method, due to its robustness for small samples.

In the $\mathbb{D}$ bin, statistically significant enhancement of the AGN fraction is observed in strongly barred galaxies $(f_{\rm AGN} = 0.21_{-0.09}^{+0.12})$ compared to unbarred systems $(f_{\rm AGN} = 0.02_{-0.02}^{+0.03})$, with $P_{\rm B} = 0.004$. A significant difference is also found between strongly and weakly barred galaxies $(f_{\rm AGN} = 0.04_{-0.02}^{+0.04}$, $P_{\rm B} = 0.019)$, while weakly barred and unbarred systems remain statistically indistinguishable.

In the $\mathbb{F}$ bin, a statistically significant enhancement of the AGN fraction is found in weakly barred galaxies, with $f_{\rm AGN} = 0.53_{-0.08}^{+0.08}$, compared to unbarred systems, $f_{\rm AGN} = 0.29_{-0.08}^{+0.10}$, with $P_{\rm B} = 0.025$. No significant differences are detected between weakly and strongly barred galaxies {$(f_{\rm AGN} = 0.38_{-0.08}^{+0.09})$} or between strongly barred and unbarred systems. This indicates that within this bin the dependence on the bar strength is not uniform across classes and does not show a consistent ordering.

A similar behavior is also observed in the $\mathbb{G}$ bin, where weakly barred galaxies show a higher AGN fraction, with $f_{\rm AGN} = 0.55_{-0.19}^{+0.18}$, compared to unbarred systems, $f_{\rm AGN} = 0.20_{-0.09}^{+0.13}$. However, the difference is only marginally significant, with $P_{\rm B} = 0.055$, which places the result at the statistical significance threshold and weakens the evidence of a bar-related enhancement of AGN activity in this bin.

In the $\mathbb{H}$ bin, no statistically significant differences are detected between bar classes. Although barred galaxies show slightly higher nominal AGN fractions, the large uncertainties prevent a statistically robust assessment of any bar--AGN connection in this bin.

Overall, our results indicate that the role of bars in AGN activity depends on stellar mass. A statistically significant enhancement of the AGN fraction is found in strongly barred galaxies at intermediate stellar masses, in agreement with previous studies \citep[e.g.,][]{2024MNRAS.532.2320G}. In this regime, weakly barred galaxies show weaker or no significant enhancement, suggesting that the effect is primarily associated with strong bars.
No statistically significant dependence on bar presence is found in our highest-mass bin. This is consistent with the results of \citet{2012ApJ...750..141L} and suggests that bar-driven gas inflow may be less efficient in massive spiral galaxies.

Our results suggest that the influence of bars on AGN activity is not universal across the spiral population. Instead, it appears to be most effective at intermediate stellar masses, while at higher masses other dynamical processes may play a more dominant role in regulating gas inflow. In massive systems, the presence of more prominent bulges and dynamically hotter stellar components can reduce the efficiency of bar-driven torques and angular momentum transport \citep[e.g.,][]{2003MNRAS.341.1179A,2004ARA&A..42..603K}, which is consistent with the absence of a significant bar dependence at the highest masses in our sample.

To further validate our findings, we repeated the analysis by varying the adopted stellar-mass and color thresholds of
the bins within the representative uncertainties shown in Fig.~\ref{mass_color_diagram}.
In all cases, we recovered the same qualitative trends and comparable levels of statistical significance across different bar classes.
These consistency checks demonstrate that our results are robust against reasonable variations in the adopted stellar-mass and color cuts.

As an independent robustness test, we also repeated the above analysis using the stellar mass and color weighting procedure of \citet{2024MNRAS.532.2320G}.
The weighted analysis yielded the same qualitative conclusions as our matched-control analysis.
The methodology and detailed results are presented in Appendix~\ref{app:garland}.

\subsection{Influence of arm classes}
\label{sec:DW}

We next consider spiral arms as another important dynamical component of disk galaxies.
Large-scale spiral structure is extensively studied in the context of disk galaxy dynamics and star formation \citep[e.g.,][]{1982MNRAS.201.1035E, 1987ApJ...314....3E, 2010MNRAS.403..625D,2016ARA&A..54..667S},
whereas its connection to AGN activity remains less well established.
Spiral arms can induce gas inflow through gravitational torques and
shock compression, leading to enhanced central gas concentrations
and elevated star formation activity
\citep[e.g.,][]{1969ApJ...158..123R, 1972ApL....11...41K, 2016MNRAS.460.2472B, 2021ApJ...917...88Y}.
Recent observations further suggest that, similarly to bars,
spiral arms may facilitate the transport of cold gas toward the central
kiloparsec, thereby increasing the molecular gas content and potentially
contributing to AGN fueling \citep[e.g.,][]{2022A&A...666A.175Y}.
We therefore examined the AGN fraction as a function of spiral ACs
in order to assess whether nuclear activity is systematically linked
to spiral structure.

\begin{table}
\centering
\caption{Pairwise comparisons between different ACs.}
\label{tab:lastArm_fisher}

\tabcolsep=5pt

\begin{tabular}{l c c l c c c}
\hline\hline
\multicolumn{2}{c}{Subsample I} &
vs.&
\multicolumn{2}{c}{Subsample II} &
$P_{\rm B}$ &
$P_{\rm F}$ \\
\cline{1-2} \cline{4-5}
AC & $f_{\rm AGN}$ &
&
AC & $f_{\rm AGN}$ &
&
\\
\hline
FL &
$0.11^{+0.02}_{-0.02}$ &
vs.&
MA &
$0.23^{+0.02}_{-0.02}$ &
$< \textbf{0.001}$ &
$< \textbf{0.001}$
\\

MA &
$0.23^{+0.02}_{-0.02}$ &
vs.&
GD &
$0.27^{+0.04}_{-0.04}$ &
0.314 &
0.326
\\

FL &
$0.11^{+0.02}_{-0.02}$ &
vs.&
GD &
$0.27^{+0.04}_{-0.04}$ &
$< \textbf{0.001}$ &
$< \textbf{0.001}$
\\

FL &
$0.11^{+0.02}_{-0.02}$ &
vs.&
GM &
$0.24^{+0.02}_{-0.02}$ &
$< \textbf{0.001}$ &
$< \textbf{0.001}$
\\
\hline
\end{tabular}

\tablefoot{
The uncertainties on $f_{\rm AGN}$ correspond to binomial confidence
intervals, calculated as described in
Table~\ref{tab:fisher_baronly}.
The $P$ values from Barnard's and Fisher's exact tests are denoted as
$P_{\rm B}$ and $P_{\rm F}$, respectively.
Statistically significant differences are highlighted in bold.
}
\end{table}

Notably, GD galaxies exhibit higher AGN fractions than FL systems.
A similar trend is observed when comparing MA and FL galaxies (see Table~\ref{tab:lastArm_fisher}).
In contrast, the AGN fractions of MA and GD galaxies are statistically consistent with each other.
The similarity of the AGN fractions in GD and MA galaxies
is most likely related to the observational fact that the stellar mass (and color)
distributions of galaxies with these ACs are very similar,
in contrast to the FL population, which exhibits markedly different properties (Fig.~\ref{mass_color_distribution}),
thus suggesting that stellar mass and color are the primary drivers of the observed AGN trends.
As discussed in Sect.~\ref{SSandR5},
this similarity is further supported by the KS and AD tests applied to the corresponding stellar-mass (and color) distributions,
which do not reject the null hypothesis for the GD and MA classes.
This indicates that GD and MA galaxies are statistically consistent with being drawn from the same parent distributions in both stellar mass and color.
Taken together,
the similarities between GD and MA galaxies in stellar mass,
color, and AGN fraction, together with the increased sample size
obtained in the subsequent analysis, motivate combining these two spiral ACs into
a single category. This grouping also has a physical motivation,
as both GD and MA galaxies are generally associated with organized spiral patterns in which spiral density waves are thought to play an important role,
unlike FL galaxies, whose spiral structure is predominantly attributed to
local gravitational instabilities \citep[e.g.,][]{2011ApJ...737...32E}.

\begin{figure}
\centering
\includegraphics[width=0.9\hsize]{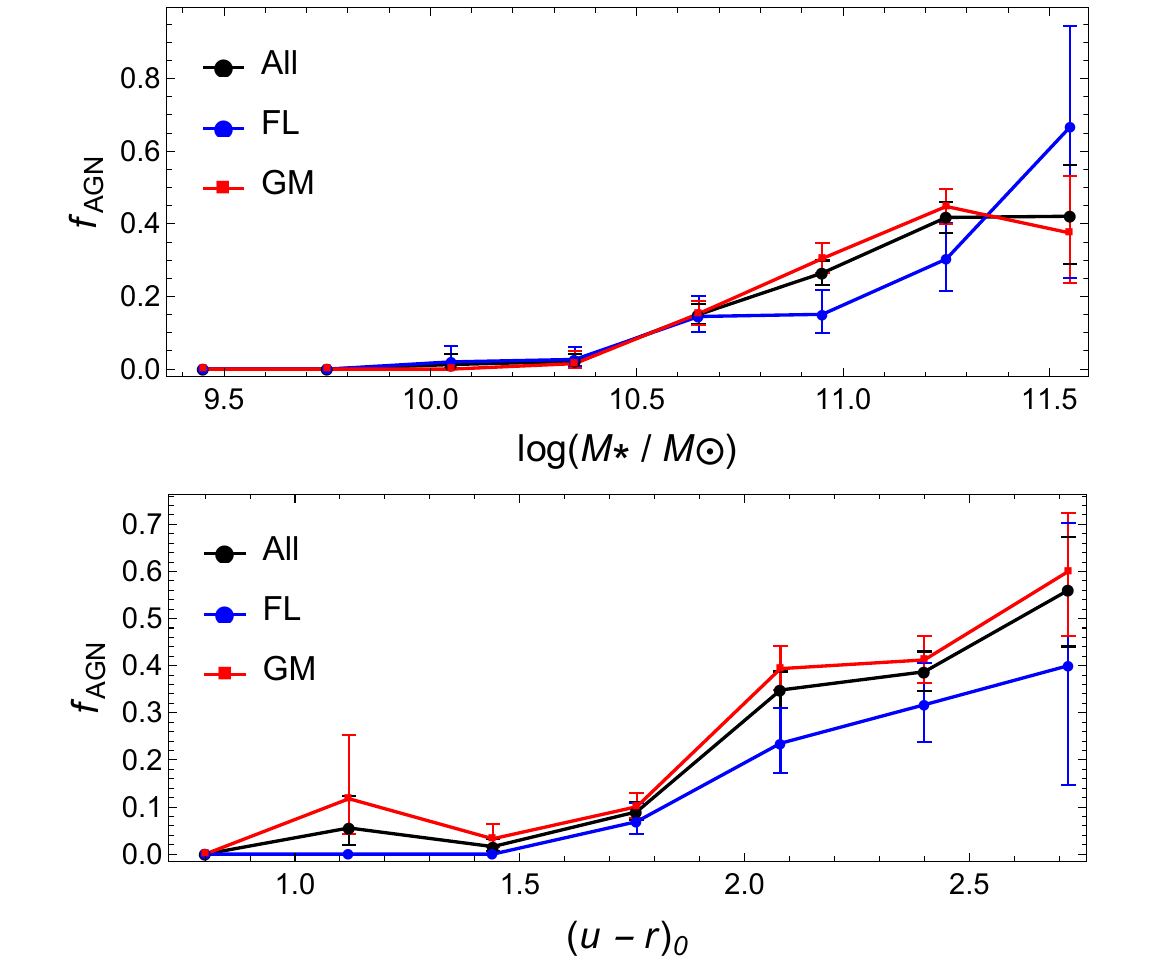}
\caption{Fraction of AGNs as a function of stellar mass
(upper panel) and rest-frame $(u-r)_0$ color
(bottom panel) for different spiral ACs.
The black, blue, and red curves correspond to all, FL, and GM galaxies, respectively.
Error bars represent binomial confidence intervals on the AGN fraction in each bin.}
\label{AGN_mass_color2}
\end{figure}

In Table~\ref{tab:lastArm_fisher},
we also present pairwise comparisons of the AGN fractions
between GM and FL galaxies.
The $P$ values obtained from the statistical tests
indicate a significant enhancement of the AGN fraction in the GM subsample.
Fig.~\ref{AGN_mass_color2} presents
the AGN fraction as a function of stellar mass and color for GM and FL galaxies.
Similar to the results of Sect.~\ref{sec:bar},
stellar mass (and color) appears to drive the primary trend
in AGN activity, while spiral AC may introduce smaller but systematic differences in the AGN fraction at fixed stellar mass and color.

To investigate whether the differences shown in Table~\ref{tab:lastArm_fisher} persist after simultaneously controlling for host galaxy stellar mass and color,
we applied the same $(u-r)_0$ -- $\log(M_*/M_\odot)$ binning
scheme to the spiral ACs as adopted for the bar classes
in Sect.~\ref{sec:bar}. For all eight bins shown in
Fig.~\ref{mass_color_diagram}, two-sample KS and AD tests
were performed to compare the stellar-mass (and color)
distributions of the FL and GM galaxies.
The null hypothesis is not rejected in bins $\mathbb{D}$, $\mathbb{E}$, $\mathbb{F}$, $\mathbb{G}$, and $\mathbb{H}$,
indicating that the FL and GM subsamples within these bins
are statistically consistent in terms of both stellar mass
and color. These bins were therefore retained for the subsequent
analysis of the AGN fractions in FL and GM galaxies.

No statistically significant differences in AGN fraction were detected between the two spiral ACs in any individual bin $(P_{\rm B} > 0.09)$.
Nevertheless, the $f_{\rm AGN}$--mass and $f_{\rm AGN}$--color
relations exhibit a mild enhancement of the AGN fraction
in GM galaxies over a restricted parameter range (see Fig.~\ref{AGN_mass_color2}).
Although the null hypothesis is not rejected in
the individual-bin analysis, the measured AGN fractions in
the intermediate stellar-mass bins $\mathbb{E}$, $\mathbb{F}$, and $\mathbb{G}$
are systematically higher for GM galaxies
($f_{\rm AGN} = 0.37_{-0.06}^{+0.06}$, $0.45_{-0.06}^{+0.06}$, and $0.42_{-0.08}^{+0.08}$, respectively)
than for FL galaxies
($f_{\rm AGN} = 0.24_{-0.08}^{+0.10}$, $0.29_{-0.09}^{+0.11}$, and $0.22_{-0.09}^{+0.12}$, respectively).
This behavior suggests the presence of a weak but systematic trend,
which is likely diluted by the limited number statistics within the individual bins.

When the adjacent bins $\mathbb{E}$ and $\mathbb{F}$ in Fig.~\ref{mass_color_diagram},
corresponding to the intermediate-mass regime,
are combined in order to increase the statistical power,
the KS and AD tests applied to the stellar-mass (and color)
distributions do not reject the null hypothesis between FL and GM galaxies,
supporting the validity of merging the two bins.
Among all adjacent-bin combinations, only the $\mathbb{E} + \mathbb{F}$ merger satisfied these requirements.
The Barnard's test performed on the combined sample ($\mathbb{E} + \mathbb{F}$) then yields a statistically significant difference in AGN fraction ($P_{\rm B} = 0.038$) between the FL $(f_{\rm AGN} = 0.26_{-0.06}^{+0.07})$ and GM $(f_{\rm AGN} = 0.41_{-0.04}^{+0.04})$ classes.
This result indicates that the enhancement is primarily
concentrated within this stellar-mass regime and becomes
statistically detectable only after increasing the statistical
power through bin combination.
Overall, spiral arm structure does not exhibit a robust
independent influence on AGN activity once stellar mass
and color are controlled, although a weak and localized
enhancement in the intermediate-mass regime cannot be excluded (e.g., in $\mathbb{E} + \mathbb{F}$ bin).

The robustness of the spiral arm analysis was also assessed using the weighting procedure of \citet{2024MNRAS.532.2320G}. The weighted analysis reproduces the same qualitative trend, although the difference between the GM and FL classes remains only marginally significant after stellar mass and color matching (see Appendix~\ref{app:garland}).
Since bars and spiral arms are dynamically interconnected
components of disk galaxies \citep[e.g.,][]{2009AJ....137.4487B,2022A&A...666A.175Y,2024AJ....168...12S},
in the next section we investigate their combined effects
in order to obtain a more complete interpretation of
the observed trends in AGN activity.

\subsection{Influence of bar and arm classes in combination}
\label{sec:combined}

\begin{table*}
\centering
\caption{Pairwise comparisons between different spiral ACs and bar strength categories.}
\label{tab:fisher_results}

\tabcolsep=3.8pt
\renewcommand{\arraystretch}{1.1}

\begin{tabular}{l l c c c l c c c c}
\hline\hline
Effects &
\multicolumn{3}{c}{Subsample I} &
vs &
\multicolumn{3}{c}{Subsample II} &
$P_{\rm B}$ &
$P_{\rm F}$ \\
\cline{2-4} \cline{6-8}
&
AC (Bar) &
$N_{\rm AGN}/N_{\rm non-AGN}$ &
$f_{\rm AGN}$ &
&
AC (Bar) &
$N_{\rm AGN}/N_{\rm non-AGN}$ &
$f_{\rm AGN}$ &
&
\\
\hline

Bar effect
& FL (unbar) & 13/131 & $0.09^{+0.03}_{-0.02}$ & vs
& FL (strong) & 11/47 & $0.19^{+0.07}_{-0.05}$
& \textbf{0.049} & 0.057 \\

& FL (unbar) & 13/131 & $0.09^{+0.03}_{-0.02}$ & vs
& FL (weak) & 9/89 & $0.09^{+0.04}_{-0.03}$
& 0.985 & 1.000 \\

& FL (weak) & 9/89 & $0.09^{+0.04}_{-0.03}$ & vs
& FL (strong) & 11/47 & $0.19^{+0.07}_{-0.05}$
& 0.079 & 0.088 \\

Density wave effect
& FL (unbar) & 13/131 & $0.09^{+0.03}_{-0.02}$ & vs
& GM (unbar) & 20/152 & $0.12^{+0.03}_{-0.03}$
& 0.458 & 0.468 \\

Mixed effect
& GM (unbar) & 20/152 & $0.12^{+0.03}_{-0.03}$ & vs
& GM (strong) & 65/102 & $0.39^{+0.04}_{-0.04}$
& $< \textbf{0.001}$ & $< \textbf{0.001}$ \\

& GM (unbar) & 20/152 & $0.12^{+0.03}_{-0.03}$ & vs
& GM (weak) & 47/157 & $0.23^{+0.03}_{-0.03}$
& \textbf{0.004} & \textbf{0.004} \\

& GM (weak) & 47/157 & $0.23^{+0.03}_{-0.03}$ & vs
& GM (strong) & 65/102 & $0.39^{+0.04}_{-0.04}$
& \textbf{0.001} & \textbf{0.001} \\

& FL (strong) & 11/47 & $0.19^{+0.07}_{-0.05}$ & vs
& GM (strong) & 65/102 & $0.39^{+0.04}_{-0.04}$
& \textbf{0.006} & \textbf{0.006} \\

& FL (weak) & 9/89 & $0.09^{+0.04}_{-0.03}$ & vs
& GM (weak) & 47/157 & $0.23^{+0.03}_{-0.03}$
& \textbf{0.004} & \textbf{0.004} \\

& FL (unbar) & 13/131 & $0.09^{+0.03}_{-0.02}$ & vs
& GM (strong) & 65/102 & $0.39^{+0.04}_{-0.04}$
& $< \textbf{0.001}$ & $< \textbf{0.001}$ \\

& FL (unbar) & 13/131 & $0.09^{+0.03}_{-0.02}$ & vs
& GM (weak) & 47/157 & $0.23^{+0.03}_{-0.03}$
& \textbf{0.001} & \textbf{0.001} \\

& FL (strong) & 11/47 & $0.19^{+0.07}_{-0.05}$ & vs
& GM (weak) & 47/157 & $0.23^{+0.03}_{-0.03}$
& 0.512 & 0.593 \\

& FL (strong) & 11/47 & $0.19^{+0.07}_{-0.05}$ & vs
& GM (unbar) & 20/152 & $0.12^{+0.03}_{-0.03}$
& 0.157 & 0.183 \\

& FL (weak) & 9/89 & $0.09^{+0.04}_{-0.03}$ & vs
& GM (unbar) & 20/152 & $0.12^{+0.03}_{-0.03}$
& 0.538 & 0.683 \\

& FL (weak) & 9/89 & $0.09^{+0.04}_{-0.03}$ & vs
& GM (strong) & 65/102 & $0.39^{+0.04}_{-0.04}$
& $< \textbf{0.001}$ & $< \textbf{0.001}$ \\

\hline
\end{tabular}

\tablefoot{
The terms $N_{\rm AGN}$ and $N_{\rm non-AGN}$ denote the actual numbers of
AGN and non-AGN galaxies in each subsample, respectively.
The uncertainties on $f_{\rm AGN}$ correspond to binomial confidence
intervals, calculated as described in
Table~\ref{tab:fisher_baronly}.
The $P$ values from Barnard's and Fisher's exact tests are denoted as
$P_{\rm B}$ and $P_{\rm F}$, respectively.
Statistically significant differences are highlighted in bold.
}
\end{table*}

Table~\ref{tab:fisher_results} presents the AGN fractions for galaxies when considering separately
the effects of bars and spiral ACs,
as well as their combined contribution.
The pairwise comparisons show that bar strength alone in FL galaxies,
where large-scale density wave effects are expected to be weak or absent,
provides only limited evidence of an enhancement of AGN activity
(e.g., the comparison between unbarred and strongly barred FL galaxies).
Conversely, spiral AC alone in unbarred galaxies,
where bar-driven effects are absent,
does not produce statistically significant differences in AGN fraction
(i.e., between unbarred FL and GM galaxies).

When bar strength and spiral AC are considered simultaneously,
several comparisons yield larger differences in AGN fraction than those obtained when each parameter is analyzed separately.
Within this combined framework, before stellar mass and color control, GM galaxies with increasing bar strength
generally exhibit higher AGN fractions than weakly barred or unbarred GM systems,
suggesting that bars are more directly associated with enhanced AGN activity in these galaxies.
On the other hand, while spiral structure does not appear to exert a strong independent impact,
but may modulate the observed trends when combined with bars.
In particular, the AGN fraction is significantly enhanced in barred GM galaxies compared to barred FL systems.
Importantly, the AGN fraction reaches its highest value in strongly barred GM galaxies,
whereas the lowest AGN fraction is observed in unbarred FL systems (see Table~\ref{tab:fisher_results}).

We now turn to the analysis performed within
the four mass- and color-controlled bins
$(\mathbb{D, F, G,}$ and $\mathbb{H})$,
restricting the statistical evaluation to pairwise comparisons
in which each subsample contains more than 15 galaxies.
This criterion ensures that all Barnard tests are
based on statistically meaningful sample sizes.
In the $\mathbb{D}$ bin, a single subsample,
namely strongly barred FL galaxies,
falls below this threshold and is therefore excluded from the comparison set.

In the $\mathbb{D}$ bin, no significant dependence of the AGN fraction on bar strength
is found within the FL class, nor is any significant dependence on spiral AC detected among unbarred galaxies.
However, when both structural components are considered jointly,
a significant difference emerges between unbarred FL and strongly barred GM galaxies
($f_{\rm AGN} = 0.04^{+0.03}_{-0.02}$ versus $0.28^{+0.07}_{-0.05}$, $P_{\rm B} = 0.034$).
Within the GM population, strongly barred galaxies exhibit higher AGN fractions
than both unbarred
($0.02^{+0.03}_{-0.03}$, $P_{\rm B} = 0.003$)
and weakly barred
($0.04^{+0.03}_{-0.03}$, $P_{\rm B} = 0.013$) systems.
These results suggest that, in this stellar-mass regime, the enhancement of AGN activity
is most pronounced when strong bars are present within galaxies hosting large-scale density wave spiral structure.

In the $\mathbb{F}$ bin, all FL subsamples fall below the adopted sample-size threshold
and are therefore excluded from the analysis, leaving only GM systems for comparison.
Within this restricted sample, a statistically significant difference is found only between unbarred and weakly barred GM galaxies ($f_{\rm AGN}=0.27^{+0.13}_{-0.10}$ versus $0.56^{+0.09}_{-0.09}$, $P_{\rm B}=0.036$). The corresponding AGN fraction for strongly barred GM galaxies is $0.42^{+0.10}_{-0.09}$. All other pairwise comparisons remain consistent with the null hypothesis.

In the $\mathbb{G}$ and $\mathbb{H}$ bins,
all subsamples fall below the adopted minimum sample-size threshold.
Consequently, no statistically admissible pairwise comparisons can be performed within these bins,
and no reliable constraints on the dependence of AGN activity on bar strength or spiral AC can be derived.

Finally, we repeated the combined arm--bar analysis using the weighting procedure of \citet{2024MNRAS.532.2320G}. The weighted analysis confirms the principal conclusions of the matched-control analysis (see Appendix~\ref{app:garland}).

Overall, we find that some trends in AGN fraction become more apparent when bar strength and
spiral structure are considered jointly rather than separately.
This suggests that a full assessment of the connection between AGN activity and galaxy morphology may require accounting for
the combined influence of bars and spiral structure.

This interpretation is broadly consistent with both theoretical and observational studies
showing that non-axisymmetric structures,
such as bars and spiral arms (density waves),
contribute to angular momentum redistribution and the transport of gas toward
the central regions of galaxies \citep[e.g.,][]{2003ASPC..290..411C,2003ApJ...589..774M,2010MNRAS.407.1529H,2021ApJ...917...88Y,2022A&A...666A.175Y,2026ApJ..1003...25L}.
In this framework, bars and spiral structure may act in a dynamically coupled rather than
fully independent manner, jointly regulating the efficiency of gas inflow and,
consequently, the triggering of AGN activity.

\section{Conclusions}
\label{sec:concl}

Using spatially resolved spectroscopy from the MaNGA survey,
we constructed a homogeneous sample of 843 morphologically undisturbed Sa--Sd galaxies $(0.026 \leq z \leq 0.1$ and $i \leq 70^\circ)$ with consistently determined bar strengths, stellar masses, colors, and nuclear activity classifications compiled from the available literature (\citealt{2022ApJS..259...35A}; \citetalias{2022ApJS..262...36S,2023A&A...674A..85A,2025MNRAS.544.1056V}).
A new aspect of this work is that we visually classified the spiral ACs for the entire galaxy sample using optical imaging data.
For the detailed morphological analysis, we used mosaics including RGB images together with residual maps obtained after subtracting the best-fitting bulge and disk surface-brightness models.
The classifications were performed following the classical AC scheme \citep[e.g.,][]{1987ApJ...314....3E,2011ApJ...737...32E, 2015ApJS..217...32B},
separating galaxies into FL, MA, and GD (or ${\rm GD+MA=GM}$) spirals on the basis of the continuity, symmetry, and large-scale organization of their spiral structure.
This enabled us to investigate, in a systematic manner, the connection between AGN activity and large-scale non-axisymmetric structures in galaxies while considering the roles of bars and spiral arms separately and, for the first time, their combined influence.
Particular attention was given to controlling for the known dependencies of AGN fraction on stellar mass and galaxy color. Our main conclusions are summarized as follows:

\begin{itemize}

\item[-]
Before controlling for stellar mass and color,
we found a dependence of AGN activity on bar strength.
The AGN fraction increased systematically from
$f_{\rm AGN} = 0.10^{+0.02}_{-0.02}$ in unbarred galaxies
to $0.18^{+0.02}_{-0.02}$ in weakly barred systems and to
$0.34^{+0.03}_{-0.03}$ in strongly barred galaxies
(Table~\ref{tab:fisher_baronly}).
Both Fisher's exact and Barnard's tests indicated that these differences
were statistically significant, particularly for comparisons involving
strongly barred galaxies.

\item[-]
After controlling for stellar mass and color, we found that the
dependence of AGN activity on bar strength remained significant in the intermediate-mass regime
($10.5 < \log(M_\star/M_\odot) \leq 11.0$).
In this mass range, strongly barred galaxies exhibited systematically
higher AGN fractions than unbarred systems in both bluer and
redder subsamples.
These results supported a scenario in which bars contribute to the
transport of gas toward the central regions of galaxies,
thereby enhancing AGN activity
\citep[e.g.,][]{2009ASPC..419..402H, 2012ApJS..198....4O, 2013A&A...549A.141A, 2015MNRAS.448.3442G, 2024MNRAS.532.2320G,2026ApJ..1003...25L}.
By contrast, at higher stellar masses
($\log(M_\star/M_\odot) > 11.0$),
the AGN fraction showed no statistically significant dependence
on bar class.
These results suggest that the efficiency of bar-driven gas inflow
decreases in massive galaxies, likely because other dynamical
processes, such as the stabilizing influence of massive bulges and
dynamically hot stellar components, become increasingly important
\citep[e.g.,][]{2003MNRAS.341.1179A,2004ARA&A..42..603K}.

\item[-]
Before controlling for stellar mass and color, we found that
GM (GD $+$ MA) galaxies exhibited systematically higher AGN fractions than
FL systems (Table~\ref{tab:lastArm_fisher}).
However, after restricting the analysis to subsamples with
statistically consistent stellar-mass and color distributions,
most of these apparent differences became statistically insignificant.
Only a weak and localized enhancement of the AGN fraction
in GM versus FL galaxies remained detectable within
the intermediate-mass regime.
Overall, these results suggest that spiral arm structure alone
does not exert a strong independent influence on AGN activity
and that much of the apparent global dependence is driven by
underlying correlations with stellar mass and galaxy color.

\item[-]
For the first time, we analyzed bar strength and spiral arm structure jointly and found that before controlling for stellar mass and galaxy color, several differences in the AGN fraction become more pronounced than when each morphological parameter is considered separately (Table~\ref{tab:fisher_results}).
The highest AGN fraction in the complete sample was observed in
strongly barred GM galaxies
($f_{\rm AGN} = 0.39^{+0.04}_{-0.04}$),
while the lowest value corresponded to unbarred FL systems
($f_{\rm AGN} = 0.09^{+0.03}_{-0.02}$).

\item[-]
After controlling for stellar mass and color,
the statistically most robust enhancement was obtained in the
intermediate-mass bin, where strongly barred GM galaxies
showed significantly higher AGN fractions than unbarred FL systems
($f_{\rm AGN} = 0.28^{+0.07}_{-0.05}$ versus
$0.04^{+0.03}_{-0.02}$).
Within the GM population itself, strongly barred galaxies also
exhibited significantly enhanced AGN fractions compared to both
unbarred and weakly barred GM systems.
In general, these results suggested that bars and spiral structures may
act in a dynamically coupled manner, jointly contributing to
the redistribution of angular momentum and gas transport toward the
central regions of galaxies
\citep[e.g.,][]{2021ApJ...917...88Y,2022A&A...666A.175Y,2026ApJ..1003...25L}.

\end{itemize}

Overall, our findings highlight the importance of simultaneously accounting
for stellar mass and color together with the combined large-scale
morphological structure of galaxies, such as bars and spiral arms,
when investigating the physical mechanisms responsible for
triggering nuclear activity.
Importantly, our conclusions remain robust against reasonable variations in the adopted stellar-mass and color thresholds used to define the different
subsamples as well as when the stricter criterion of EW(H$\alpha$) $>$ 3\,\AA\,was applied (see Appendix~\ref{app:stricterEW}).
However, one of the main limitations of this study is the decrease in statistical power when the galaxy sample is simultaneously subdivided according to stellar mass, color, bar strength, and spiral AC.
In addition, the alternative stellar mass and color weighting procedure of \citet{2024MNRAS.532.2320G} yields consistent results for the bar, spiral arm,
and combined arm--bar analyses, further supporting the robustness of our principal conclusions.

Future studies combining the greatly expanded spectroscopic samples from DESI \citep{2026AJ....171..285D} with the deep wide-field imaging provided by Euclid and the Vera C. Rubin Observatory \citep{2026A&A...711A...1E,2019ApJ...873..111I} will overcome current statistical limitations and enable a more detailed investigation of the interplay between bars, spiral structure, and AGN activity across a wide range of galaxy environments and cosmic time.
Moreover, combining detailed structural classifications with measurements of molecular gas content, gas kinematics, and central star formation activity \citep[e.g.,][]{2024ApJ...973..116D,2025A&A...698A.296R} may provide stronger constraints on the physical mechanisms responsible for angular momentum transport and AGN fueling.

\section*{Data availability}

The database of 843 galaxies, including their MaNGA identifiers, bar strengths, spiral ACs, nuclear activities, stellar masses, $(u-r)_0$ colors, morphological $t$-types, and redshifts,
is only available in electronic form at the CDS via anonymous
ftp to \href{http://cdsarc.u-strasbg.fr}{cdsarc.u-strasbg.fr}
(\href{ftp://130.79.128.5}{130.79.128.5}) or via
\href{http://cdsweb.u-strasbg.fr/cgi-bin/qcat?J/A+A/}
{http://cdsweb.u-strasbg.fr/cgi-bin/qcat?J/A+A/}.

\begin{acknowledgements}

We thank the anonymous referee for their thoughtful comments and constructive suggestions, which have helped us improve the paper.
The research was supported by the Higher Education and Science Committee of MESCS
RA (Research project \textnumero~24LCG--1C021).

\end{acknowledgements}

\bibliographystyle{aa}
\bibliography{references}

@preamble{"\newcommand{\SortNoop}[1]{}"}

@ARTICLE{2022ApJS..262...36S,
       author = {{S{\'a}nchez}, S.~F. and {Barrera-Ballesteros}, J.~K. and {Lacerda}, E. and {Mej{\'\i}a-Narvaez}, A. and {Camps-Fari{\~n}a}, A. and {Bruzual}, Gustavo and {Espinosa-Ponce}, C. and {Rodr{\'\i}guez-Puebla}, A. and {Calette}, A.~R. and {Ibarra-Medel}, H. and {Avila-Reese}, V. and {Hernandez-Toledo}, H. and {Bershady}, M.~A. and {Cano-Diaz}, M. and {Munguia-Cordova}, A.~M.},
        title = "{SDSS-IV MaNGA: pyPipe3D Analysis Release for 10,000 Galaxies}",
      journal = {\apjs},
         year = 2022,
        month = oct,
       volume = {262},
       number = {2},
          eid = {36},
        pages = {36},
          doi = {10.3847/1538-4365/ac7b8f},
archivePrefix = {arXiv},
       eprint = {2206.07062},
 primaryClass = {astro-ph.GA},
       adsurl = {https://ui.adsabs.harvard.edu/abs/2022ApJS..262...36S}
}

@ARTICLE{2006MNRAS.373.1389C,
       author = {{Conselice}, Christopher J.},
        title = "{The fundamental properties of galaxies and a new galaxy classification system}",
      journal = {\mnras},
         year = 2006,
        month = dec,
       volume = {373},
       number = {4},
        pages = {1389-1408},
          doi = {10.1111/j.1365-2966.2006.11114.x},
archivePrefix = {arXiv},
       eprint = {astro-ph/0610016},
 primaryClass = {astro-ph},
       adsurl = {https://ui.adsabs.harvard.edu/abs/2006MNRAS.373.1389C}
}

@ARTICLE{2002ApJS..143...73E,
       author = {{Eskridge}, Paul B. and {Frogel}, Jay A. and {Pogge}, Richard W. and {Quillen}, Alice C. and {Berlind}, Andreas A. and {Davies}, Roger L. and {DePoy}, D.~L. and {Gilbert}, Karoline M. and {Houdashelt}, Mark L. and {Kuchinski}, Leslie E. and {Ram{\'\i}rez}, Solange V. and {Sellgren}, K. and {Stutz}, Amelia and {Terndrup}, Donald M. and {Tiede}, Glenn P.},
        title = "{Near-Infrared and Optical Morphology of Spiral Galaxies}",
      journal = {\apjs},
         year = 2002,
        month = nov,
       volume = {143},
       number = {1},
        pages = {73-111},
          doi = {10.1086/342340},
archivePrefix = {arXiv},
       eprint = {astro-ph/0206320},
 primaryClass = {astro-ph},
       adsurl = {https://ui.adsabs.harvard.edu/abs/2002ApJS..143...73E}
}

@ARTICLE{2018RMxAA..54..217S,
       author = {{S{\'a}nchez}, S.~F. and {Avila-Reese}, V. and {Hernandez-Toledo}, H. and {Cortes-Su{\'a}rez}, E. and {Rodr{\'\i}guez-Puebla}, A. and {Ibarra-Medel}, H. and {Cano-D{\'\i}az}, M. and {Barrera-Ballesteros}, J.~K. and {Negrete}, C.~A. and {Calette}, A.~R. and {de Lorenzo-C{\'a}ceres}, A. and {Ortega-Minakata}, R.~A. and {Aquino}, E. and {Valenzuela}, O. and {Clemente}, J.~C. and {Storchi-Bergmann}, T. and {Riffel}, R. and {Schimoia}, J. and {Riffel}, R.~A. and {Rembold}, S.~B. and {Brownstein}, J.~R. and {Pan}, K. and {Yates}, R. and {Mallmann}, N. and {Bitsakis}, T.},
        title = "{SDSS IV MaNGA - Properties of AGN Host Galaxies}",
      journal = {\rmxaa},
         year = 2018,
        month = apr,
       volume = {54},
        pages = {217-260},
          doi = {10.48550/arXiv.1709.05438},
archivePrefix = {arXiv},
       eprint = {1709.05438},
 primaryClass = {astro-ph.GA},
       adsurl = {https://ui.adsabs.harvard.edu/abs/2018RMxAA..54..217S}
}

@ARTICLE{2010MNRAS.403.1036C,
       author = {{Cid Fernandes}, R. and {Stasi{\'n}ska}, G. and {Schlickmann}, M.~S. and {Mateus}, A. and {Vale Asari}, N. and {Schoenell}, W. and {Sodr{\'e}}, L.},
        title = "{Alternative diagnostic diagrams and the `forgotten' population of weak line galaxies in the SDSS}",
      journal = {\mnras},
         year = 2010,
        month = apr,
       volume = {403},
       number = {2},
        pages = {1036-1053},
          doi = {10.1111/j.1365-2966.2009.16185.x},
archivePrefix = {arXiv},
       eprint = {0912.1643},
 primaryClass = {astro-ph.CO},
       adsurl = {https://ui.adsabs.harvard.edu/abs/2010MNRAS.403.1036C}
}

@ARTICLE{2017MNRAS.472.4382R,
       author = {{Rembold}, Sandro B. and {Shimoia}, J{\'a}derson S. and {Storchi-Bergmann}, Thaisa and {Riffel}, Rog{\'e}rio and {Riffel}, Rogemar A. and {Mallmann}, N{\'\i}colas D. and {do Nascimento}, Jana{\'\i}na C. and {Moreira}, Thales N. and {Ilha}, Gabriele S. and {Machado}, Alice D. and {Cirolini}, Rafael and {da Costa}, Luiz N. and {Maia}, Marcio A.~G. and {Santiago}, Bas{\'\i}lio X. and {Schneider}, Donald P. and {Wylezalek}, Dominika and {Bizyaev}, Dmitry and {Pan}, Kaike and {M{\"u}ller-S{\'a}nchez}, Francisco},
        title = "{The first 62 AGNs observed with SDSS-IV MaNGA - I. Their characterization and definition of a control sample}",
      journal = {\mnras},
         year = 2017,
        month = dec,
       volume = {472},
       number = {4},
        pages = {4382-4403},
          doi = {10.1093/mnras/stx2264},
archivePrefix = {arXiv},
       eprint = {1709.10086},
 primaryClass = {astro-ph.GA},
       adsurl = {https://ui.adsabs.harvard.edu/abs/2017MNRAS.472.4382R}
}

@ARTICLE{2017AJ....154...28B,
       author = {{Blanton}, Michael R. and {Bershady}, Matthew A. and {Abolfathi}, Bela and {Albareti}, Franco D. and {Allende Prieto}, Carlos and {Almeida}, Andres and {Alonso-Garc{\'\i}a}, Javier and {Anders}, Friedrich and {Anderson}, Scott F. and {Andrews}, Brett and {Aquino-Ort{\'\i}z}, Erik and {Arag{\'o}n-Salamanca}, Alfonso and {Argudo-Fern{\'a}ndez}, Maria and {Armengaud}, Eric and {Aubourg}, Eric and {Avila-Reese}, Vladimir and {Badenes}, Carles and {Bailey}, Stephen and {Barger}, Kathleen A. and {Barrera-Ballesteros}, Jorge and {Bartosz}, Curtis and {Bates}, Dominic and {Baumgarten}, Falk and {Bautista}, Julian and {Beaton}, Rachael and {Beers}, Timothy C. and {Belfiore}, Francesco and {Bender}, Chad F. and {Berlind}, Andreas A. and {Bernardi}, Mariangela and {Beutler}, Florian and {Bird}, Jonathan C. and {Bizyaev}, Dmitry and {Blanc}, Guillermo A. and {Blomqvist}, Michael and {Bolton}, Adam S. and {Boquien}, M{\'e}d{\'e}ric and {Borissova}, Jura and {van den Bosch}, Remco and {Bovy}, Jo and {Brandt}, William N. and {Brinkmann}, Jonathan and {Brownstein}, Joel R. and {Bundy}, Kevin and {Burgasser}, Adam J. and {Burtin}, Etienne and {Busca}, Nicol{\'a}s G. and {Cappellari}, Michele and {Delgado Carigi}, Maria Leticia and {Carlberg}, Joleen K. and {Carnero Rosell}, Aurelio and {Carrera}, Ricardo and {Chanover}, Nancy J. and {Cherinka}, Brian and {Cheung}, Edmond and {G{\'o}mez Maqueo Chew}, Yilen and {Chiappini}, Cristina and {Choi}, Peter Doohyun and {Chojnowski}, Drew and {Chuang}, Chia-Hsun and {Chung}, Haeun and {Cirolini}, Rafael Fernando and {Clerc}, Nicolas and {Cohen}, Roger E. and {Comparat}, Johan and {da Costa}, Luiz and {Cousinou}, Marie-Claude and {Covey}, Kevin and {Crane}, Jeffrey D. and {Croft}, Rupert A.~C. and {Cruz-Gonzalez}, Irene and {Garrido Cuadra}, Daniel and {Cunha}, Katia and {Damke}, Guillermo J. and {Darling}, Jeremy and {Davies}, Roger and {Dawson}, Kyle and {de la Macorra}, Axel and {Dell'Agli}, Flavia and {De Lee}, Nathan and {Delubac}, Timoth{\'e}e and {Di Mille}, Francesco and {Diamond-Stanic}, Aleks and {Cano-D{\'\i}az}, Mariana and {Donor}, John and {Downes}, Juan Jos{\'e} and {Drory}, Niv and {du Mas des Bourboux}, H{\'e}lion and {Duckworth}, Christopher J. and {Dwelly}, Tom and {Dyer}, Jamie and {Ebelke}, Garrett and {Eigenbrot}, Arthur D. and {Eisenstein}, Daniel J. and {Emsellem}, Eric and {Eracleous}, Mike and {Escoffier}, Stephanie and {Evans}, Michael L. and {Fan}, Xiaohui and {Fern{\'a}ndez-Alvar}, Emma and {Fernandez-Trincado}, J.~G. and {Feuillet}, Diane K. and {Finoguenov}, Alexis and {Fleming}, Scott W. and {Font-Ribera}, Andreu and {Fredrickson}, Alexander and {Freischlad}, Gordon and {Frinchaboy}, Peter M. and {Fuentes}, Carla E. and {Galbany}, Llu{\'\i}s and {Garcia-Dias}, R. and {Garc{\'\i}a-Hern{\'a}ndez}, D.~A. and {Gaulme}, Patrick and {Geisler}, Doug and {Gelfand}, Joseph D. and {Gil-Mar{\'\i}n}, H{\'e}ctor and {Gillespie}, Bruce A. and {Goddard}, Daniel and {Gonzalez-Perez}, Violeta and {Grabowski}, Kathleen and {Green}, Paul J. and {Grier}, Catherine J. and {Gunn}, James E. and {Guo}, Hong and {Guy}, Julien and {Hagen}, Alex and {Hahn}, ChangHoon and {Hall}, Matthew and {Harding}, Paul and {Hasselquist}, Sten and {Hawley}, Suzanne L. and {Hearty}, Fred and {Gonzalez Hern{\'a}ndez}, Jonay I. and {Ho}, Shirley and {Hogg}, David W. and {Holley-Bockelmann}, Kelly and {Holtzman}, Jon A. and {Holzer}, Parker H. and {Huehnerhoff}, Joseph and {Hutchinson}, Timothy A. and {Hwang}, Ho Seong and {Ibarra-Medel}, H{\'e}ctor J. and {da Silva Ilha}, Gabriele and {Ivans}, Inese I. and {Ivory}, KeShawn and {Jackson}, Kelly and {Jensen}, Trey W. and {Johnson}, Jennifer A. and {Jones}, Amy and {J{\"o}nsson}, Henrik and {Jullo}, Eric and {Kamble}, Vikrant and {Kinemuchi}, Karen and {Kirkby}, David and {Kitaura}, Francisco-Shu and {Klaene}, Mark and {Knapp}, Gillian R. and {Kneib}, Jean-Paul and {Kollmeier}, Juna A. and {Lacerna}, Ivan and {Lane}, Richard R. and {Lang}, Dustin and {Law}, David R. and {Lazarz}, Daniel and {Lee}, Youngbae and {Le Goff}, Jean-Marc and {Liang}, Fu-Heng and {Li}, Cheng and {Li}, Hongyu and {Lian}, Jianhui and {Lima}, Marcos and {Lin}, Lihwai and {Lin}, Yen-Ting and {Bertran de Lis}, Sara and {Liu}, Chao and {de Icaza Lizaola}, Miguel Angel C. and {Long}, Dan and {Lucatello}, Sara and {Lundgren}, Britt and {MacDonald}, Nicholas K. and {Deconto Machado}, Alice and {MacLeod}, Chelsea L. and {Mahadevan}, Suvrath and {Geimba Maia}, Marcio Antonio and {Maiolino}, Roberto and {Majewski}, Steven R. and {Malanushenko}, Elena and {Malanushenko}, Viktor and {Manchado}, Arturo and {Mao}, Shude and {Maraston}, Claudia and {Marques-Chaves}, Rui and {Masseron}, Thomas and {Masters}, Karen L. and {McBride}, Cameron K. and {McDermid}, Richard M. and {McGrath}, Brianne and {McGreer}, Ian D. and {Medina Pe{\~n}a}, Nicol{\'a}s and {Melendez}, Matthew},
        title = "{Sloan Digital Sky Survey IV: Mapping the Milky Way, Nearby Galaxies, and the Distant Universe}",
      journal = {\aj},
         year = 2017,
        month = jul,
       volume = {154},
       number = {1},
          eid = {28},
        pages = {28},
          doi = {10.3847/1538-3881/aa7567},
archivePrefix = {arXiv},
       eprint = {1703.00052},
 primaryClass = {astro-ph.GA},
       adsurl = {https://ui.adsabs.harvard.edu/abs/2017AJ....154...28B}
}

@ARTICLE{2015ApJ...798....7B,
       author = {{Bundy}, Kevin and {Bershady}, Matthew A. and {Law}, David R. and {Yan}, Renbin and {Drory}, Niv and {MacDonald}, Nicholas and {Wake}, David A. and {Cherinka}, Brian and {S{\'a}nchez-Gallego}, Jos{\'e} R. and {Weijmans}, Anne-Marie and {Thomas}, Daniel and {Tremonti}, Christy and {Masters}, Karen and {Coccato}, Lodovico and {Diamond-Stanic}, Aleksandar M. and {Arag{\'o}n-Salamanca}, Alfonso and {Avila-Reese}, Vladimir and {Badenes}, Carles and {Falc{\'o}n-Barroso}, J{\'e}sus and {Belfiore}, Francesco and {Bizyaev}, Dmitry and {Blanc}, Guillermo A. and {Bland-Hawthorn}, Joss and {Blanton}, Michael R. and {Brownstein}, Joel R. and {Byler}, Nell and {Cappellari}, Michele and {Conroy}, Charlie and {Dutton}, Aaron A. and {Emsellem}, Eric and {Etherington}, James and {Frinchaboy}, Peter M. and {Fu}, Hai and {Gunn}, James E. and {Harding}, Paul and {Johnston}, Evelyn J. and {Kauffmann}, Guinevere and {Kinemuchi}, Karen and {Klaene}, Mark A. and {Knapen}, Johan H. and {Leauthaud}, Alexie and {Li}, Cheng and {Lin}, Lihwai and {Maiolino}, Roberto and {Malanushenko}, Viktor and {Malanushenko}, Elena and {Mao}, Shude and {Maraston}, Claudia and {McDermid}, Richard M. and {Merrifield}, Michael R. and {Nichol}, Robert C. and {Oravetz}, Daniel and {Pan}, Kaike and {Parejko}, John K. and {Sanchez}, Sebastian F. and {Schlegel}, David and {Simmons}, Audrey and {Steele}, Oliver and {Steinmetz}, Matthias and {Thanjavur}, Karun and {Thompson}, Benjamin A. and {Tinker}, Jeremy L. and {van den Bosch}, Remco C.~E. and {Westfall}, Kyle B. and {Wilkinson}, David and {Wright}, Shelley and {Xiao}, Ting and {Zhang}, Kai},
        title = "{Overview of the SDSS-IV MaNGA Survey: Mapping nearby Galaxies at Apache Point Observatory}",
      journal = {\apj},
         year = 2015,
        month = jan,
       volume = {798},
       number = {1},
          eid = {7},
        pages = {7},
          doi = {10.1088/0004-637X/798/1/7},
archivePrefix = {arXiv},
       eprint = {1412.1482},
 primaryClass = {astro-ph.GA},
       adsurl = {https://ui.adsabs.harvard.edu/abs/2015ApJ...798....7B}
}

@ARTICLE{2022ApJS..259...35A,
       author = {{Abdurro'uf} and {Accetta}, Katherine and {Aerts}, Conny and {Silva Aguirre}, V{\'\i}ctor and {Ahumada}, Romina and {Ajgaonkar}, Nikhil and {Filiz Ak}, N. and {Alam}, Shadab and {Allende Prieto}, Carlos and {Almeida}, Andr{\'e}s and {Anders}, Friedrich and {Anderson}, Scott F. and {Andrews}, Brett H. and {Anguiano}, Borja and {Aquino-Ort{\'\i}z}, Erik and {Arag{\'o}n-Salamanca}, Alfonso and {Argudo-Fern{\'a}ndez}, Maria and {Ata}, Metin and {Aubert}, Marie and {Avila-Reese}, Vladimir and {Badenes}, Carles and {Barb{\'a}}, Rodolfo H. and {Barger}, Kat and {Barrera-Ballesteros}, Jorge K. and {Beaton}, Rachael L. and {Beers}, Timothy C. and {Belfiore}, Francesco and {Bender}, Chad F. and {Bernardi}, Mariangela and {Bershady}, Matthew A. and {Beutler}, Florian and {Bidin}, Christian Moni and {Bird}, Jonathan C. and {Bizyaev}, Dmitry and {Blanc}, Guillermo A. and {Blanton}, Michael R. and {Boardman}, Nicholas Fraser and {Bolton}, Adam S. and {Boquien}, M{\'e}d{\'e}ric and {Borissova}, Jura and {Bovy}, Jo and {Brandt}, W.~N. and {Brown}, Jordan and {Brownstein}, Joel R. and {Brusa}, Marcella and {Buchner}, Johannes and {Bundy}, Kevin and {Burchett}, Joseph N. and {Bureau}, Martin and {Burgasser}, Adam and {Cabang}, Tuesday K. and {Campbell}, Stephanie and {Cappellari}, Michele and {Carlberg}, Joleen K. and {Wanderley}, F{\'a}bio Carneiro and {Carrera}, Ricardo and {Cash}, Jennifer and {Chen}, Yan-Ping and {Chen}, Wei-Huai and {Cherinka}, Brian and {Chiappini}, Cristina and {Choi}, Peter Doohyun and {Chojnowski}, S. Drew and {Chung}, Haeun and {Clerc}, Nicolas and {Cohen}, Roger E. and {Comerford}, Julia M. and {Comparat}, Johan and {da Costa}, Luiz and {Covey}, Kevin and {Crane}, Jeffrey D. and {Cruz-Gonzalez}, Irene and {Culhane}, Connor and {Cunha}, Katia and {Dai}, Y. Sophia and {Damke}, Guillermo and {Darling}, Jeremy and {Davidson}, Jr., James W. and {Davies}, Roger and {Dawson}, Kyle and {De Lee}, Nathan and {Diamond-Stanic}, Aleksandar M. and {Cano-D{\'\i}az}, Mariana and {S{\'a}nchez}, Helena Dom{\'\i}nguez and {Donor}, John and {Duckworth}, Chris and {Dwelly}, Tom and {Eisenstein}, Daniel J. and {Elsworth}, Yvonne P. and {Emsellem}, Eric and {Eracleous}, Mike and {Escoffier}, Stephanie and {Fan}, Xiaohui and {Farr}, Emily and {Feng}, Shuai and {Fern{\'a}ndez-Trincado}, Jos{\'e} G. and {Feuillet}, Diane and {Filipp}, Andreas and {Fillingham}, Sean P. and {Frinchaboy}, Peter M. and {Fromenteau}, Sebastien and {Galbany}, Llu{\'\i}s and {Garc{\'\i}a}, Rafael A. and {Garc{\'\i}a-Hern{\'a}ndez}, D.~A. and {Ge}, Junqiang and {Geisler}, Doug and {Gelfand}, Joseph and {G{\'e}ron}, Tobias and {Gibson}, Benjamin J. and {Goddy}, Julian and {Godoy-Rivera}, Diego and {Grabowski}, Kathleen and {Green}, Paul J. and {Greener}, Michael and {Grier}, Catherine J. and {Griffith}, Emily and {Guo}, Hong and {Guy}, Julien and {Hadjara}, Massinissa and {Harding}, Paul and {Hasselquist}, Sten and {Hayes}, Christian R. and {Hearty}, Fred and {Hern{\'a}ndez}, Jes{\'u}s and {Hill}, Lewis and {Hogg}, David W. and {Holtzman}, Jon A. and {Horta}, Danny and {Hsieh}, Bau-Ching and {Hsu}, Chin-Hao and {Hsu}, Yun-Hsin and {Huber}, Daniel and {Huertas-Company}, Marc and {Hutchinson}, Brian and {Hwang}, Ho Seong and {Ibarra-Medel}, H{\'e}ctor J. and {Chitham}, Jacob Ider and {Ilha}, Gabriele S. and {Imig}, Julie and {Jaekle}, Will and {Jayasinghe}, Tharindu and {Ji}, Xihan and {Johnson}, Jennifer A. and {Jones}, Amy and {J{\"o}nsson}, Henrik and {Katkov}, Ivan and {Khalatyan}, Dr., Arman and {Kinemuchi}, Karen and {Kisku}, Shobhit and {Knapen}, Johan H. and {Kneib}, Jean-Paul and {Kollmeier}, Juna A. and {Kong}, Miranda and {Kounkel}, Marina and {Kreckel}, Kathryn and {Krishnarao}, Dhanesh and {Lacerna}, Ivan and {Lane}, Richard R. and {Langgin}, Rachel and {Lavender}, Ramon and {Law}, David R. and {Lazarz}, Daniel and {Leung}, Henry W. and {Leung}, Ho-Hin and {Lewis}, Hannah M. and {Li}, Cheng and {Li}, Ran and {Lian}, Jianhui and {Liang}, Fu-Heng and {Lin}, Lihwai and {Lin}, Yen-Ting and {Lin}, Sicheng and {Lintott}, Chris and {Long}, Dan and {Longa-Pe{\~n}a}, Pen{\'e}lope and {L{\'o}pez-Cob{\'a}}, Carlos and {Lu}, Shengdong and {Lundgren}, Britt F. and {Luo}, Yuanze and {Mackereth}, J. Ted and {de la Macorra}, Axel and {Mahadevan}, Suvrath and {Majewski}, Steven R. and {Manchado}, Arturo and {Mandeville}, Travis and {Maraston}, Claudia and {Margalef-Bentabol}, Berta and {Masseron}, Thomas and {Masters}, Karen L. and {Mathur}, Savita and {McDermid}, Richard M. and {Mckay}, Myles and {Merloni}, Andrea and {Merrifield}, Michael and {Meszaros}, Szabolcs and {Miglio}, Andrea and {Di Mille}, Francesco and {Minniti}, Dante and {Minsley}, Rebecca and {Monachesi}, Antonela},
        title = "{The Seventeenth Data Release of the Sloan Digital Sky Surveys: Complete Release of MaNGA, MaStar, and APOGEE-2 Data}",
      journal = {\apjs},
         year = 2022,
        month = apr,
       volume = {259},
       number = {2},
          eid = {35},
        pages = {35},
          doi = {10.3847/1538-4365/ac4414},
archivePrefix = {arXiv},
       eprint = {2112.02026},
 primaryClass = {astro-ph.GA},
       adsurl = {https://ui.adsabs.harvard.edu/abs/2022ApJS..259...35A}
}

@ARTICLE{2022MNRAS.509.3966W,
       author = {{Walmsley}, Mike and {Lintott}, Chris and {G{\'e}ron}, Tobias and {Kruk}, Sandor and {Krawczyk}, Coleman and {Willett}, Kyle W. and {Bamford}, Steven and {Kelvin}, Lee S. and {Fortson}, Lucy and {Gal}, Yarin and {Keel}, William and {Masters}, Karen L. and {Mehta}, Vihang and {Simmons}, Brooke D. and {Smethurst}, Rebecca and {Smith}, Lewis and {Baeten}, Elisabeth M. and {Macmillan}, Christine},
        title = "{Galaxy Zoo DECaLS: Detailed visual morphology measurements from volunteers and deep learning for 314 000 galaxies}",
      journal = {\mnras},
         year = 2022,
        month = jan,
       volume = {509},
       number = {3},
        pages = {3966-3988},
          doi = {10.1093/mnras/stab2093},
archivePrefix = {arXiv},
       eprint = {2102.08414},
 primaryClass = {astro-ph.GA},
       adsurl = {https://ui.adsabs.harvard.edu/abs/2022MNRAS.509.3966W}
}

@ARTICLE{2015ApJS..217...32B,
       author = {{Buta}, Ronald J. and {Sheth}, Kartik and {Athanassoula}, E. and {Bosma}, A. and {Knapen}, Johan H. and {Laurikainen}, Eija and {Salo}, Heikki and {Elmegreen}, Debra and {Ho}, Luis C. and {Zaritsky}, Dennis and {Courtois}, Helene and {Hinz}, Joannah L. and {Mu{\~n}oz-Mateos}, Juan-Carlos and {Kim}, Taehyun and {Regan}, Michael W. and {Gadotti}, Dimitri A. and {Gil de Paz}, Armando and {Laine}, Jarkko and {Men{\'e}ndez-Delmestre}, Kar{\'\i}n and {Comer{\'o}n}, S{\'e}bastien and {Erroz Ferrer}, Santiago and {Seibert}, Mark and {Mizusawa}, Trisha and {Holwerda}, Benne and {Madore}, Barry F.},
        title = "{A Classical Morphological Analysis of Galaxies in the Spitzer Survey of Stellar Structure in Galaxies (S4G)}",
      journal = {\apjs},
         year = 2015,
        month = apr,
       volume = {217},
       number = {2},
          eid = {32},
        pages = {32},
          doi = {10.1088/0067-0049/217/2/32},
archivePrefix = {arXiv},
       eprint = {1501.00454},
 primaryClass = {astro-ph.GA},
       adsurl = {https://ui.adsabs.harvard.edu/abs/2015ApJS..217...32B}
}

@ARTICLE{2023A&A...671A.141M,
       author = {{Mart{\'\i}nez-Delgado}, David and {Cooper}, Andrew P. and {Rom{\'a}n}, Javier and {Pillepich}, Annalisa and {Erkal}, Denis and {Pearson}, Sarah and {Moustakas}, John and {Laporte}, Chervin F.~P. and {Laine}, Seppo and {Akhlaghi}, Mohammad and {Lang}, Dustin and {Makarov}, Dmitry and {Borlaff}, Alejandro S. and {Donatiello}, Giuseppe and {Pearson}, William J. and {Mir{\'o}-Carretero}, Juan and {Cuillandre}, Jean-Charles and {Dom{\'\i}nguez}, Helena and {Roca-F{\`a}brega}, Santi and {Frenk}, Carlos S. and {Schmidt}, Judy and {G{\'o}mez-Flechoso}, Mar{\'\i}a A. and {Guzman}, Rafael and {Libeskind}, Noam I. and {Dey}, Arjun and {Weaver}, Benjamin A. and {Schlegel}, David and {Myers}, Adam D. and {Valdes}, Frank G.},
        title = "{Hidden depths in the local Universe: The Stellar Stream Legacy Survey}",
      journal = {\aap},
         year = 2023,
        month = mar,
       volume = {671},
          eid = {A141},
        pages = {A141},
          doi = {10.1051/0004-6361/202245011},
archivePrefix = {arXiv},
       eprint = {2104.06071},
 primaryClass = {astro-ph.GA},
       adsurl = {https://ui.adsabs.harvard.edu/abs/2023A&A...671A.141M}
}

@ARTICLE{2023JCAP...08..044S,
       author = {{Sarkar}, Suman and {Narayanan}, Ganesh and {Banerjee}, Arunima},
        title = "{Analyzing the cosmic web environment in the vicinity of grand-design and flocculent spirals with local geometric index}",
      journal = {\jcap},
         year = 2023,
        month = aug,
       volume = {2023},
       number = {8},
          eid = {044},
        pages = {044},
          doi = {10.1088/1475-7516/2023/08/044},
archivePrefix = {arXiv},
       eprint = {2302.08087},
 primaryClass = {astro-ph.GA},
       adsurl = {https://ui.adsabs.harvard.edu/abs/2023JCAP...08..044S}
}

@ARTICLE{2011ApJ...737...32E,
       author = {{Elmegreen}, Debra Meloy and {Elmegreen}, Bruce G. and {Yau}, Andrew and {Athanassoula}, E. and {Bosma}, Albert and {Buta}, Ronald J. and {Helou}, George and {Ho}, Luis C. and {Gadotti}, Dimitri A. and {Knapen}, Johan H. and {Laurikainen}, Eija and {Madore}, Barry F. and {Masters}, Karen L. and {Meidt}, Sharon E. and {Men{\'e}ndez-Delmestre}, Kar{\'\i}n and {Regan}, Michael W. and {Salo}, Heikki and {Sheth}, Kartik and {Zaritsky}, Dennis and {Aravena}, Manuel and {Skibba}, Ramin and {Hinz}, Joannah L. and {Laine}, Jarkko and {Gil de Paz}, Armando and {Mu{\~n}oz-Mateos}, Juan-Carlos and {Seibert}, Mark and {Mizusawa}, Trisha and {Kim}, Taehyun and {Erroz Ferrer}, Santiago},
        title = "{Grand Design and Flocculent Spirals in the Spitzer Survey of Stellar Structure in Galaxies (S$^{4}$G)}",
      journal = {\apj},
         year = 2011,
        month = aug,
       volume = {737},
       number = {1},
          eid = {32},
        pages = {32},
          doi = {10.1088/0004-637X/737/1/32},
archivePrefix = {arXiv},
       eprint = {1106.4840},
 primaryClass = {astro-ph.GA},
       adsurl = {https://ui.adsabs.harvard.edu/abs/2011ApJ...737...32E}
}

@ARTICLE{2018ApJ...869...29Y,
       author = {{Yu}, Si-Yue and {Ho}, Luis C.},
        title = "{Dependence of the Spiral Arms Pitch Angle on Wavelength as a Test of the Density Wave Theory}",
      journal = {\apj},
         year = 2018,
        month = dec,
       volume = {869},
       number = {1},
          eid = {29},
        pages = {29},
          doi = {10.3847/1538-4357/aaeacd},
archivePrefix = {arXiv},
       eprint = {1810.08979},
 primaryClass = {astro-ph.GA},
       adsurl = {https://ui.adsabs.harvard.edu/abs/2018ApJ...869...29Y}
}

@ARTICLE{2020ApJ...900..150Y,
       author = {{Yu}, Si-Yue and {Ho}, Luis C.},
        title = "{The Statistical Properties of Spiral Arms in Nearby Disk Galaxies}",
      journal = {\apj},
         year = 2020,
        month = sep,
       volume = {900},
       number = {2},
          eid = {150},
        pages = {150},
          doi = {10.3847/1538-4357/abac5b},
       adsurl = {https://ui.adsabs.harvard.edu/abs/2020ApJ...900..150Y}
}

@ARTICLE{2013MNRAS.435.2835W,
       author = {{Willett}, Kyle W. and {Lintott}, Chris J. and {Bamford}, Steven P. and {Masters}, Karen L. and {Simmons}, Brooke D. and {Casteels}, Kevin R.~V. and {Edmondson}, Edward M. and {Fortson}, Lucy F. and {Kaviraj}, Sugata and {Keel}, William C. and {Melvin}, Thomas and {Nichol}, Robert C. and {Raddick}, M. Jordan and {Schawinski}, Kevin and {Simpson}, Robert J. and {Skibba}, Ramin A. and {Smith}, Arfon M. and {Thomas}, Daniel},
        title = "{Galaxy Zoo 2: detailed morphological classifications for 304 122 galaxies from the Sloan Digital Sky Survey}",
      journal = {\mnras},
         year = 2013,
        month = nov,
       volume = {435},
       number = {4},
        pages = {2835-2860},
          doi = {10.1093/mnras/stt1458},
archivePrefix = {arXiv},
       eprint = {1308.3496},
 primaryClass = {astro-ph.CO},
       adsurl = {https://ui.adsabs.harvard.edu/abs/2013MNRAS.435.2835W}
}

@ARTICLE{2021MNRAS.507.3923M,
       author = {{Masters}, Karen L. and {Krawczyk}, Coleman and {Shamsi}, Shoaib and {Todd}, Alexander and {Finnegan}, Daniel and {Bershady}, Matthew and {Bundy}, Kevin and {Cherinka}, Brian and {Fraser-McKelvie}, Amelia and {Krishnarao}, Dhanesh and {Kruk}, Sandor and {Lane}, Richard R. and {Law}, David and {Lintott}, Chris and {Merrifield}, Michael and {Simmons}, Brooke and {Weijmans}, Anne-Marie and {Yan}, Renbin},
        title = "{Galaxy Zoo: 3D - crowdsourced bar, spiral, and foreground star masks for MaNGA target galaxies}",
      journal = {\mnras},
         year = 2021,
        month = nov,
       volume = {507},
       number = {3},
        pages = {3923-3935},
          doi = {10.1093/mnras/stab2282},
archivePrefix = {arXiv},
       eprint = {2108.02065},
 primaryClass = {astro-ph.GA},
       adsurl = {https://ui.adsabs.harvard.edu/abs/2021MNRAS.507.3923M}
}

@ARTICLE{2024AJ....168...12S,
       author = {{Smith}, Beverly J. and {Watson}, Matthew and {Giroux}, Mark L. and {Struck}, Curtis},
        title = "{Grand Design versus Multiarmed Spiral Galaxies: Dependence on Galaxy Structure}",
      journal = {\aj},
         year = 2024,
        month = jul,
       volume = {168},
       number = {1},
          eid = {12},
        pages = {12},
          doi = {10.3847/1538-3881/ad46fb},
archivePrefix = {arXiv},
       eprint = {2405.01516},
 primaryClass = {astro-ph.GA},
       adsurl = {https://ui.adsabs.harvard.edu/abs/2024AJ....168...12S}
}

@ARTICLE{2018MNRAS.481..566K,
       author = {{Karapetyan}, A.~G. and {Hakobyan}, A.~A. and {Barkhudaryan}, L.~V. and {Mamon}, G.~A. and {Kunth}, D. and {Adibekyan}, V. and {Turatto}, M.},
        title = "{The impact of spiral density waves on the distribution of supernovae}",
      journal = {\mnras},
         year = 2018,
        month = nov,
       volume = {481},
       number = {1},
        pages = {566-577},
          doi = {10.1093/mnras/sty2291},
archivePrefix = {arXiv},
       eprint = {1808.03099},
 primaryClass = {astro-ph.GA},
       adsurl = {https://ui.adsabs.harvard.edu/abs/2018MNRAS.481..566K}
}

@ARTICLE{1990NYASA.596...40E,
       author = {{Elmegreen}, B.~G.},
        title = "{Grand design, multiple arm, and flocculent spiral galaxies}",
      journal = {Annals of the New York Academy of Sciences},
         year = 1990,
        month = jun,
       volume = {596},
        pages = {40-52},
          doi = {10.1111/j.1749-6632.1990.tb27410.x},
       adsurl = {https://ui.adsabs.harvard.edu/abs/1990NYASA.596...40E}
}

@ARTICLE{2019ApJ...872...97L,
       author = {{Lee}, Yun Hee and {Ann}, Hong Bae and {Park}, Myeong-Gu},
        title = "{Bar Fraction in Early- and Late-type Spirals}",
      journal = {\apj},
         year = 2019,
        month = feb,
       volume = {872},
       number = {1},
          eid = {97},
        pages = {97},
          doi = {10.3847/1538-4357/ab0024},
archivePrefix = {arXiv},
       eprint = {1901.05183},
 primaryClass = {astro-ph.GA},
       adsurl = {https://ui.adsabs.harvard.edu/abs/2019ApJ...872...97L}
}

@ARTICLE{1982MNRAS.201.1035E,
       author = {{Elmegreen}, D.~M. and {Elmegreen}, B.~G. and {Dressler}, A.},
        title = "{Flocculent and grand design spiral arm structure in cluster galaxies}",
      journal = {\mnras},
         year = 1982,
        month = dec,
       volume = {201},
        pages = {1035-1039},
          doi = {10.1093/mnras/201.4.1035},
       adsurl = {https://ui.adsabs.harvard.edu/abs/1982MNRAS.201.1035E}
}

@ARTICLE{2017MNRAS.471.1070B,
       author = {{Bittner}, Adrian and {Gadotti}, Dimitri A. and {Elmegreen}, Bruce G. and {Athanassoula}, Evangelie and {Elmegreen}, Debra M. and {Bosma}, Albert and {Mu{\~n}oz-Mateos}, Juan-Carlos},
        title = "{How do spiral arm contrasts relate to bars, disc breaks and other fundamental galaxy properties?}",
      journal = {\mnras},
         year = 2017,
        month = oct,
       volume = {471},
       number = {1},
        pages = {1070-1087},
          doi = {10.1093/mnras/stx1646},
archivePrefix = {arXiv},
       eprint = {1706.09904},
 primaryClass = {astro-ph.GA},
       adsurl = {https://ui.adsabs.harvard.edu/abs/2017MNRAS.471.1070B}
}

@ARTICLE{2013JKAS...46..141A,
       author = {{Ann}, Hong Bae and {Lee}, Hyun-Rok},
        title = "{Spiral Arm Morphology of Nearby Galaxies}",
      journal = {Journal of Korean Astronomical Society},
         year = 2013,
        month = jun,
       volume = {46},
       number = {3},
        pages = {141-149},
          doi = {10.5303/JKAS.2013.46.3.141},
       adsurl = {https://ui.adsabs.harvard.edu/abs/2013JKAS...46..141A}
}

@ARTICLE{2019AJ....157..168D,
       author = {{Dey}, Arjun and {Schlegel}, David J. and {Lang}, Dustin and {Blum}, Robert and {Burleigh}, Kaylan and {Fan}, Xiaohui and {Findlay}, Joseph R. and {Finkbeiner}, Doug and {Herrera}, David and {Juneau}, St{\'e}phanie and {Landriau}, Martin and {Levi}, Michael and {McGreer}, Ian and {Meisner}, Aaron and {Myers}, Adam D. and {Moustakas}, John and {Nugent}, Peter and {Patej}, Anna and {Schlafly}, Edward F. and {Walker}, Alistair R. and {Valdes}, Francisco and {Weaver}, Benjamin A. and {Y{\`e}che}, Christophe and {Zou}, Hu and {Zhou}, Xu and {Abareshi}, Behzad and {Abbott}, T.~M.~C. and {Abolfathi}, Bela and {Aguilera}, C. and {Alam}, Shadab and {Allen}, Lori and {Alvarez}, A. and {Annis}, James and {Ansarinejad}, Behzad and {Aubert}, Marie and {Beechert}, Jacqueline and {Bell}, Eric F. and {BenZvi}, Segev Y. and {Beutler}, Florian and {Bielby}, Richard M. and {Bolton}, Adam S. and {Brice{\~n}o}, C{\'e}sar and {Buckley-Geer}, Elizabeth J. and {Butler}, Karen and {Calamida}, Annalisa and {Carlberg}, Raymond G. and {Carter}, Paul and {Casas}, Ricard and {Castander}, Francisco J. and {Choi}, Yumi and {Comparat}, Johan and {Cukanovaite}, Elena and {Delubac}, Timoth{\'e}e and {DeVries}, Kaitlin and {Dey}, Sharmila and {Dhungana}, Govinda and {Dickinson}, Mark and {Ding}, Zhejie and {Donaldson}, John B. and {Duan}, Yutong and {Duckworth}, Christopher J. and {Eftekharzadeh}, Sarah and {Eisenstein}, Daniel J. and {Etourneau}, Thomas and {Fagrelius}, Parker A. and {Farihi}, Jay and {Fitzpatrick}, Mike and {Font-Ribera}, Andreu and {Fulmer}, Leah and {G{\"a}nsicke}, Boris T. and {Gaztanaga}, Enrique and {George}, Koshy and {Gerdes}, David W. and {Gontcho}, Satya Gontcho A. and {Gorgoni}, Claudio and {Green}, Gregory and {Guy}, Julien and {Harmer}, Diane and {Hernandez}, M. and {Honscheid}, Klaus and {Huang}, Lijuan Wendy and {James}, David J. and {Jannuzi}, Buell T. and {Jiang}, Linhua and {Joyce}, Richard and {Karcher}, Armin and {Karkar}, Sonia and {Kehoe}, Robert and {Kneib}, Jean-Paul and {Kueter-Young}, Andrea and {Lan}, Ting-Wen and {Lauer}, Tod R. and {Le Guillou}, Laurent and {Le Van Suu}, Auguste and {Lee}, Jae Hyeon and {Lesser}, Michael and {Perreault Levasseur}, Laurence and {Li}, Ting S. and {Mann}, Justin L. and {Marshall}, Robert and {Mart{\'\i}nez-V{\'a}zquez}, C.~E. and {Martini}, Paul and {du Mas des Bourboux}, H{\'e}lion and {McManus}, Sean and {Meier}, Tobias Gabriel and {M{\'e}nard}, Brice and {Metcalfe}, Nigel and {Mu{\~n}oz-Guti{\'e}rrez}, Andrea and {Najita}, Joan and {Napier}, Kevin and {Narayan}, Gautham and {Newman}, Jeffrey A. and {Nie}, Jundan and {Nord}, Brian and {Norman}, Dara J. and {Olsen}, Knut A.~G. and {Paat}, Anthony and {Palanque-Delabrouille}, Nathalie and {Peng}, Xiyan and {Poppett}, Claire L. and {Poremba}, Megan R. and {Prakash}, Abhishek and {Rabinowitz}, David and {Raichoor}, Anand and {Rezaie}, Mehdi and {Robertson}, A.~N. and {Roe}, Natalie A. and {Ross}, Ashley J. and {Ross}, Nicholas P. and {Rudnick}, Gregory and {Safonova}, Sasha and {Saha}, Abhijit and {S{\'a}nchez}, F. Javier and {Savary}, Elodie and {Schweiker}, Heidi and {Scott}, Adam and {Seo}, Hee-Jong and {Shan}, Huanyuan and {Silva}, David R. and {Slepian}, Zachary and {Soto}, Christian and {Sprayberry}, David and {Staten}, Ryan and {Stillman}, Coley M. and {Stupak}, Robert J. and {Summers}, David L. and {Sien Tie}, Suk and {Tirado}, H. and {Vargas-Maga{\~n}a}, Mariana and {Vivas}, A. Katherina and {Wechsler}, Risa H. and {Williams}, Doug and {Yang}, Jinyi and {Yang}, Qian and {Yapici}, Tolga and {Zaritsky}, Dennis and {Zenteno}, A. and {Zhang}, Kai and {Zhang}, Tianmeng and {Zhou}, Rongpu and {Zhou}, Zhimin},
        title = "{Overview of the DESI Legacy Imaging Surveys}",
      journal = {\aj},
         year = 2019,
        month = may,
       volume = {157},
       number = {5},
          eid = {168},
        pages = {168},
          doi = {10.3847/1538-3881/ab089d},
archivePrefix = {arXiv},
       eprint = {1804.08657},
 primaryClass = {astro-ph.IM},
       adsurl = {https://ui.adsabs.harvard.edu/abs/2019AJ....157..168D}
}

@ARTICLE{1987ApJ...314....3E,
       author = {{Elmegreen}, Debra Meloy and {Elmegreen}, Bruce G.},
        title = "{Arm Classifications for Spiral Galaxies}",
      journal = {\apj},
         year = 1987,
        month = mar,
       volume = {314},
        pages = {3},
          doi = {10.1086/165034},
       adsurl = {https://ui.adsabs.harvard.edu/abs/1987ApJ...314....3E}
}

@ARTICLE{1989ApJ...342..677E,
       author = {{Elmegreen}, Bruce G. and {Elmegreen}, Debra Meloy},
        title = "{On the Relative Frequency of Flocculent and Grand Design Spiral Structures in Barred Galaxies}",
      journal = {\apj},
         year = 1989,
        month = jul,
       volume = {342},
        pages = {677},
          doi = {10.1086/167628},
       adsurl = {https://ui.adsabs.harvard.edu/abs/1989ApJ...342..677E}
}

@ARTICLE{2016A&A...590A..44G,
       author = {{Gonz{\'a}lez Delgado}, R.~M. and {Cid Fernandes}, R. and {P{\'e}rez}, E. and {Garc{\'\i}a-Benito}, R. and {L{\'o}pez Fern{\'a}ndez}, R. and {Lacerda}, E.~A.~D. and {Cortijo-Ferrero}, C. and {de Amorim}, A.~L. and {Vale Asari}, N. and {S{\'a}nchez}, S.~F. and {Walcher}, C.~J. and {Wisotzki}, L. and {Mast}, D. and {Alves}, J. and {Ascasibar}, Y. and {Bland-Hawthorn}, J. and {Galbany}, L. and {Kennicutt}, R.~C. and {M{\'a}rquez}, I. and {Masegosa}, J. and {Moll{\'a}}, M. and {S{\'a}nchez-Bl{\'a}zquez}, P. and {V{\'\i}lchez}, J.~M.},
        title = "{Star formation along the Hubble sequence. Radial structure of the star formation of CALIFA galaxies}",
      journal = {\aap},
         year = 2016,
        month = may,
       volume = {590},
          eid = {A44},
        pages = {A44},
          doi = {10.1051/0004-6361/201628174},
archivePrefix = {arXiv},
       eprint = {1603.00874},
 primaryClass = {astro-ph.GA},
       adsurl = {https://ui.adsabs.harvard.edu/abs/2016A&A...590A..44G}
}

@ARTICLE{2022A&A...666A.175Y,
       author = {{Yu}, Si-Yue and {Kalinova}, Veselina and {Colombo}, Dario and {Bolatto}, Alberto D. and {Wong}, Tony and {Levy}, Rebecca C. and {Villanueva}, Vicente and {S{\'a}nchez}, Sebasti{\'a}n F. and {Ho}, Luis C. and {Vogel}, Stuart N. and {Teuben}, Peter and {Rubio}, M{\'o}nica},
        title = "{The EDGE-CALIFA survey: The role of spiral arms and bars in driving central molecular gas concentrations}",
      journal = {\aap},
         year = 2022,
        month = oct,
       volume = {666},
          eid = {A175},
        pages = {A175},
          doi = {10.1051/0004-6361/202244306},
archivePrefix = {arXiv},
       eprint = {2208.14950},
 primaryClass = {astro-ph.GA},
       adsurl = {https://ui.adsabs.harvard.edu/abs/2022A&A...666A.175Y}
}

@ARTICLE{2024MNRAS.532.2320G,
       author = {{Garland}, Izzy L. and {Walmsley}, Mike and {Silcock}, Maddie S. and {Potts}, Leah M. and {Smith}, Josh and {Simmons}, Brooke D. and {Lintott}, Chris J. and {Smethurst}, Rebecca J. and {Dawson}, James M. and {Keel}, William C. and {Kruk}, Sandor and {Mantha}, Kameswara Bharadwaj and {Masters}, Karen L. and {O'Ryan}, David and {Popp}, J{\"u}rgen J. and {Thorne}, Matthew R.},
        title = "{Galaxy Zoo DESI: large-scale bars as a secular mechanism for triggering AGNs}",
      journal = {\mnras},
         year = 2024,
        month = aug,
       volume = {532},
       number = {2},
        pages = {2320-2330},
          doi = {10.1093/mnras/stae1620},
archivePrefix = {arXiv},
       eprint = {2406.20096},
 primaryClass = {astro-ph.GA},
       adsurl = {https://ui.adsabs.harvard.edu/abs/2024MNRAS.532.2320G}
}

@ARTICLE{2024A&A...687A.293Q,
       author = {{Querejeta}, Miguel and {Leroy}, Adam K. and {Meidt}, Sharon E. and {Schinnerer}, Eva and {Belfiore}, Francesco and {Emsellem}, Eric and {Klessen}, Ralf S. and {Sun}, Jiayi and {Sormani}, Mattia and {Be{\v{s}}li{\'c}}, Ivana and {Cao}, Yixian and {Chevance}, M{\'e}lanie and {Colombo}, Dario and {Dale}, Daniel A. and {Garc{\'\i}a-Burillo}, Santiago and {Glover}, Simon C.~O. and {Grasha}, Kathryn and {Groves}, Brent and {Koch}, Eric. W. and {Neumann}, Lukas and {Pan}, Hsi-An and {Pessa}, Ismael and {Pety}, J{\'e}r{\^o}me and {Pinna}, Francesca and {Ramambason}, Lise and {Razza}, Alessandro and {Romanelli}, Andrea and {Rosolowsky}, Erik and {Ruiz-Garc{\'\i}a}, Marina and {S{\'a}nchez-Bl{\'a}zquez}, Patricia and {Smith}, Rowan and {Stuber}, Sophia and {Ubeda}, Leonardo and {Usero}, Antonio and {Williams}, Thomas G.},
        title = "{Do spiral arms enhance star formation efficiency?}",
      journal = {\aap},
         year = 2024,
        month = jul,
       volume = {687},
          eid = {A293},
        pages = {A293},
          doi = {10.1051/0004-6361/202449733},
archivePrefix = {arXiv},
       eprint = {2405.05364},
 primaryClass = {astro-ph.GA},
       adsurl = {https://ui.adsabs.harvard.edu/abs/2024A&A...687A.293Q}
}

@ARTICLE{2025ApJ...983...57W,
       author = {{Wisz}, M.~E. and {Masters}, Karen L. and {Daniel}, Kathryne J. and {Stark}, David V. and {Belfiore}, Francesco},
        title = "{The Impacts of Bars, Spirals, and Bulge Size on Gas-phase Metallicity Gradients in MaNGA Galaxies}",
      journal = {\apj},
         year = 2025,
        month = apr,
       volume = {983},
       number = {1},
          eid = {57},
        pages = {57},
          doi = {10.3847/1538-4357/adbb6f},
archivePrefix = {arXiv},
       eprint = {2502.10922},
 primaryClass = {astro-ph.GA},
       adsurl = {https://ui.adsabs.harvard.edu/abs/2025ApJ...983...57W}
}

@ARTICLE{2023A&A...674A..85A,
       author = {{Alb{\'a}n}, M. and {Wylezalek}, D.},
        title = "{Classifying the full SDSS-IV MaNGA Survey using optical diagnostic diagrams: Presentation of AGN catalogs in flexible apertures}",
      journal = {\aap},
         year = 2023,
        month = jun,
       volume = {674},
          eid = {A85},
        pages = {A85},
          doi = {10.1051/0004-6361/202245437},
archivePrefix = {arXiv},
       eprint = {2302.08519},
 primaryClass = {astro-ph.GA},
       adsurl = {https://ui.adsabs.harvard.edu/abs/2023A&A...674A..85A}
}

@ARTICLE{1989Natur.338...45S,
       author = {{Shlosman}, Isaac and {Frank}, Juhan and {Begelman}, Mitchell C.},
        title = "{Bars within bars: a mechanism for fuelling active galactic nuclei}",
      journal = {\nat},
         year = 1989,
        month = mar,
       volume = {338},
       number = {6210},
        pages = {45-47},
          doi = {10.1038/338045a0},
       adsurl = {https://ui.adsabs.harvard.edu/abs/1989Natur.338...45S}
}

@ARTICLE{2004ARA&A..42..603K,
       author = {{Kormendy}, John and {Kennicutt}, Jr., Robert C.},
        title = "{Secular Evolution and the Formation of Pseudobulges in Disk Galaxies}",
      journal = {\araa},
         year = 2004,
        month = sep,
       volume = {42},
       number = {1},
        pages = {603-683},
          doi = {10.1146/annurev.astro.42.053102.134024},
archivePrefix = {arXiv},
       eprint = {astro-ph/0407343},
 primaryClass = {astro-ph},
       adsurl = {https://ui.adsabs.harvard.edu/abs/2004ARA&A..42..603K}
}

@ARTICLE{2006AJ....131.2332G,
       author = {{Gunn}, James E. and {Siegmund}, Walter A. and {Mannery}, Edward J. and {Owen}, Russell E. and {Hull}, Charles L. and {Leger}, R. French and {Carey}, Larry N. and {Knapp}, Gillian R. and {York}, Donald G. and {Boroski}, William N. and {Kent}, Stephen M. and {Lupton}, Robert H. and {Rockosi}, Constance M. and {Evans}, Michael L. and {Waddell}, Patrick and {Anderson}, John E. and {Annis}, James and {Barentine}, John C. and {Bartoszek}, Larry M. and {Bastian}, Steven and {Bracker}, Stephen B. and {Brewington}, Howard J. and {Briegel}, Charles I. and {Brinkmann}, Jon and {Brown}, Yorke J. and {Carr}, Michael A. and {Czarapata}, Paul C. and {Drennan}, Craig C. and {Dombeck}, Thomas and {Federwitz}, Glenn R. and {Gillespie}, Bruce A. and {Gonzales}, Carlos and {Hansen}, Sten U. and {Harvanek}, Michael and {Hayes}, Jeffrey and {Jordan}, Wendell and {Kinney}, Ellyne and {Klaene}, Mark and {Kleinman}, S.~J. and {Kron}, Richard G. and {Kresinski}, Jurek and {Lee}, Glenn and {Limmongkol}, Siriluk and {Lindenmeyer}, Carl W. and {Long}, Daniel C. and {Loomis}, Craig L. and {McGehee}, Peregrine M. and {Mantsch}, Paul M. and {Neilsen}, Jr., Eric H. and {Neswold}, Richard M. and {Newman}, Peter R. and {Nitta}, Atsuko and {Peoples}, Jr., John and {Pier}, Jeffrey R. and {Prieto}, Peter S. and {Prosapio}, Angela and {Rivetta}, Claudio and {Schneider}, Donald P. and {Snedden}, Stephanie and {Wang}, Shu-i.},
        title = "{The 2.5 m Telescope of the Sloan Digital Sky Survey}",
      journal = {\aj},
         year = 2006,
        month = apr,
       volume = {131},
       number = {4},
        pages = {2332-2359},
          doi = {10.1086/500975},
archivePrefix = {arXiv},
       eprint = {astro-ph/0602326},
 primaryClass = {astro-ph},
       adsurl = {https://ui.adsabs.harvard.edu/abs/2006AJ....131.2332G}
}

@ARTICLE{1981PASP...93....5B,
       author = {{Baldwin}, J.~A. and {Phillips}, M.~M. and {Terlevich}, R.},
        title = "{Classification parameters for the emission-line spectra of extragalactic objects.}",
      journal = {\pasp},
         year = 1981,
        month = feb,
       volume = {93},
        pages = {5-19},
          doi = {10.1086/130766},
       adsurl = {https://ui.adsabs.harvard.edu/abs/1981PASP...93....5B}
}

@ARTICLE{2021MNRAS.507.4389G,
       author = {{G{\'e}ron}, Tobias and {Smethurst}, R.~J. and {Lintott}, Chris and {Kruk}, Sandor and {Masters}, Karen L. and {Simmons}, Brooke and {Stark}, David V.},
        title = "{Galaxy zoo: stronger bars facilitate quenching in star-forming galaxies}",
      journal = {\mnras},
         year = 2021,
        month = nov,
       volume = {507},
       number = {3},
        pages = {4389-4408},
          doi = {10.1093/mnras/stab2064},
archivePrefix = {arXiv},
       eprint = {2107.06913},
 primaryClass = {astro-ph.GA},
       adsurl = {https://ui.adsabs.harvard.edu/abs/2021MNRAS.507.4389G}
}

@ARTICLE{2010ApJS..186..427N,
       author = {{Nair}, Preethi B. and {Abraham}, Roberto G.},
        title = "{A Catalog of Detailed Visual Morphological Classifications for 14,034 Galaxies in the Sloan Digital Sky Survey}",
      journal = {\apjs},
         year = 2010,
        month = feb,
       volume = {186},
       number = {2},
        pages = {427-456},
          doi = {10.1088/0067-0049/186/2/427},
archivePrefix = {arXiv},
       eprint = {1001.2401},
 primaryClass = {astro-ph.CO},
       adsurl = {https://ui.adsabs.harvard.edu/abs/2010ApJS..186..427N}
}

@INPROCEEDINGS{2020IAUS..353..140B,
       author = {{Bittner}, A. and {Gadotti}, D.~A. and {Elmegreen}, B.~G. and {Athanassoula}, E. and {Elmegreen}, D.~M. and {Bosma}, A. and {Mu{\~n}oz-Mateos}, J.},
        title = "{The sequence of spiral arm classes: Observational signatures of persistent spiral density waves in grand-design galaxies}",
    booktitle = {Galactic Dynamics in the Era of Large Surveys},
         year = 2020,
       editor = {{Valluri}, Monica and {Sellwood}, J.~A.},
       series = {IAU Symposium},
       volume = {353},
        month = jan,
        pages = {140-143},
          doi = {10.1017/S1743921319008160},
archivePrefix = {arXiv},
       eprint = {1910.01139},
 primaryClass = {astro-ph.GA},
       adsurl = {https://ui.adsabs.harvard.edu/abs/2020IAUS..353..140B}
}

@ARTICLE{2011MNRAS.418.1055M,
       author = {{Masters}, Karen L. and {Maraston}, Claudia and {Nichol}, Robert C. and {Thomas}, Daniel and {Beifiori}, Alessadra and {Bundy}, Kevin and {Edmondson}, Edward M. and {Higgs}, Tim D. and {Leauthaud}, Alexie and {Mandelbaum}, Rachel and {Pforr}, Janine and {Ross}, Ashley J. and {Ross}, Nicholas P. and {Schneider}, Donald P. and {Skibba}, Ramin and {Tinker}, Jeremy and {Tojeiro}, Rita and {Wake}, David A. and {Brinkmann}, Jon and {Weaver}, Benjamin A.},
        title = "{The morphology of galaxies in the Baryon Oscillation Spectroscopic Survey}",
      journal = {\mnras},
         year = 2011,
        month = dec,
       volume = {418},
       number = {2},
        pages = {1055-1070},
          doi = {10.1111/j.1365-2966.2011.19557.x},
archivePrefix = {arXiv},
       eprint = {1106.3331},
 primaryClass = {astro-ph.CO},
       adsurl = {https://ui.adsabs.harvard.edu/abs/2011MNRAS.418.1055M}
}

@ARTICLE{2012ApJ...750..141L,
       author = {{Lee}, Gwang-Ho and {Woo}, Jong-Hak and {Lee}, Myung Gyoon and {Hwang}, Ho Seong and {Lee}, Jong Chul and {Sohn}, Jubee and {Lee}, Jong Hwan},
        title = "{Do Bars Trigger Activity in Galactic Nuclei?}",
      journal = {\apj},
         year = 2012,
        month = may,
       volume = {750},
       number = {2},
          eid = {141},
        pages = {141},
          doi = {10.1088/0004-637X/750/2/141},
archivePrefix = {arXiv},
       eprint = {1203.1693},
 primaryClass = {astro-ph.CO},
       adsurl = {https://ui.adsabs.harvard.edu/abs/2012ApJ...750..141L}
}

@ARTICLE{2013A&A...549A.141A,
       author = {{Alonso}, M.~S. and {Coldwell}, G. and {Lambas}, D.~G.},
        title = "{Effect of bars in AGN host galaxies and black hole activity}",
      journal = {\aap},
         year = 2013,
        month = jan,
       volume = {549},
          eid = {A141},
        pages = {A141},
          doi = {10.1051/0004-6361/201220117},
archivePrefix = {arXiv},
       eprint = {1211.5156},
 primaryClass = {astro-ph.CO},
       adsurl = {https://ui.adsabs.harvard.edu/abs/2013A&A...549A.141A}
}

@ARTICLE{2014RvMP...86....1S,
       author = {{Sellwood}, J.~A.},
        title = "{Secular evolution in disk galaxies}",
      journal = {Reviews of Modern Physics},
         year = 2014,
        month = jan,
       volume = {86},
       number = {1},
        pages = {1-46},
          doi = {10.1103/RevModPhys.86.1},
archivePrefix = {arXiv},
       eprint = {1310.0403},
 primaryClass = {astro-ph.GA},
       adsurl = {https://ui.adsabs.harvard.edu/abs/2014RvMP...86....1S}
}

@ARTICLE{2023MNRAS.521.1775G,
       author = {{G{\'e}ron}, Tobias and {Smethurst}, Rebecca J. and {Lintott}, Chris and {Kruk}, Sandor and {Masters}, Karen L. and {Simmons}, Brooke and {Mantha}, Kameswara Bharadwaj and {Walmsley}, Mike and {Garma-Oehmichen}, L. and {Drory}, Niv and {Lane}, Richard R.},
        title = "{Galaxy Zoo: kinematics of strongly and weakly barred galaxies}",
      journal = {\mnras},
         year = 2023,
        month = may,
       volume = {521},
       number = {2},
        pages = {1775-1793},
          doi = {10.1093/mnras/stad501},
archivePrefix = {arXiv},
       eprint = {2302.05464},
 primaryClass = {astro-ph.GA},
       adsurl = {https://ui.adsabs.harvard.edu/abs/2023MNRAS.521.1775G}
}

@ARTICLE{2025A&A...699A.204M,
       author = {{Marels}, V. and {Mesa}, V. and {Jaque Arancibia}, M. and {Alonso}, S. and {Coldwell}, G. and {Damke}, G. and {Contreras Rojas}, V.},
        title = "{The role of bars in triggering active galactic nucleus galaxies}",
      journal = {\aap},
         year = 2025,
        month = jul,
       volume = {699},
          eid = {A204},
        pages = {A204},
          doi = {10.1051/0004-6361/202554961},
archivePrefix = {arXiv},
       eprint = {2505.23958},
 primaryClass = {astro-ph.GA},
       adsurl = {https://ui.adsabs.harvard.edu/abs/2025A&A...699A.204M}
}

@ARTICLE{2025MNRAS.544.1056V,
       author = {{V{\'a}zquez-Mata}, J.~A. and {Hern{\'a}ndez-Toledo}, H.~M. and {Avila-Reese}, V. and {Rodr{\'\i}guez-Puebla}, A. and {Mart{\'\i}nez-V{\'a}zquez}, L.~A. and {Herrera-Endoqui}, M. and {Lacerna}, I. and {Mascherpa}, L.~C. and {Morell}, D.~F.},
        title = "{Visual morphological classification of the full MaNGA DR17 sample: a general characterization}",
      journal = {\mnras},
         year = 2025,
        month = nov,
       volume = {544},
       number = {1},
        pages = {1056-1084},
          doi = {10.1093/mnras/staf1784},
archivePrefix = {arXiv},
       eprint = {2510.12792},
 primaryClass = {astro-ph.GA},
       adsurl = {https://ui.adsabs.harvard.edu/abs/2025MNRAS.544.1056V}
}

@ARTICLE{2024ApJ...973..116D,
       author = {{DiGiorgio Zanger}, Brian and {Westfall}, Kyle B. and {Bundy}, Kevin and {Drory}, Niv and {Bershady}, Matthew A. and {Campbell}, Stephanie and {Weijmans}, Anne-Marie and {Masters}, Karen L. and {Stark}, David and {Law}, David},
        title = "{The Strength of Bisymmetric Modes in SDSS-IV/MaNGA Barred Galaxy Kinematics}",
      journal = {\apj},
         year = 2024,
        month = oct,
       volume = {973},
       number = {2},
          eid = {116},
        pages = {116},
          doi = {10.3847/1538-4357/ad6606},
archivePrefix = {arXiv},
       eprint = {2407.11908},
 primaryClass = {astro-ph.GA},
       adsurl = {https://ui.adsabs.harvard.edu/abs/2024ApJ...973..116D}
}

@ARTICLE{2024ApJ...973..129G,
       author = {{G{\'e}ron}, Tobias and {Smethurst}, R.~J. and {Lintott}, Chris and {Masters}, Karen L. and {Garland}, I.~L. and {Mengistu}, Petra and {O'Ryan}, David and {Simmons}, B.~D.},
        title = "{The Effects of Bar Strength and Kinematics on Galaxy Evolution: Slow Strong Bars Affect Their Hosts the Most}",
      journal = {\apj},
         year = 2024,
        month = oct,
       volume = {973},
       number = {2},
          eid = {129},
        pages = {129},
          doi = {10.3847/1538-4357/ad66b7},
archivePrefix = {arXiv},
       eprint = {2405.05960},
 primaryClass = {astro-ph.GA},
       adsurl = {https://ui.adsabs.harvard.edu/abs/2024ApJ...973..129G}
}

@ARTICLE{2019MNRAS.489.1338P,
       author = {{Peterken}, Thomas and {Fraser-McKelvie}, Amelia and {Arag{\'o}n-Salamanca}, Alfonso and {Merrifield}, Michael and {Kraljic}, Katarina and {Knapen}, Johan H. and {Riffel}, Rog{\'e}rio and {Brownstein}, Joel and {Drory}, Niv},
        title = "{Time-slicing spiral galaxies with SDSS-IV MaNGA}",
      journal = {\mnras},
         year = 2019,
        month = oct,
       volume = {489},
       number = {1},
        pages = {1338-1343},
          doi = {10.1093/mnras/stz2204},
archivePrefix = {arXiv},
       eprint = {1908.05013},
 primaryClass = {astro-ph.GA},
       adsurl = {https://ui.adsabs.harvard.edu/abs/2019MNRAS.489.1338P}
}

@ARTICLE{2025A&A...698A.296R,
       author = {{Romanelli}, Andrea and {Chevance}, M{\'e}lanie and {Kruijssen}, J.~M. Diederik and {Ramambason}, Lise and {Querejeta}, Miguel and {Boquien}, Mederic and {Dale}, Daniel A. and {den Brok}, Jakob and {Glover}, Simon C.~O. and {Grasha}, Kathryn and {Hughes}, Annie and {Kim}, Jaeyeon and {Longmore}, Steven and {Meidt}, Sharon E. and {Mendez-Delgado}, Jos{\'e} Eduardo and {Neumann}, Lukas and {Pety}, J{\'e}r{\^o}me and {Schinnerer}, Eva and {Smith}, Rowan and {Sun}, Jiayi and {Williams}, Thomas G.},
        title = "{The impact of spiral arms on the star formation life cycle}",
      journal = {\aap},
         year = 2025,
        month = jun,
       volume = {698},
          eid = {A296},
        pages = {A296},
          doi = {10.1051/0004-6361/202553895},
archivePrefix = {arXiv},
       eprint = {2505.10908},
 primaryClass = {astro-ph.GA},
       adsurl = {https://ui.adsabs.harvard.edu/abs/2025A&A...698A.296R}
}

@ARTICLE{2026A&A...707A.152L,
       author = {{La Marca}, A. and {Nardone}, M.~T. and {Wang}, L. and {Margalef-Bentabol}, B. and {Kruk}, S. and {Trager}, S.~C.},
        title = "{Galactic bars and active galactic nucleus fuelling in the second half of cosmic history}",
      journal = {\aap},
         year = 2026,
        month = mar,
       volume = {707},
          eid = {A152},
        pages = {A152},
          doi = {10.1051/0004-6361/202556236},
archivePrefix = {arXiv},
       eprint = {2510.23522},
 primaryClass = {astro-ph.GA},
       adsurl = {https://ui.adsabs.harvard.edu/abs/2026A&A...707A.152L}
}

@INPROCEEDINGS{2003ASPC..290..411C,
       author = {{Combes}, F.},
        title = "{AGN Fueling: The observational point of view}",
    booktitle = {Active Galactic Nuclei: From Central Engine to Host Galaxy},
         year = 2003,
       editor = {{Collin}, Suzy and {Combes}, Francoise and {Shlosman}, Isaac},
       series = {Astronomical Society of the Pacific Conference Series},
       volume = {290},
        month = jan,
        pages = {411},
          doi = {10.48550/arXiv.astro-ph/0210232},
archivePrefix = {arXiv},
       eprint = {astro-ph/0210232},
 primaryClass = {astro-ph},
       adsurl = {https://ui.adsabs.harvard.edu/abs/2003ASPC..290..411C}
}

@ARTICLE{2000ApJ...529...93K,
       author = {{Knapen}, Johan H. and {Shlosman}, Isaac and {Peletier}, Reynier F.},
        title = "{A Subarcsecond Resolution Near-Infrared Study of Seyfert and ``Normal'' Galaxies. II. Morphology}",
      journal = {\apj},
         year = 2000,
        month = jan,
       volume = {529},
       number = {1},
        pages = {93-100},
          doi = {10.1086/308266},
archivePrefix = {arXiv},
       eprint = {astro-ph/9907379},
 primaryClass = {astro-ph},
       adsurl = {https://ui.adsabs.harvard.edu/abs/2000ApJ...529...93K}
}

@ARTICLE{2002ApJ...567...97L,
       author = {{Laine}, Seppo and {Shlosman}, Isaac and {Knapen}, Johan H. and {Peletier}, Reynier F.},
        title = "{Nested and Single Bars in Seyfert and Non-Seyfert Galaxies}",
      journal = {\apj},
         year = 2002,
        month = mar,
       volume = {567},
       number = {1},
        pages = {97-117},
          doi = {10.1086/323964},
archivePrefix = {arXiv},
       eprint = {astro-ph/0108029},
 primaryClass = {astro-ph},
       adsurl = {https://ui.adsabs.harvard.edu/abs/2002ApJ...567...97L}
}

@ARTICLE{2003ApJ...589..774M,
       author = {{Martini}, Paul and {Regan}, Michael W. and {Mulchaey}, John S. and {Pogge}, Richard W.},
        title = "{Circumnuclear Dust in Nearby Active and Inactive Galaxies. II. Bars, Nuclear Spirals, and the Fueling of Active Galactic Nuclei}",
      journal = {\apj},
         year = 2003,
        month = jun,
       volume = {589},
       number = {2},
        pages = {774-782},
          doi = {10.1086/374685},
archivePrefix = {arXiv},
       eprint = {astro-ph/0212391},
 primaryClass = {astro-ph},
       adsurl = {https://ui.adsabs.harvard.edu/abs/2003ApJ...589..774M}
}

@ARTICLE{2010MNRAS.407.1529H,
       author = {{Hopkins}, Philip F. and {Quataert}, Eliot},
        title = "{How do massive black holes get their gas?}",
      journal = {\mnras},
         year = 2010,
        month = sep,
       volume = {407},
       number = {3},
        pages = {1529-1564},
          doi = {10.1111/j.1365-2966.2010.17064.x},
archivePrefix = {arXiv},
       eprint = {0912.3257},
 primaryClass = {astro-ph.CO},
       adsurl = {https://ui.adsabs.harvard.edu/abs/2010MNRAS.407.1529H}
}

@ARTICLE{2015MNRAS.448.3442G,
       author = {{Galloway}, Melanie A. and {Willett}, Kyle W. and {Fortson}, Lucy F. and {Cardamone}, Carolin N. and {Schawinski}, Kevin and {Cheung}, Edmond and {Lintott}, Chris J. and {Masters}, Karen L. and {Melvin}, Thomas and {Simmons}, Brooke D.},
        title = "{Galaxy Zoo: the effect of bar-driven fuelling on the presence of an active galactic nucleus in disc galaxies}",
      journal = {\mnras},
         year = 2015,
        month = apr,
       volume = {448},
       number = {4},
        pages = {3442-3454},
          doi = {10.1093/mnras/stv235},
archivePrefix = {arXiv},
       eprint = {1502.01033},
 primaryClass = {astro-ph.GA},
       adsurl = {https://ui.adsabs.harvard.edu/abs/2015MNRAS.448.3442G}
}

@ARTICLE{2024MNRAS.527.3366K,
       author = {{Kataria}, Sandeep Kumar and {Vivek}, M.},
        title = "{How does the presence of bar affects the fueling of supermassive black holes? An IllustrisTNG100 perspective}",
      journal = {\mnras},
         year = 2024,
        month = jan,
       volume = {527},
       number = {2},
        pages = {3366-3380},
          doi = {10.1093/mnras/stad3383},
archivePrefix = {arXiv},
       eprint = {2311.00040},
 primaryClass = {astro-ph.GA},
       adsurl = {https://ui.adsabs.harvard.edu/abs/2024MNRAS.527.3366K}
}

@INPROCEEDINGS{2009ASPC..419..402H,
       author = {{Hao}, L. and {Jogee}, S. and {Barazza}, F.~D. and {Marinova}, I. and {Shen}, J.},
        title = "{Bars in Starbursts and AGNs - A Quantitative Reexamination}",
    booktitle = {Galaxy Evolution: Emerging Insights and Future Challenges},
         year = 2009,
       editor = {{Jogee}, S. and {Marinova}, I. and {Hao}, L. and {Blanc}, G.~A.},
       series = {Astronomical Society of the Pacific Conference Series},
       volume = {419},
        month = dec,
        pages = {402},
          doi = {10.48550/arXiv.0910.3960},
archivePrefix = {arXiv},
       eprint = {0910.3960},
 primaryClass = {astro-ph.CO},
       adsurl = {https://ui.adsabs.harvard.edu/abs/2009ASPC..419..402H}
}

@ARTICLE{2012ApJS..198....4O,
       author = {{Oh}, Sree and {Oh}, Kyuseok and {Yi}, Sukyoung K.},
        title = "{Bar Effects on Central Star Formation and Active Galactic Nucleus Activity}",
      journal = {\apjs},
         year = 2012,
        month = jan,
       volume = {198},
       number = {1},
          eid = {4},
        pages = {4},
          doi = {10.1088/0067-0049/198/1/4},
archivePrefix = {arXiv},
       eprint = {1111.3623},
 primaryClass = {astro-ph.GA},
       adsurl = {https://ui.adsabs.harvard.edu/abs/2012ApJS..198....4O}
}

@ARTICLE{Fisher1922,
  author = {{Fisher}, R.~A.},
  title = "{On the interpretation of $\chi^2$ from contingency tables, and the calculation of $P$}",
  journal = {J. R. Stat. Soc.},
  year = 1922,
  volume = {85},
  pages = {87-94},
  doi = {10.2307/2340521}
}

@ARTICLE{1945Natur.156..177B,
       author = {{Barnard}, G.~A.},
        title = "{A New Test for 2 {\texttimes} 2 Tables}",
      journal = {\nat},
         year = 1945,
        month = aug,
       volume = {156},
       number = {3954},
        pages = {177},
          doi = {10.1038/156177a0},
       adsurl = {https://ui.adsabs.harvard.edu/abs/1945Natur.156..177B}
}

@ARTICLE{2011PASA...28..128C,
       author = {{Cameron}, Ewan},
        title = "{On the Estimation of Confidence Intervals for Binomial Population Proportions in Astronomy: The Simplicity and Superiority of the Bayesian Approach}",
      journal = {\pasa},
         year = 2011,
        month = jun,
       volume = {28},
       number = {2},
        pages = {128-139},
          doi = {10.1071/AS10046},
archivePrefix = {arXiv},
       eprint = {1012.0566},
 primaryClass = {astro-ph.IM},
       adsurl = {https://ui.adsabs.harvard.edu/abs/2011PASA...28..128C}
}

@ARTICLE{2012ApJ...746...90A,
       author = {{Aird}, James and {Coil}, Alison L. and {Moustakas}, John and {Blanton}, Michael R. and {Burles}, Scott M. and {Cool}, Richard J. and {Eisenstein}, Daniel J. and {Smith}, M. Stephen M. and {Wong}, Kenneth C. and {Zhu}, Guangtun},
        title = "{PRIMUS: The Dependence of AGN Accretion on Host Stellar Mass and Color}",
      journal = {\apj},
         year = 2012,
        month = feb,
       volume = {746},
       number = {1},
          eid = {90},
        pages = {90},
          doi = {10.1088/0004-637X/746/1/90},
archivePrefix = {arXiv},
       eprint = {1107.4368},
 primaryClass = {astro-ph.CO},
       adsurl = {https://ui.adsabs.harvard.edu/abs/2012ApJ...746...90A}
}

@ARTICLE{2003MNRAS.341.1179A,
       author = {{Athanassoula}, E.},
        title = "{What determines the strength and the slowdown rate of bars?}",
      journal = {\mnras},
         year = 2003,
        month = jun,
       volume = {341},
       number = {4},
        pages = {1179-1198},
          doi = {10.1046/j.1365-8711.2003.06473.x},
archivePrefix = {arXiv},
       eprint = {astro-ph/0302519},
 primaryClass = {astro-ph},
       adsurl = {https://ui.adsabs.harvard.edu/abs/2003MNRAS.341.1179A}
}

@ARTICLE{2010MNRAS.403..625D,
       author = {{Dobbs}, C.~L. and {Theis}, C. and {Pringle}, J.~E. and {Bate}, M.~R.},
        title = "{Simulations of the grand design galaxy M51: a case study for analysing tidally induced spiral structure}",
      journal = {\mnras},
         year = 2010,
        month = apr,
       volume = {403},
       number = {2},
        pages = {625-645},
          doi = {10.1111/j.1365-2966.2009.16161.x},
archivePrefix = {arXiv},
       eprint = {0912.1201},
 primaryClass = {astro-ph.GA},
       adsurl = {https://ui.adsabs.harvard.edu/abs/2010MNRAS.403..625D}
}

@ARTICLE{2021ApJ...917...88Y,
       author = {{Yu}, Si-Yue and {Ho}, Luis C. and {Wang}, Jing},
        title = "{Spiral Structure Boosts Star Formation in Disk Galaxies}",
      journal = {\apj},
         year = 2021,
        month = aug,
       volume = {917},
       number = {2},
          eid = {88},
        pages = {88},
          doi = {10.3847/1538-4357/ac0c77},
archivePrefix = {arXiv},
       eprint = {2106.09715},
 primaryClass = {astro-ph.GA},
       adsurl = {https://ui.adsabs.harvard.edu/abs/2021ApJ...917...88Y}
}

@ARTICLE{1969ApJ...158..123R,
       author = {{Roberts}, W.~W.},
        title = "{Large-Scale Shock Formation in Spiral Galaxies and its Implications on Star Formation}",
      journal = {\apj},
         year = 1969,
        month = oct,
       volume = {158},
        pages = {123},
          doi = {10.1086/150177},
       adsurl = {https://ui.adsabs.harvard.edu/abs/1969ApJ...158..123R}
}

@ARTICLE{1972ApL....11...41K,
       author = {{Kalnajs}, A.~J.},
        title = "{The Damping of the Galactic Density Waves by their Induced Shocks}",
      journal = {\aplett},
         year = 1972,
        month = may,
       volume = {11},
        pages = {41},
       adsurl = {https://ui.adsabs.harvard.edu/abs/1972ApL....11...41K}
}

@ARTICLE{2016MNRAS.460.2472B,
       author = {{Baba}, J. and {Morokuma-Matsui}, K. and {Miyamoto}, Y. and {Egusa}, F. and {Kuno}, N.},
        title = "{Gas velocity patterns in simulated galaxies: observational diagnostics of spiral structure theories}",
      journal = {\mnras},
         year = 2016,
        month = aug,
       volume = {460},
       number = {3},
        pages = {2472-2481},
          doi = {10.1093/mnras/stw987},
archivePrefix = {arXiv},
       eprint = {1604.06879},
 primaryClass = {astro-ph.GA},
       adsurl = {https://ui.adsabs.harvard.edu/abs/2016MNRAS.460.2472B}
}

@ARTICLE{2010ApJ...714L.260N,
       author = {{Nair}, Preethi B. and {Abraham}, Roberto G.},
        title = "{On the Fraction of Barred Spiral Galaxies}",
      journal = {\apjl},
         year = 2010,
        month = may,
       volume = {714},
       number = {2},
        pages = {L260-L264},
          doi = {10.1088/2041-8205/714/2/L260},
archivePrefix = {arXiv},
       eprint = {1004.0684},
 primaryClass = {astro-ph.CO},
       adsurl = {https://ui.adsabs.harvard.edu/abs/2010ApJ...714L.260N}
}

@ARTICLE{2011A&A...532A..75D,
       author = {{de Lapparent}, V. and {Baillard}, A. and {Bertin}, E.},
        title = "{The EFIGI catalogue of 4458 nearby galaxies with morphology. II. Statistical properties along the Hubble sequence}",
      journal = {\aap},
         year = 2011,
        month = aug,
       volume = {532},
          eid = {A75},
        pages = {A75},
          doi = {10.1051/0004-6361/201016424},
archivePrefix = {arXiv},
       eprint = {1103.5735},
 primaryClass = {astro-ph.CO},
       adsurl = {https://ui.adsabs.harvard.edu/abs/2011A&A...532A..75D}
}

@ARTICLE{2010MNRAS.405.1409C,
       author = {{Chilingarian}, Igor V. and {Melchior}, Anne-Laure and {Zolotukhin}, Ivan Yu.},
        title = "{Analytical approximations of K-corrections in optical and near-infrared bands}",
      journal = {\mnras},
         year = 2010,
        month = jul,
       volume = {405},
       number = {3},
        pages = {1409-1420},
          doi = {10.1111/j.1365-2966.2010.16506.x},
archivePrefix = {arXiv},
       eprint = {1002.2360},
 primaryClass = {astro-ph.IM},
       adsurl = {https://ui.adsabs.harvard.edu/abs/2010MNRAS.405.1409C}
}

@ARTICLE{2014MNRAS.440..889S,
       author = {{Schawinski}, Kevin and {Urry}, C. Megan and {Simmons}, Brooke D. and {Fortson}, Lucy and {Kaviraj}, Sugata and {Keel}, William C. and {Lintott}, Chris J. and {Masters}, Karen L. and {Nichol}, Robert C. and {Sarzi}, Marc and {Skibba}, Ramin and {Treister}, Ezequiel and {Willett}, Kyle W. and {Wong}, O. Ivy and {Yi}, Sukyoung K.},
        title = "{The green valley is a red herring: Galaxy Zoo reveals two evolutionary pathways towards quenching of star formation in early- and late-type galaxies}",
      journal = {\mnras},
         year = 2014,
        month = may,
       volume = {440},
       number = {1},
        pages = {889-907},
          doi = {10.1093/mnras/stu327},
archivePrefix = {arXiv},
       eprint = {1402.4814},
 primaryClass = {astro-ph.GA},
       adsurl = {https://ui.adsabs.harvard.edu/abs/2014MNRAS.440..889S}
}

@ARTICLE{2026ApJ..1003...25L,
       author = {{Liu}, Jianfei and {Zhou}, Zhimin},
        title = "{Investigating the Effects of Bars on Star Formation and Nuclear Activity of Galaxies Using DESI Survey Data}",
      journal = {\apj},
         year = 2026,
        month = may,
       volume = {1003},
       number = {1},
          eid = {25},
        pages = {25},
          doi = {10.3847/1538-4357/ae626a},
archivePrefix = {arXiv},
       eprint = {2605.04537},
 primaryClass = {astro-ph.GA},
       adsurl = {https://ui.adsabs.harvard.edu/abs/2026ApJ..1003...25L}
}

@ARTICLE{2009AJ....137.4487B,
       author = {{Buta}, Ronald J. and {Knapen}, Johan H. and {Elmegreen}, Bruce G. and {Salo}, Heikki and {Laurikainen}, Eija and {Elmegreen}, Debra Meloy and {Puerari}, Iv{\^a}nio and {Block}, David L.},
        title = "{Do Bars Drive Spiral Density Waves?}",
      journal = {\aj},
         year = 2009,
        month = may,
       volume = {137},
       number = {5},
        pages = {4487-4516},
          doi = {10.1088/0004-6256/137/5/4487},
archivePrefix = {arXiv},
       eprint = {0903.2008},
 primaryClass = {astro-ph.CO},
       adsurl = {https://ui.adsabs.harvard.edu/abs/2009AJ....137.4487B}
}

@ARTICLE{2026AJ....171..285D,
       author = {{DESI Collaboration} and {Abdul Karim}, M. and {Adame}, A.~G. and {Aguado}, D. and {Aguilar}, J. and {Ahlen}, S. and {Alam}, S. and {Aldering}, G. and {Alexander}, D.~M. and {Alfarsy}, R. and {Allen}, L. and {Allende Prieto}, C. and {Alves}, O. and {Anand}, A. and {Andrade}, U. and {Armengaud}, E. and {Avila}, S. and {Aviles}, A. and {Awan}, H. and {Bailey}, S. and {Baleato Lizancos}, A. and {Ballester}, O. and {Bault}, A. and {Bautista}, J. and {Bean}, R. and {Behera}, J. and {BenZvi}, S. and {Beraldo e Silva}, L. and {Bermejo-Climent}, J.~R. and {Beutler}, F. and {Bianchi}, D. and {Blake}, C. and {Blum}, R. and {Bolton}, A.~S. and {Bonici}, M. and {Brieden}, S. and {Brodzeller}, A. and {Brooks}, D. and {Buckley-Geer}, E. and {Burtin}, E. and {Bystr{\"o}m}, A. and {Canning}, R. and {Carnero Rosell}, A. and {Carr}, A. and {Carrilho}, P. and {Casas}, L. and {Castander}, F.~J. and {Cereskaite}, R. and {Cervantes-Cota}, J.~L. and {Chaussidon}, E. and {Chaves-Montero}, J. and {Chen}, S. and {Chen}, X. and {Circosta}, C. and {Claybaugh}, T. and {Cole}, S. and {Cooper}, A.~P. and {Cousinou}, M.-C. and {Cuceu}, A. and {Davis}, T.~M. and {Dawson}, K.~S. and {de Belsunce}, R. and {de la Cruz}, R. and {de la Macorra}, A. and {de Mattia}, A. and {Deiosso}, N. and {Della Costa}, J. and {Demina}, R. and {Demirbozan}, U. and {DeRose}, J. and {Dey}, A. and {Dey}, B. and {Ding}, J. and {Ding}, Z. and {Doel}, P. and {Douglass}, K. and {Dowicz}, M. and {Ebina}, H. and {Edelstein}, J. and {Eisenstein}, D.~J. and {Elbers}, W. and {Emas}, N. and {Escoffier}, S. and {Fagrelius}, P. and {Fan}, X. and {Fanning}, K. and {Favole}, G. and {Fawcett}, V.~A. and {Fern{\'a}ndez-Garc{\'\i}a}, E. and {Ferraro}, S. and {Findlay}, N. and {Font-Ribera}, A. and {Forero-Romero}, J.~E. and {Forero-S{\'a}nchez}, D. and {Frenk}, C.~S. and {G{\"a}nsicke}, B.~T. and {Galbany}, L. and {Garc{\'\i}a-Bellido}, J. and {Garcia-Quintero}, C. and {Garrison}, L.~H. and {Gazta{\~n}aga}, E. and {Gil-Mar{\'\i}n}, H. and {Gloudemans}, A. and {Gnedin}, O.~Y. and {Gontcho A Gontcho}, S. and {Gonzalez}, D. and {Gonzalez-Morales}, A.~X. and {Gonzalez-Perez}, V. and {Gordon}, C. and {Graur}, O. and {Green}, D. and {Gruen}, D. and {Gsponer}, R. and {Guandalin}, C. and {Gutierrez}, G. and {Guy}, J. and {Hahn}, C. and {Han}, J.~J. and {Han}, J. and {He}, S. and {Herrera-Alcantar}, H.~K. and {Heydenreich}, S. and {Honscheid}, K. and {Hou}, J. and {Howlett}, C. and {Huterer}, D. and {Ir{\v{s}}i{\v{c}}}, V. and {Ishak}, M. and {Jacques}, A. and {Jiang}, L. and {Jimenez}, J. and {Jing}, Y.~P. and {Joachimi}, B. and {Joudaki}, S. and {Joyce}, R. and {Jullo}, E. and {Juneau}, S. and {Kara{\c{c}}ayl{\i}}, N.~G. and {Karim}, T. and {Kehoe}, R. and {Kent}, S. and {Khederlarian}, A. and {Kirkby}, D. and {Kisner}, T. and {Kitaura}, F.-S. and {Kizhuprakkat}, N. and {Kong}, H. and {Koposov}, S.~E. and {Kremin}, A. and {Krolewski}, A. and {Lahav}, O. and {Lai}, Y. and {Lamman}, C. and {Lan}, T.-W. and {Landriau}, M. and {Lang}, D. and {Lange}, J.~U. and {Lasker}, J. and {Le Goff}, J.~M. and {Le Guillou}, L. and {Leauthaud}, A. and {Levi}, M.~E. and {Li}, S. and {Li}, T.~S. and {Liu}, W. and {Lodha}, K. and {Lokken}, M. and {Luo}, Y. and {Magneville}, C. and {Manera}, M. and {Manser}, C.~J. and {Margala}, D. and {Martini}, P. and {Maus}, M. and {McCullough}, J. and {McDonald}, P. and {Medina}, G.~E. and {Medina-Varela}, L. and {Meisner}, A. and {Mena-Fern{\'a}ndez}, J. and {Menegas}, A. and {Meneses-Rizo}, J. and {Mezcua}, M. and {Miquel}, R. and {Montero-Camacho}, P. and {Moon}, J. and {Moustakas}, J. and {Mu{\~n}oz-Guti{\'e}rrez}, A. and {Mu noz-Santos}, D. and {Myers}, A.~D. and {Myles}, J. and {Nadathur}, S. and {Najita}, J. and {Napolitano}, L. and {Newman}, J.~A. and {Nikakhtar}, F. and {Nikutta}, R. and {Niz}, G. and {Noriega}, H.~E. and {Nugent}, P.},
        title = "{Data Release 1 of the Dark Energy Spectroscopic Instrument}",
      journal = {\aj},
         year = 2026,
        month = may,
       volume = {171},
       number = {5},
          eid = {285},
        pages = {285},
          doi = {10.3847/1538-3881/ae4c43},
archivePrefix = {arXiv},
       eprint = {2503.14745},
 primaryClass = {astro-ph.CO},
       adsurl = {https://ui.adsabs.harvard.edu/abs/2026AJ....171..285D}
}

@ARTICLE{2026A&A...711A...1E,
       author = {{Euclid Collaboration} and {Aussel}, H. and {Tereno}, I. and {Schirmer}, M. and {Alguero}, G. and {Altieri}, B. and {Balbinot}, E. and {de Boer}, T. and {Casenove}, P. and {Corcho-Caballero}, P. and {Furusawa}, H. and {Furusawa}, J. and {Hudson}, M.~J. and {Jahnke}, K. and {Libet}, G. and {Macias-Perez}, J. and {Masoumzadeh}, N. and {Mohr}, J.~J. and {Odier}, J. and {Scott}, D. and {Vassallo}, T. and {Verdoes Kleijn}, G. and {Zacchei}, A. and {Aghanim}, N. and {Amara}, A. and {Andreon}, S. and {Auricchio}, N. and {Awan}, S. and {Azzollini}, R. and {Baccigalupi}, C. and {Baldi}, M. and {Balestra}, A. and {Bardelli}, S. and {Basset}, A. and {Battaglia}, P. and {Belikov}, A.~N. and {Bender}, R. and {Biviano}, A. and {Bonchi}, A. and {Bonino}, D. and {Branchini}, E. and {Brescia}, M. and {Brinchmann}, J. and {Camera}, S. and {Ca{\~n}as-Herrera}, G. and {Capobianco}, V. and {Carbone}, C. and {Cardone}, V.~F. and {Carretero}, J. and {Casas}, S. and {Castander}, F.~J. and {Castellano}, M. and {Castignani}, G. and {Cavuoti}, S. and {Chambers}, K.~C. and {Cimatti}, A. and {Colodro-Conde}, C. and {Congedo}, G. and {Conselice}, C.~J. and {Conversi}, L. and {Copin}, Y. and {Courbin}, F. and {Courtois}, H.~M. and {Cropper}, M. and {Cuby}, J.-G. and {Da Silva}, A. and {da Silva}, R. and {Degaudenzi}, H. and {de Jong}, J.~T.~A. and {De Lucia}, G. and {Di Giorgio}, A.~M. and {Dinis}, J. and {Dolding}, C. and {Dole}, H. and {Douspis}, M. and {Dubath}, F. and {Duncan}, C.~A.~J. and {Dupac}, X. and {Dusini}, S. and {Ealet}, A. and {Escoffier}, S. and {Fabricius}, M. and {Farina}, M. and {Farinelli}, R. and {Faustini}, F. and {Ferriol}, S. and {Fotopoulou}, S. and {Fourmanoit}, N. and {Frailis}, M. and {Franceschi}, E. and {Franzetti}, P. and {Galeotta}, S. and {George}, K. and {Gillard}, W. and {Gillis}, B. and {Giocoli}, C. and {G{\'o}mez-Alvarez}, P. and {Gracia-Carpio}, J. and {Granett}, B.~R. and {Grazian}, A. and {Grupp}, F. and {Guzzo}, L. and {Gwyn}, S. and {Haugan}, S.~V.~H. and {Herent}, O. and {Hoar}, J. and {Hoekstra}, H. and {Holliman}, M.~S. and {Holmes}, W. and {Hook}, I.~M. and {Hormuth}, F. and {Hornstrup}, A. and {Hudelot}, P. and {Ili{\'c}}, S. and {Jhabvala}, M. and {Joachimi}, B. and {Keih{\"a}nen}, E. and {Kermiche}, S. and {Kiessling}, A. and {Kubik}, B. and {Kuijken}, K. and {K{\"u}mmel}, M. and {Kunz}, M. and {Kurki-Suonio}, H. and {Lahav}, O. and {Le Boulc'h}, Q. and {Le Brun}, A.~M.~C. and {Le Mignant}, D. and {Liebing}, P. and {Ligori}, S. and {Lilje}, P.~B. and {Lindholm}, V. and {Lloro}, I. and {Mainetti}, G. and {Maino}, D. and {Maiorano}, E. and {Mansutti}, O. and {Marcin}, S. and {Marggraf}, O. and {Markovic}, K. and {Martinelli}, M. and {Martinet}, N. and {Marulli}, F. and {Massey}, R. and {Maurogordato}, S. and {McCracken}, H.~J. and {Medinaceli}, E. and {Mei}, S. and {Melchior}, M. and {Mellier}, Y. and {Meneghetti}, M. and {Merlin}, E. and {Meylan}, G. and {Mora}, A. and {Moresco}, M. and {Morris}, P.~W. and {Moscardini}, L. and {Mourre}, S. and {Nakajima}, R. and {Neissner}, C. and {Nichol}, R.~C. and {Niemi}, S.-M. and {Nightingale}, J.~W. and {Nutma}, T. and {Padilla}, C. and {Paltani}, S. and {Pasian}, F. and {Peacock}, J.~A. and {Pedersen}, K. and {Percival}, W.~J. and {Pettorino}, V. and {Pires}, S. and {Polenta}, G. and {Pollack}, J.~E. and {Poncet}, M. and {Popa}, L.~A. and {Pozzetti}, L. and {Racca}, G.~D. and {Raison}, F. and {Rebolo}, R. and {Renzi}, A. and {Rhodes}, J. and {Riccio}, G. and {Rix}, H.-W. and {Romelli}, E. and {Roncarelli}, M. and {Rossetti}, E. and {Rusholme}, B. and {Saglia}, R. and {Sakr}, Z. and {S{\'a}nchez}, A.~G. and {Sapone}, D. and {Sartoris}, B. and {Sauvage}, M. and {Schewtschenko}, J.~A. and {Schneider}, P. and {Scodeggio}, M. and {Secroun}, A. and {Sefusatti}, E. and {Seidel}, G.},
        title = "{Euclid Quick Data Release (Q1): I. Data release overview}",
      journal = {\aap},
         year = 2026,
        month = jun,
       volume = {711},
          eid = {A1},
        pages = {A1},
          doi = {10.1051/0004-6361/202554610},
archivePrefix = {arXiv},
       eprint = {2503.15302},
 primaryClass = {astro-ph.GA},
       adsurl = {https://ui.adsabs.harvard.edu/abs/2026A&A...711A...1E}
}

@ARTICLE{2019ApJ...873..111I,
       author = {{Ivezi{\'c}}, {\v{Z}}eljko and {Kahn}, Steven M. and {Tyson}, J. Anthony and {Abel}, Bob and {Acosta}, Emily and {Allsman}, Robyn and {Alonso}, David and {AlSayyad}, Yusra and {Anderson}, Scott F. and {Andrew}, John and {Angel}, James Roger P. and {Angeli}, George Z. and {Ansari}, Reza and {Antilogus}, Pierre and {Araujo}, Constanza and {Armstrong}, Robert and {Arndt}, Kirk T. and {Astier}, Pierre and {Aubourg}, {\'E}ric and {Auza}, Nicole and {Axelrod}, Tim S. and {Bard}, Deborah J. and {Barr}, Jeff D. and {Barrau}, Aurelian and {Bartlett}, James G. and {Bauer}, Amanda E. and {Bauman}, Brian J. and {Baumont}, Sylvain and {Bechtol}, Ellen and {Bechtol}, Keith and {Becker}, Andrew C. and {Becla}, Jacek and {Beldica}, Cristina and {Bellavia}, Steve and {Bianco}, Federica B. and {Biswas}, Rahul and {Blanc}, Guillaume and {Blazek}, Jonathan and {Blandford}, Roger D. and {Bloom}, Josh S. and {Bogart}, Joanne and {Bond}, Tim W. and {Booth}, Michael T. and {Borgland}, Anders W. and {Borne}, Kirk and {Bosch}, James F. and {Boutigny}, Dominique and {Brackett}, Craig A. and {Bradshaw}, Andrew and {Brandt}, William Nielsen and {Brown}, Michael E. and {Bullock}, James S. and {Burchat}, Patricia and {Burke}, David L. and {Cagnoli}, Gianpietro and {Calabrese}, Daniel and {Callahan}, Shawn and {Callen}, Alice L. and {Carlin}, Jeffrey L. and {Carlson}, Erin L. and {Chandrasekharan}, Srinivasan and {Charles-Emerson}, Glenaver and {Chesley}, Steve and {Cheu}, Elliott C. and {Chiang}, Hsin-Fang and {Chiang}, James and {Chirino}, Carol and {Chow}, Derek and {Ciardi}, David R. and {Claver}, Charles F. and {Cohen-Tanugi}, Johann and {Cockrum}, Joseph J. and {Coles}, Rebecca and {Connolly}, Andrew J. and {Cook}, Kem H. and {Cooray}, Asantha and {Covey}, Kevin R. and {Cribbs}, Chris and {Cui}, Wei and {Cutri}, Roc and {Daly}, Philip N. and {Daniel}, Scott F. and {Daruich}, Felipe and {Daubard}, Guillaume and {Daues}, Greg and {Dawson}, William and {Delgado}, Francisco and {Dellapenna}, Alfred and {de Peyster}, Robert and {de Val-Borro}, Miguel and {Digel}, Seth W. and {Doherty}, Peter and {Dubois}, Richard and {Dubois-Felsmann}, Gregory P. and {Durech}, Josef and {Economou}, Frossie and {Eifler}, Tim and {Eracleous}, Michael and {Emmons}, Benjamin L. and {Fausti Neto}, Angelo and {Ferguson}, Henry and {Figueroa}, Enrique and {Fisher-Levine}, Merlin and {Focke}, Warren and {Foss}, Michael D. and {Frank}, James and {Freemon}, Michael D. and {Gangler}, Emmanuel and {Gawiser}, Eric and {Geary}, John C. and {Gee}, Perry and {Geha}, Marla and {Gessner}, Charles J.~B. and {Gibson}, Robert R. and {Gilmore}, D. Kirk and {Glanzman}, Thomas and {Glick}, William and {Goldina}, Tatiana and {Goldstein}, Daniel A. and {Goodenow}, Iain and {Graham}, Melissa L. and {Gressler}, William J. and {Gris}, Philippe and {Guy}, Leanne P. and {Guyonnet}, Augustin and {Haller}, Gunther and {Harris}, Ron and {Hascall}, Patrick A. and {Haupt}, Justine and {Hernandez}, Fabio and {Herrmann}, Sven and {Hileman}, Edward and {Hoblitt}, Joshua and {Hodgson}, John A. and {Hogan}, Craig and {Howard}, James D. and {Huang}, Dajun and {Huffer}, Michael E. and {Ingraham}, Patrick and {Innes}, Walter R. and {Jacoby}, Suzanne H. and {Jain}, Bhuvnesh and {Jammes}, Fabrice and {Jee}, M. James and {Jenness}, Tim and {Jernigan}, Garrett and {Jevremovi{\'c}}, Darko and {Johns}, Kenneth and {Johnson}, Anthony S. and {Johnson}, Margaret W.~G. and {Jones}, R. Lynne and {Juramy-Gilles}, Claire and {Juri{\'c}}, Mario and {Kalirai}, Jason S. and {Kallivayalil}, Nitya J. and {Kalmbach}, Bryce and {Kantor}, Jeffrey P. and {Karst}, Pierre and {Kasliwal}, Mansi M. and {Kelly}, Heather and {Kessler}, Richard and {Kinnison}, Veronica and {Kirkby}, David and {Knox}, Lloyd and {Kotov}, Ivan V. and {Krabbendam}, Victor L. and {Krughoff}, K. Simon and {Kub{\'a}nek}, Petr and {Kuczewski}, John and {Kulkarni}, Shri and {Ku}, John and {Kurita}, Nadine R. and {Lage}, Craig S. and {Lambert}, Ron and {Lange}, Travis and {Langton}, J. Brian and {Le Guillou}, Laurent and {Levine}, Deborah and {Liang}, Ming and {Lim}, Kian-Tat and {Lintott}, Chris J. and {Long}, Kevin E. and {Lopez}, Margaux and {Lotz}, Paul J. and {Lupton}, Robert H. and {Lust}, Nate B. and {MacArthur}, Lauren A. and {Mahabal}, Ashish and {Mandelbaum}, Rachel and {Markiewicz}, Thomas W. and {Marsh}, Darren S. and {Marshall}, Philip J. and {Marshall}, Stuart and {May}, Morgan and {McKercher}, Robert and {McQueen}, Michelle and {Meyers}, Joshua and {Migliore}, Myriam and {Miller}, Michelle and {Mills}, David J.},
        title = "{LSST: From Science Drivers to Reference Design and Anticipated Data Products}",
      journal = {\apj},
         year = 2019,
        month = mar,
       volume = {873},
       number = {2},
          eid = {111},
        pages = {111},
          doi = {10.3847/1538-4357/ab042c},
archivePrefix = {arXiv},
       eprint = {0805.2366},
 primaryClass = {astro-ph},
       adsurl = {https://ui.adsabs.harvard.edu/abs/2019ApJ...873..111I}
}

@ARTICLE{2016ARA&A..54..667S,
       author = {{Shu}, Frank H.},
        title = "{Six Decades of Spiral Density Wave Theory}",
      journal = {\araa},
         year = 2016,
        month = sep,
       volume = {54},
        pages = {667-724},
          doi = {10.1146/annurev-astro-081915-023426},
       adsurl = {https://ui.adsabs.harvard.edu/abs/2016ARA&A..54..667S}
}

@ARTICLE{2022MNRAS.517L.132K,
       author = {{Karapetyan}, Arpine G.},
        title = "{Constraining Type Ia supernovae via their distances from spiral arms}",
      journal = {\mnras},
         year = 2022,
        month = nov,
       volume = {517},
       number = {1},
        pages = {L132-L137},
          doi = {10.1093/mnrasl/slac121},
archivePrefix = {arXiv},
       eprint = {2209.14796},
 primaryClass = {astro-ph.GA},
       adsurl = {https://ui.adsabs.harvard.edu/abs/2022MNRAS.517L.132K}
}

@ARTICLE{2024A&A...692A.159S,
       author = {{Semczuk}, Marcin and {Dehnen}, Walter and {Sch{\"o}nrich}, Ralph and {Athanassoula}, E.},
        title = "{Pattern speed evolution of barred galaxies in TNG50}",
      journal = {\aap},
         year = 2024,
        month = dec,
       volume = {692},
          eid = {A159},
        pages = {A159},
          doi = {10.1051/0004-6361/202451521},
archivePrefix = {arXiv},
       eprint = {2407.11154},
 primaryClass = {astro-ph.GA},
       adsurl = {https://ui.adsabs.harvard.edu/abs/2024A&A...692A.159S}
}

\begin{appendix}
\nolinenumbers

\section{Analysis using the stricter criterion of EW(H$\alpha$) $>$ 3\,\AA}
\label{app:stricterEW}

\begin{table}[h!]
\caption{Overall AGN fractions obtained using
EW(H$\alpha$) $>3\,\AA$.}
\label{tab:ew3_overall}
\centering
\begin{tabular}{lrrc}
\hline\hline
Class & $N_{\rm gal}$ & $N_{\rm AGN}$ & $f_{\rm AGN}$ \\
\hline
Unbarred        & 298 & 15 & $0.05^{+0.02}_{-0.01}$ \\
Weakly barred   & 269 & 23 & $0.09^{+0.02}_{-0.01}$ \\
Strongly barred & 198 & 49 & $0.25^{+0.03}_{-0.03}$ \\
\hline
FL              & 281 & 14 & $0.05^{+0.02}_{-0.01}$ \\
GM              & 484 & 73 & $0.15^{+0.02}_{-0.02}$ \\
\hline
\end{tabular}
\end{table}

To assess the effect of the adopted emission-line EW
threshold, we repeated the entire analysis using the more conservative
criterion of EW(H$\alpha$) $>$ 3\,\AA\,(\citetalias{2023A&A...674A..85A}),
instead of the fiducial value of
EW(H$\alpha$) $>$ 1.5\,\AA \,\citep{2010MNRAS.403.1036C}.
The stricter criterion reduces the final
sample from 843 to 765 galaxies and the number of AGN hosts from 165
to 87, that is, by nearly a factor of two. The corresponding sample
sizes and overall AGN fractions are listed in Table~\ref{tab:ew3_overall}.
The stellar-mass--color matching procedure was repeated for the reduced sample before performing the subsequent statistical analyses.

For the bar analysis, only bins $\mathbb{D}$ and $\mathbb{G}$ satisfy the matching criteria.
In bin $\mathbb{D}$, the AGN fractions are $0.03^{+0.03}_{-0.01}$ ($2/80$), $0.02^{+0.03}_{-0.01}$ ($1/64$), and $0.21^{+0.10}_{-0.06}$ ($5/24$) for unbarred, weakly barred, and strongly barred galaxies, respectively. Strongly barred galaxies retain significantly higher AGN fractions than both unbarred galaxies ($P_{\rm B}=0.027$) and weakly barred galaxies ($P_{\rm B}=0.027$), whereas the weakly barred--unbarred difference is not significant. Thus, the principal bar-related result remains unchanged when the stricter EW(H$\alpha$) criterion is applied.

In bin $\mathbb{G}$, the corresponding AGN fractions are
$0.00^{+0.10}_{-0.00}$ ($0/16$),
$0.29^{+0.20}_{-0.11}$ ($2/7$), and
$0.21^{+0.09}_{-0.05}$ ($7/34$) for unbarred, weakly barred,
and strongly barred galaxies, respectively, but the statistical evidence is weak because of the small subsample sizes. The weakly barred--unbarred comparison gives $P_{\rm B}=0.036$, while the other pairwise differences are not statistically significant.

The AGN fraction remains higher for GM than for FL galaxies in bins $\mathbb{D-G}$, while the trend is reversed in
bin $\mathbb{H}$. After the renewed mass--color validation, none of the individual matched bins shows a statistically significant FL--GM difference according to Barnard's exact test.

For the combined $\mathbb{E+F}$ comparison, the AGN fractions are
$0.13^{+0.06}_{-0.03}$ ($7/52$) for FL galaxies and
$0.26^{+0.04}_{-0.03}$ ($37/140$) for GM galaxies.
Although the same qualitative trend is preserved, the difference
is only marginally significant ($P_{\rm B}=0.060$).

Finally, we repeated the combined arm--bar analysis. In bin $\mathbb{D}$, the AGN
fractions of unbarred, weakly barred, and strongly barred GM galaxies
are $0.02^{+0.04}_{-0.01}$ ($1/54$),
$0.02^{+0.04}_{-0.01}$ ($1/48$), and
$0.28^{+0.12}_{-0.08}$ ($5/18$), respectively. Strongly barred
GM galaxies retain significantly higher AGN fractions than both
unbarred GM galaxies ($P_{\rm B}=0.006$) and
weakly barred GM galaxies ($P_{\rm B}=0.008$).
The difference between unbarred FL and strongly barred GM galaxies
remains marginally significant ($P_{\rm B}=0.057$).

Overall, adopting EW(H$\alpha$) $>3\,\AA$ substantially reduces the
sample size and nearly halves the number of AGN hosts. Nevertheless,
the principal bar-related results remain statistically
significant. The higher AGN fraction of GM galaxies relative to FL galaxies
is also qualitatively preserved, although the combined $\mathbb{E+F}$ comparison
is now only marginally significant.
The principal joint bar--spiral arm result remains marginally significant.
This indicates that the
weaker spiral arm result is more sensitive to the reduction in
sample size than to the adopted EW(H$\alpha$) threshold.

\section{External validation of the visual arm classification}
\label{app:validation}

\begin{figure}[h!]
\centering
\includegraphics[width=0.95\hsize]{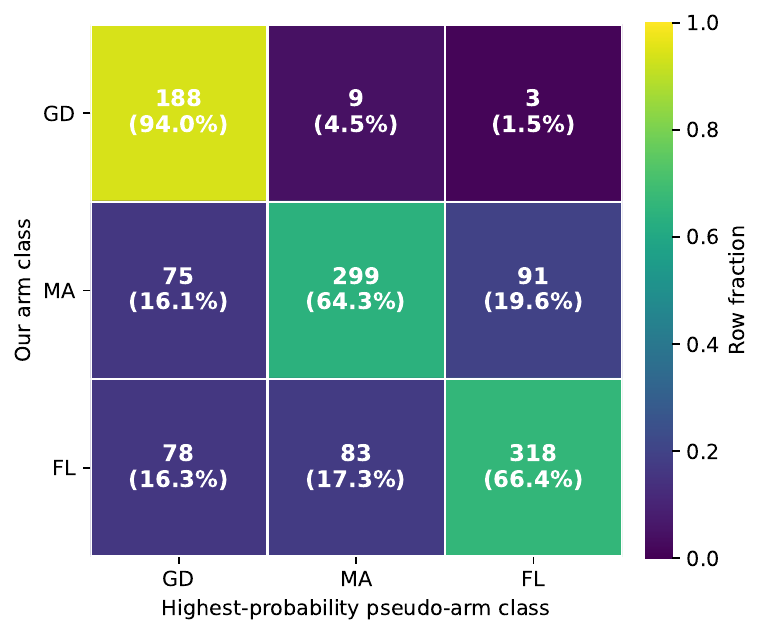}
\caption{
Confusion matrix comparing our visual ACs with the Galaxy Zoo pseudo-ACs for 1,144 matched galaxies. Each cell gives the number of galaxies and the corresponding row-normalized percentage. Rows represent our visual classifications, while columns correspond to the pseudo-ACs derived from the Galaxy Zoo arm-multiplicity probabilities of \citet{2021MNRAS.507.3923M}.
}
\label{matrix}
\end{figure}

As an independent external validation of our visual spiral ACs, we compared them with the Galaxy Zoo arm-multiplicity probabilities reported by \citet{2021MNRAS.507.3923M}. Since Galaxy Zoo characterizes galaxies according to the number of visible spiral arms rather than according to the Elmegreen ACs, we defined three approximate pseudo-ACs:
\[
\begin{aligned}
& P_{\rm GD} = P({\rm 2\ arms}),\\
& P_{\rm MA} = P({\rm 3\ arms}) + P({\rm 4\ arms}) + P({\rm >4\ arms}),\\
& P_{\rm FL} = P({\rm 1\ arm}) + P({\rm cannot\ tell}).
\end{aligned}
\]
These pseudo-ACs are intended only as an approximate statistical representation of the Elmegreen ACs because arm multiplicity and spiral arm morphology are related but non-equivalent quantities.

Of the 1,200 galaxies with visual arm classifications,
1,144 were matched to the Galaxy Zoo catalog and used for the comparison.
For each galaxy, the pseudo-AC with the highest combined probability was adopted. There are 13 cases in which two pseudo-ACs have equal maximum probabilities. When our visual class is among the tied classes, the corresponding pseudo-AC was adopted. Otherwise, the tied class closest to our visual classification along the GD--MA--FL sequence was selected.
In the single case involving an MA galaxy with equal GD and FL probabilities, GD was adopted because GD and MA were combined into
the GM category in the final analysis.

Figure~\ref{matrix} presents the confusion matrix between our visual arm classifications and the resulting pseudo-ACs. The overall agreement between the two classification schemes is 805/1,144 (70.4\%). The corresponding row-normalized agreement fractions are 94.0\% for GD, 64.3\% for MA, and 66.4\% for FL galaxies.

The mean values of $P_{\rm GD}, P_{\rm MA}, P_{\rm FL}$ are $0.832, 0.096, 0.072$ for our GD galaxies, $0.203, 0.533, 0.264$ for our MA galaxies, and $0.227, 0.233, 0.540$ for our FL galaxies.
In each visually classified subsample, the largest mean pseudo-arm probability corresponds to the expected class.
We used the Kruskal--Wallis test to assess whether the Galaxy Zoo probabilities $P_{\rm GD}$, $P_{\rm MA}$, and $P_{\rm FL}$ follow the same distribution across our visually classified GD, MA, and FL subsamples. The null hypothesis is rejected for all three probabilities ($P<2\times10^{-90}$), demonstrating that the Galaxy Zoo vote distributions differ significantly among our visual ACs.

The imperfect object-by-object agreement is expected because Galaxy Zoo arm multiplicity and the Elmegreen arm classification describe related but intrinsically different aspects of spiral structure. Galaxy Zoo primarily quantifies the number of visible spiral arms, whereas the Elmegreen classification additionally incorporates arm continuity, symmetry, and global spiral organization. Therefore, perfect one-to-one agreement is not expected, while the observed level of agreement demonstrates good statistical consistency between the two classification schemes.

As an additional robustness check, we repeated the spiral arm and
combined bar--spiral arm analyses presented in
Sects.~\ref{sec:DW} and \ref{sec:combined}, respectively, using the
Galaxy Zoo pseudo-ACs. Of the 843 galaxies in our redshift-limited
sample, 806 have usable Galaxy Zoo arm-multiplicity information and
were included in these analyses. The principal qualitative trends and
the statistical significance of all other comparisons are preserved.
The only exception is the combined $\mathbb{E}+\mathbb{F}$ comparison
in Sect.~\ref{sec:DW}, for which the difference between the GM and FL
classes changes from statistically significant
($P_{\rm B}=0.038$) when using our visual ACs to marginally significant
($P_{\rm B}=0.078$) when using the pseudo-ACs. Recall that, as noted
above, the pseudo-ACs are not equivalent to the Elmegreen ACs because
arm multiplicity and spiral arm morphology represent related but
distinct properties.

 \section{Representativeness of the classified sample}
\label{app:representative}

To assess whether the galaxies with secure visual spiral arm classifications constitute a representative subset of the parent sample, we compared the 1,200 classified galaxies with the 1,742 galaxies for which no reliable arm classification could be assigned, as well as with the full parent sample of 2,942 galaxies, in terms of stellar mass, rest-frame $(u-r)_0$ color, bar class, and AGN incidence.
The classified, unclassified, and full samples have mean (median) stellar masses of 10.52 (10.56), 10.39 (10.38), and 10.44 (10.45), with corresponding standard deviations of 0.55, 0.61, and 0.59 dex. Their mean (median) Galactic-extinction- and K-corrected $(u-r)_0$ colors are 1.81 (1.75), 1.91 (1.88), and 1.87 (1.81), with corresponding standard deviations of 0.39, 0.55, and 0.49. Importantly, the differences in the mean stellar masses and colors are small compared with the intrinsic dispersions of the corresponding distributions.
In addition, two-sample KS and AD tests comparing the stellar-mass and $(u-r)_0$
color distributions of the classified and unclassified samples yield
$P>0.1$ in all cases, indicating no statistically significant
differences between the samples in either stellar mass or color.

The AGN fractions are nearly identical in the three samples, amounting to 15.5\%, 15.2\%, and 15.3\% for the classified, unclassified, and full samples, respectively. Similarly, the bar-class distributions are essentially unchanged. The fractions of strongly barred galaxies are 25.58\%, 25.60\%, and 25.60\%; those of weakly barred galaxies are 37.92\%, 37.89\%, and 37.90\%; and those of unbarred galaxies are 36.50\%, 36.51\%, and 36.51\%, respectively.
In addition, Barnard's exact tests show that the relative numbers of
strongly barred, weakly barred and unbarred galaxies, as well as those of AGN and non-AGN
galaxies, do not differ significantly between the classified and
unclassified samples, with the corresponding $P$ values being close
to unity.

Overall, these results demonstrate that the sample of 1,200 galaxies with secure spiral arm classifications is broadly representative of the full parent sample of 2,942 galaxies.
We therefore do not expect the selection of galaxies with reliable arm classifications to affect the principal conclusions of the paper.

\section{Robustness tests using the weighting procedure of Garland et al. (2024)}
\label{app:garland}

To assess the robustness of our conclusions against an alternative approach for controlling the dependence of AGN incidence on stellar mass and galaxy color, we repeated the principal analyses using a weighting procedure similar to that proposed by \citet{2024MNRAS.532.2320G}.
Unlike our matched-control analysis, this method assigns statistical weights to galaxies such that the compared morphological (bar and arm classifications) subsamples reproduce the same stellar-mass--color distribution.

For each morphological (bar and arm) class $k$, galaxies belonging to stellar-mass--color bin $b$ are assigned the weight
\[
w_{kb}=\frac{N_b/N}{N_{kb}/N_k},
\]
where $N_b/N$ is the fraction of the combined comparison sample in bin $b$, and $N_{kb}/N_k$ is the corresponding fraction within morphological (bar and arm) class $k$. The weighted AGN fraction is then calculated as
\[
f_{{\rm AGN},k}^{\rm w}
=
\frac{\sum_i w_iA_i}{\sum_i w_i},
\]
where $A_i=1$ for AGN hosts and $A_i=0$ otherwise.
Uncertainties were estimated from 10,000 bootstrap realizations, with the weights recalculated independently in each realization.
Because this procedure corrects differences in the relative occupancy of the stellar-mass--color bins, but cannot remove residual differences within an individual bin, only bins satisfying the matching criteria established in the main analysis were retained.

For the bar analysis, bins $\mathbb{D, F, G}$, and $\mathbb{H}$ satisfy the matching criteria. Bins $\mathbb{A-C}$ together contain only one AGN among 211 galaxies and therefore essentially provide no discriminatory information on the dependence of AGN incidence on bar class, whereas bin $\mathbb{E}$ is excluded because the KS and AD tests reveal significant residual differences in stellar mass or color between at least one pair of bar classes. The resulting weighted AGN fractions are $0.167^{+0.033}_{-0.032}$, $0.319^{+0.042}_{-0.042}$, and $0.328^{+0.046}_{-0.045}$ for unbarred, weakly barred, and strongly barred galaxies, respectively. Relative to unbarred galaxies, both weakly and strongly barred systems retain significantly higher AGN fractions ($\Delta f_{\rm AGN}^{\rm w}=0.153$, 95\% interval 0.050 -- 0.255, and $\Delta f_{\rm AGN}^{\rm w}=0.161$, interval 0.053 -- 0.269, respectively), whereas the difference between weakly and strongly barred galaxies is not significant ($\Delta f_{\rm AGN}^{\rm w}=0.008$, interval $-0.111$ -- 0.130).

For the spiral arm analysis, bins $\mathbb{D, E, F, G}$, and $\mathbb{H}$ satisfy the matching criteria, while bins $\mathbb{A-C}$ were excluded because they contain only one AGN. The corresponding weighted AGN fractions are $0.210^{+0.035}_{-0.034}$ for FL galaxies and $0.297^{+0.023}_{-0.023}$ for GM galaxies. The weighted difference is $\Delta f_{\rm AGN}^{\rm w}=0.087$ with a 95\% bootstrap interval of 0.005 -- 0.163. Thus, after matching the stellar-mass and color distributions, GM galaxies retain a weak enhancement of AGN incidence relative to FL galaxies, although the significance of the difference is only marginal.

Finally, the combined arm--bar analysis was performed using bins $\mathbb{D, F, G}$, and $\mathbb{H}$, which satisfy the matching criteria for all six structural classes. The resulting weighted AGN fractions are $0.184^{+0.058}_{-0.054}$, $0.303^{+0.099}_{-0.104}$, and $0.245^{+0.080}_{-0.076}$ for FL(unbar), FL(weak), and FL(strong), respectively, and $0.156^{+0.043}_{-0.041}$, $0.322^{+0.046}_{-0.047}$, and $0.364^{+0.055}_{-0.054}$ for GM(unbar), GM(weak), and GM(strong), respectively. No statistically significant dependence on bar class is detected within the FL population. In contrast, both weakly and strongly barred GM galaxies retain significantly higher AGN fractions than unbarred GM galaxies, whereas the difference between weakly and strongly barred GM galaxies is not significant. Likewise, no statistically significant GM--FL difference is found when the bar class is held fixed. Nevertheless, the strongest contrast remains between strongly barred GM galaxies and unbarred FL galaxies, with $\Delta f_{\rm AGN}^{\rm w}=0.178$ (95\% bootstrap interval: 0.029 -- 0.325).
Overall, the weighting analyses lead to the same qualitative conclusions as the matched-control analyses, demonstrating that our principal results are robust against the choice of method used to control for stellar mass and galaxy color.

\end{appendix}

\end{document}